\documentclass[12pt,a4paper]{article}
\usepackage{type1cm,amsmath,hangcaption,graphicx,indentfirst}
\usepackage[psamsfonts]{amssymb}
\numberwithin{equation}{section}

\newcommand{\C}{\mathbb{C}}
\newcommand{\R}{\mathbb{R}}
\newcommand{\sgn}{\text{sgn}}
\newcommand{\del}{\partial}
\newcommand{\str}{\text{str}}

\begin{document}
\begin{titlepage}

 \renewcommand{\thefootnote}{\fnsymbol{footnote}}
\begin{flushright}
 \begin{tabular}{l}
 arXiv:1004.1977\\ 
 \end{tabular}
\end{flushright}

 \begin{center}

 \vskip 2.5 truecm

\noindent{\large \textbf{Branes in the OSP(1$|$2) WZNW model}}\\
\vspace{1.5cm}

\noindent{ Thomas Creutzig$^a$\footnote{E-mail: creutzig@physics.unc.edu} and Yasuaki Hikida$^b$\footnote{E-mail:
hikida@phys-h.keio.ac.jp}}
\bigskip

 \vskip .6 truecm
\centerline{\it $^a$Department of Physics and Astronomy, University of North Carolina,}
\centerline{\it Phillips Hall, CB 3255, Chapel Hill, NC 27599-3255, USA}
\bigskip
\centerline{\it $^b$Department of Physics, and Research and Education
Center for Natural Sciences,}
\centerline{\it  Keio University, Hiyoshi, Yokohama 223-8521, Japan}

 \vskip .4 truecm

 \end{center}

 \vfill
\vskip 0.5 truecm

\begin{abstract}

The boundary OSP(1$|$2) WZNW model possesses two types of branes, which are localized on supersymmetric Euclidean AdS$_2$ and on two-dimensional superspheres.
We compute the coupling of closed strings to these branes with two different methods. The first one uses factorization constraints and
the other one a correspondence to boundary $\mathcal{N}=1$ super-Liouville field theory, which we proof with path integral techniques.
 We check that the results obey the Cardy condition and reproduce the semi-classical computations.
For the check we also compute the spectral density of open strings that are attached to the non-compact branes.

\end{abstract}
\vfill
\vskip 0.5 truecm

\setcounter{footnote}{0}
\renewcommand{\thefootnote}{\arabic{footnote}}
\end{titlepage}

\newpage

\tableofcontents
\section{Introduction}
\label{Intoduction}

Conformal field theories (CFTs) with target-space supersymmetry play an important role in areas as superstring theory and condensed matter physics.
In particular two-dimensional models are building blocks of superstring theories with AdS-background, which are important for the
AdS/CFT correspondence \cite{Maldacena}.
Moreover, the computation of spectral densities and transport properties in systems with random disorder involve theories with internal
supersymmetry \cite{Efetov}.
In both cases orthosymplectic supergroup symmetries appear frequently.
Especially the coset model OSP(1$|$2)/U(1) with
target-space and world-sheet supersymmetries has been recently proposed in
\cite{GHT} to describe the two-dimensional superstring,
which is holographically dual to a Hermitian matrix model \cite{TT,Matrix}.

Two-dimensional conformal field theories with
target-space supersymmetry are non-unitary and non-rational.
A class of such models are Wess-Zumino-Novikov-Witten (WZNW) models with Lie supergroup target.
Recently, a variety of methods were developed to treat these models. The idea is to reduce the model to a simpler problem that is already understood.
Most importantly the  computation of correlation functions in WZNW models of unitary (type I) supergroups can be reduced to computations on the WZNW model of its bosonic subgroup plus some free fermions \cite{Quella:2007hr}.
Unfortunately these free-field methods do not carry over to the family of orthosymplectic (type II) supergroups.
In this case an idea is to generalize the correspondence between Euclidean AdS and bosonic Liouville theory \cite{TeschnerH,TeschnerL,RT,HS}.
For this the path integral derivation of the correspondence \cite{HS} is very suitable.
Indeed it is shown that the correlation functions of OSP(N$|$2) WZNW models can be computed in terms of
$\mathcal{N}=($N$,$N$)$ super-Liouville field theory \cite{HS2}.%
\footnote{Structure constants for $\text{N}=1$ case have been obtained
explicitly in \cite{HS2} with the knowledge of those of super-Liouville field theory \cite{RS,Poghosian}, see also \cite{FH}.}

It is a natural task to extend these works to world-sheets with boundary. We will recall some of their features.
So far, the boundary GL(1$|$1) WZNW model is well understood, i.e. boundary states and correlation functions are computed
\cite{CQS,Creutzig:2008an,Creutzig:2008ek,Creutzig:2009zz}.
The computation of correlation functions can be performed in
a similar free fermion realization as in the bulk case of type I supergroup.
The difficulty in establishing the formalism is to find the appropriate boundary term of the action and it involves
additional fermionic boundary degrees of freedom.
This situation is known from world-sheet supersymmetric models as super-Liouville field theory
\cite{FH,Hosomichi:2004ph,Ahn:2003wy} and in general from matrix factorization \cite{Warner:1995ay}.
It may indicate that the OSP(N$|$2) to $\mathcal{N}=($N$,$N$)$ super-Liouville correspondence also holds for the boundary models.
Beyond the GL(1$|$1) WZNW models,
some boundary spectra of PSL(2$|$2) sigma models have been studied \cite{Quella:2007sg}.
Furthermore, it is known that branes on super groups are described by twisted super conjugacy classes \cite{Creutzig:2008ag}.

In this work we combine our experience from bulk OSP(N$|$2) WZNW models and boundary supergroup models to understand the boundary OSP(1$|$2) WZNW model.
We find that there are two classes of branes in the OSP(1$|$2) WZNW model. We call the first one (super) AdS$_2$ branes, because the geometry of their bosonic subspace is AdS$_2$.
These branes have super-dimension $2|2$. The other class are spherical ones, also of super-dimension $2|2$.
When the sphere degenerates to a point, there is as well a point-like brane as a two-dimensional one in the fermionic directions.

Branes are characterized by their coupling to closed strings. These couplings are expressed in boundary conformal field theory by bulk one-point functions
on a disk or equivalently by the boundary state.
One the other hand,  with the boundary state we can compute
the partition function of open strings ending on the branes via world-sheet duality \cite{Cardy:1989ir}.
In general its computation is more involved than in rational boundary CFTs,
since the boundary state involves a non-trivial spectral density.
This happens when the open string spectrum is continuous, which is usually the case in non-compact and non-rational theories.
The spectral density can be obtained from the reflection amplitude
describing scattering of open strings and thus contains information about this process.

In this article, we compute these quantities in two different ways.
The first one is to proceed in analogy with its bosonic cousin,
the $H_3^+$ model, or Euclidean AdS$_3$ model.
In order to compute bulk one-point functions on a disk,
we utilize two-point functions with a degenerate operator.
They can be factorized into two ways, and constraints can be
obtained by comparing the two expressions. The one-point
functions are then obtained by solving these constraints.
We test these one-point functions by careful analysis of their classical limits.
For the super AdS$_2$ branes we determine the
spectral density of open strings between them and check that it is consistent with the one-point functions via world-sheet duality.
The second approach is less direct. We extend the correspondence between $\mathcal{N}=1$ super-Liouville field theory and OSP(1$|$2) model
to the boundary case.
We can then use the results of boundary $\mathcal{N}=1$ super-Liouville field theory in, e.g., \cite{FH}.
One advantage to the previous approach is that we can easily
obtain one-point functions of bulk operators in the NSNS-sector as well.%
\footnote{In this paper we call the NS-sector as the one with the
anti-periodic boundary conditios for worldsheet fermions. If we would like to treat
fermions as the Grassmann odd coordinates of target-space, then
they should satisfy the periodic boundary condition, in other words,
they are in the R-sector.}

The article is organized as follows. We start in section 2 with some geometric considerations. In particular, we find a semi-classical expression for the boundary states. This allows us
to read off the semi-classical limit of closed string couplings to the branes.
In section 3 we review the basic properties of bulk OSP(1$|$2) WZNW model.
In particular, we show how OSP(1$|$2) symmetry restricts the form of
correlation functions generically. Previous results of two and three
point functions in \cite{HS2} are also given.
In section \ref{sec:BOSP}, they are
used to obtain bulk one-point functions in the boundary theory.
Then in section 5 we derive a correspondence between the boundary OSP(1$|$2) theory describing super AdS$_2$ branes and $\mathcal{N}=1$ super-Liouville theory with boundary.
In section 6 we verify that our results agree with world-sheet duality.
Namely, the bulk one-point functions define the boundary states, whose modular S-transformation leads to the open string partition functions.
Section 7 concludes with a summary of results and a list of interesting open problems.
The appendices contain detailed computations of correlation functions and a derivation of the action of the OSP(1$|$2) boundary theory describing
super AdS$_2$ branes.

\section{The semi-classical limit of branes}

The analysis of the branes' geometry already provides useful information of the closed string.
The boundary states contain the information of the closed string couplings to the brane. In the semi-classical limit this boundary state becomes a delta-distribution localized on the brane. The aim of this section is to rewrite these semi-classical boundary states in such a way that we can identify the semi-classical limit of the closed strings couplings to the brane.

The results are as follows.
We will find two types of branes, whose bosonic subspaces are AdS$_2$ and spherical ones. The geometry of the spherical ones are superconjugacy classes of OSP(1$|$2), while the AdS$_2$ branes are twisted superconjugacy classes. In both cases, we can characterize the branes by the position $a$ of the (twisted) super conjugacy class. The delta-distribution then has the following form
\begin{equation}\label{deltadistribution}
\delta(g-a) \ = \
 \int dj \ \int d^2ud^2\lambda \ (\Phi_h(u,\lambda|g))^*\ \langle \Phi_h(u,\lambda|z)\rangle_a\, ,
\end{equation}
where $\Phi_h(u,\lambda|g)$ are eigenfunctions of the Laplacian%
\footnote{The dual function is $(\Phi_h(u,\lambda|g))^*=\Phi_{-h+1/2}(u,\lambda|g)$.} and $\langle \Phi_h(u,\lambda|z)\rangle_a$ are the corresponding closed string couplings in the semiclassical limit.
For the AdS$_2$ branes they are precisely
\begin{equation}\label{1ptclassicalads}
\begin{split}
\langle \Phi_h(x,\lambda|z)\rangle^{\text{AdS}}_{r,+} \ \ \ &\substack{k\,\rightarrow\,\infty\\ \sim}\ \ \
|x+\bar x+\lambda\bar\lambda|^{-2h}e^{\sgn(x+\bar x)r(-2h+1/2)}\qquad\quad\text{and}\\
\langle \Phi_h(x,\lambda|z)\rangle^{\text{AdS}}_{r,-} \ \ \ &\substack{k\,\rightarrow\,\infty\\ \sim}\ \ \
\sgn(x+\bar x)|x+\bar x-\lambda\bar\lambda|^{-2h}e^{\sgn(x+\bar x)r(-2h+1/2)}\, ,
\end{split}
\end{equation}
and for the spherical ones
\begin{equation}\label{1ptclassicalsphere}
\begin{split}
\langle \Phi_h(u,\lambda|z)\rangle^\text{sphere}_{\Lambda_0,-} \ \ \ &\substack{k\,\rightarrow\,\infty\\ \sim}\ \ \
|1+u\bar u+\lambda\lambda^\sharp|^{-2h}\sinh(\Lambda_0(2h-1/2))\qquad\text{and}\\
\langle \Phi_h(u,\lambda|z)\rangle^\text{sphere}_{\Lambda_0,+} \ \ \ &\substack{k\,\rightarrow\,\infty\\ \sim}\ \ \
|1+u\bar u-\lambda\lambda^\sharp|^{-2h}\cosh(\Lambda_0(2h-1/2))\, .
\end{split}
\end{equation}
The remainder of this section is the explanation of \eqref{1ptclassicalads} and \eqref{1ptclassicalsphere}. The analysis is similar to the one of its bosonic cousin, the $H_3^+$ model \cite{PST}.

\subsection{Geometry of the branes}

In this subsection we describe the geometry of the branes in OSP(1$|$2) following the general analysis of branes on supergroups \cite{Creutzig:2008ag}.
We want to describe branes that are maximally symmetry preserving, i.e they preserve conformal symmetry as well as the Lie super algebra  current symmetry. Such branes are described by automorphisms $\omega$ of the Lie superalgebra which respect its invariant metric. The left and right moving currents are glued together along the boundary of the worldsheet with such an automorphism ensuring current and Virasoro symmetry on the boundary.
We want to describe branes that are maximally symmetry preserving, i.e they preserve conformal symmetry as well as the Lie super algebra  current symmetry. Such branes are described by automorphisms $\omega$ of the Lie superalgebra which respect its invariant metric. The left and right moving currents are glued together along the boundary of the world-sheet with such an automorphism ensuring current and Virasoro symmetry on the boundary.
Similar to Lie group WZNW models \cite{Alekseev:1998mc} these gluing conditions describe branes localized at twisted super conjugacy classes
\begin{equation}
	C^\omega_a\ = \ \{\ \omega(b)ab^{-1} \ | \ b \ \in \ G \ \}\, .
\end{equation}

We turn to OSP(1$|$2).
The Lie superalgebra osp(m$|$2n) is most conveniently expressed in a matrix representation as 
\begin{equation}
	\begin{split}
		\text{osp}(m|2n)=\bigl\{ X \ \in \text{gl}(m|2n) \ | \ X^{st}B_{m,n}+B_{m,n}X=0\bigr\} \, ,
	\end{split}
\end{equation}
where the supertranspose is given in \eqref{eq:supertranspose}
and
\begin{equation}
	\begin{split}
		B_{m,n}=\left(\begin{array}{cc}1_m & 0 \\ 0 & J_n\end{array}\right) 
\, , \   \text{where} \
			J_n=\left(\begin{array}{cc}0 & 1_n \\ -1_n & 0\end{array}\right) \, .
	\end{split}
\end{equation}
Thus osp(1$|$2) is represented by matrices as
\begin{equation}
	\begin{split}
		A\ = \ \left(\begin{array}{ccc}0 & \theta_- & \theta_+ \\
                             \theta_+ & a & b \\
                             -\theta_- & c & -a \end{array}\right)\, .
	\end{split}
\end{equation}
We obtain elements of OSP(1$|$2) by exponentiation. Moreover, we require that our OSP(1$|$2) valued fields are super-hermitean, i.e. they satisfy
$g=g^\ddag$. The super-hermitean conjugate $\ddag$ is complex conjugation concanated with supertransposition. Here complex conjugation for Grassmann numbers is the super star operation $\sharp$, see appendix \ref{app:complex}. Strictly speaking this means that our model takes values in the coset of super-hermitean matrices. This situation is analogous to Euclidean AdS$_3$ or $H_3^+$, which is the coset of SL(2) consisting of hermitean matrices, and it means that the bosonic subspace of our model is Euclidean AdS$_3$.

Let us turn to gluing automorphisms. All automorphisms of osp(1$|$2) are inner. Nonetheless, the geometry of branes corresponding to different gluing maps can differ.
The reason is that the automorphisms are not related as automorphisms of the coset of super-hermitean elements. The same situation appears in the $H_3^+$ model, and it is explained in \cite{PST}. In that case there are
essentially two different types of branes, and their geometry is AdS$_2$ and S$^2$. The branes we are going to study are their analogs in OSP(1$|$2).

The gluing map that we want to consider acts in our matrix representation by conjugation with the matrix
\begin{equation}\label{glueautomorphism}
\begin{split}
		X\ = \ \left(\begin{array}{ccc}1 & 0 & 0 \\
                             0 & 0 & 1 \\
                             0 & -1 & 0 \end{array}\right)\, .
	\end{split}
\end{equation}
Note that $X$ acts exactly as $(-1)\circ(st)$.
This automorphism has order four and its fix points form a one-dimensional space.
This implies that all branes have co-dimension one. Moreover the restriction of the twisted super conjugacy class
\begin{equation}
	C^X_a\ = \ \{\ g\,=\,XbX^{-1}ab^{-1} \ | \ b \ \in \ G \ \}\,
\end{equation}
to its bosonic subspace is Euclidean AdS$_2$. Hence we describe supersymmetric AdS branes of super-dimension 2$|$2. The geometry of branes corresponding to the gluing automorphism that is conjugation by the inverse of $X$ is also supersymmetric AdS.
The open string ends on the brane, which means that the Lie supergroup valued field $g$ describing the string is restricted to its branes world-volume, the twisted super conjugacy class. This allows us to read off the Dirichlet boundary conditions
\begin{equation}
 \str(X^{-1}g) \ = \ \str(bX^{-1}ab^{-1}) \ = \ \str(X^{-1}a) \ = \ \text{constant}\, .
\end{equation}

The second class of branes is described by (untwisted) super conjugacy classes
\begin{equation}
	C_a\ = \ \{\ bab^{-1} \ | \ b \ \in \ G \ \}\, .
\end{equation}
In this case three different types of geometries arise. The restriction of these super conjugacy classes to its bosonic subspace are two-spheres, 
which would degenerate to points at the identity. In this case also the super conjugacy class becomes zero-dimensional in the fermionic directions. If we instead choose $X^2=(-1)^F$ as gluing condition, then the twisted super conjugacy classes are also superspheres of dimension 2$|$2, but at the identity they degenerate only to a point in the bosonic direction while still extending in the fermionic ones. Note that these non-generic branes resemble those of GL(1$|$1) \cite{CQS}. In this spherical case the Dirichlet condition is
\begin{equation}
 \str(g) \ = \ \str(bab^{-1}) \ = \ \str(a) \ = \ \text{constant}\, .
\end{equation}

The Dirichlet conditions can be stated more explicitly. 
We parameterize an OSP(1$|$2) group element by a real number $\phi$, a complex number $(\gamma,\bar\gamma)$, and a complex 
Grassmann number $(\theta,\theta^\sharp)$. It then reads
\begin{equation}\label{eq:supergroupelement}
 \begin{split}
		g\ = \ \left(\begin{array}{ccc}
1+\theta\theta^\sharp e^{ - \phi } & \theta e^{ - \phi } & \theta^\sharp+\theta\bar\gamma e^{ - \phi } \\
\theta^\sharp e^{ - \phi } & e^{ - \phi } & e^{ - \phi } \bar\gamma \\
-\theta+\gamma\theta^\sharp e^{ - \phi } & \gamma e^{ - \phi } & -\theta\theta^\sharp+\gamma\bar\gamma e^{ - \phi } +e^{\phi} \end{array}\right)\, .
	\end{split}
\end{equation}
In this parameterizations super AdS branes are described by the following Dirichlet conditions
\begin{equation}\label{dirichletads1}
 \pm\theta\theta^\sharp+\gamma-\bar\gamma \ = \ ice^{\phi}\, .
\end{equation}
Here $\pm$ corresponds to conjugation by $X^{\pm1}$.
The super spherical ones satisfy
\begin{equation}\label{dirichletsphere1}
 e^{ - \phi } (\pm\theta\theta^\sharp+\gamma\bar\gamma+1)-\theta\theta^\sharp+e^{\phi} \ = \ c
\end{equation}
for some real constants $c$. Plus corresponds to conjugation with $(-1)^F$ and minus to the trivial gluing automorphism.

\subsection{The semiclassical boundary states}

In this section we rewrite the delta-distributions corresponding to the branes in terms of eigenfunctions of the Laplacian, which allows us to identify the semiclassical limit of bulk one-point functions.
We proceed as \cite{PST}. First, we need a suitable parametrization of eigenfunctions of the Laplacian. The Laplacian is obtained as the Casimir for the invariant vector fields. As such it is clear that the supergroup element $g$ \eqref{eq:supergroupelement} is an eigenfunction. The Laplacian reads explicitly
\begin{equation}
\Delta \ = \ \frac{1}{4}\del^2_\phi-\frac{1}{4}\del_\phi+e^{2\phi}\del_\gamma\del_{\bar\gamma}-\frac{1}{2}e^{\phi}(\del_\theta+\theta\del_\gamma)(\del_{\theta^\sharp}+\theta^\sharp\del_{\bar\gamma}) \, .
\end{equation}
Let $v= (\lambda,u,1)$ a complex vector, $\lambda$ Grassmann odd, and $v^\ddagger =(-\lambda^\sharp,\bar u, 1)^t$ its super-adjoint.
Then we define the field
\begin{equation}
\begin{split}
 \Phi_h(u,\lambda|g) \ &= \ (vgv^\ddagger)^{-2h} \\
 &= \ (-|\lambda-\theta|^2+e^{ - \phi } |u+\gamma+\lambda\theta|^2+e^{\phi})^{-2h}
\end{split}
\end{equation}
and compute that it is indeed an eigenfunction
\begin{equation}
 \Delta\Phi_h(u,\lambda|g) \ = \ h(h-1/2)\Phi_h(u,\lambda|g)\, .
\end{equation}
The representation label takes values $h=1/4-iP/2$ with $P$ in $\R$ and non-negative.

Let $k$ be an element of the complexification of OSP(1$|$2), then for
\begin{equation}
\begin{split}
    k\ = \ \left(\begin{array}{ccc}
A & \theta^1  & \theta^2\\
{\theta^1}^\sharp & \alpha & \beta \\
{\theta^2}^\sharp & \gamma & \delta \end{array}\right)\\
 \end{split}
\end{equation}
$\Phi_h$ transforms as follows
\begin{equation}
 \Phi_h(u,\lambda|kgk^\ddagger) \ = \ |\lambda\theta^2+u\beta+\delta|^{-4h}\Phi_h(k\cdot u,k\cdot\lambda|g) \, ,
\end{equation}
where
\begin{equation}
 k\cdot u \ = \ \frac{\lambda\theta^1+u\alpha+\gamma}{\lambda\theta^2+u\beta+\delta}
\quad\text{and}\quad
k\cdot \lambda \ = \ \frac{\lambda A+u{\theta^1}^\sharp+{\theta^2}^\sharp}{\lambda\theta^2+u\beta+\delta} \, .
\end{equation}
It is convenient to summarize $k\cdot u$ and $k\cdot \lambda$ in a vector
$v'=(k\cdot \lambda, k\cdot u, 1)$. In terms of $v$ this vector reads
\begin{equation}
 v'\ = \ \frac{vk}{\lambda\theta^2+u\beta+\delta} \, .
\end{equation}
Now suppose that $k$ satisfies
\begin{equation}\label{eq:twistinvariance}
 k^\ddagger \ = \ Xk^{-1}X^{-1}
\end{equation}
for some $X$ in OSP(1$|$2), then we get
\begin{equation}
 v'X^{-1}(v')^\ddagger \ = \ \frac{vkX^{-1}k^\ddagger v^\ddagger}{|\lambda\theta^2+u\beta+\delta|^2} \ = \ \frac{vX^{-1}v^\ddagger}{|\lambda\theta^2+u\beta+\delta|^2} \, .
\end{equation}

As a result, we get that the distribution
representation\begin{equation}
 D_\epsilon^h[f] \ = \ \int d^2ud^2\lambda \ |vX^{-1}v^\ddagger|^{2h-1} \text{sgn}^\epsilon (ivX^{-1}v^\ddagger)\int dk\, \Phi_h(u,\lambda|k)f(k)
\end{equation}
is invariant under conjugation, that is
\begin{equation}
  D_\epsilon^h[f] \ = \  D_\epsilon^h[T_gf]
\end{equation}
where $T_gf(k)=f(g^{-1}k(g^{-1})^\ddagger)$ for all $g$ satisfying \eqref{eq:twistinvariance}.
This means that we found a distribution which does not depend on the position in the orbit of the twisted super conjugacy class $C^X_a$.
Especially we can view the distribution as a distribution in one variable labeling the position of the twisted super conjugacy class,
i.e.
\begin{equation}
  D_\epsilon^h[f(k)] \ = \ D_\epsilon^h[f(a)] \, .
\end{equation}
Now $f=f(a)$ is a function on the position of the
twisted super conjugacy class.
Since $\Phi_h$ is an eigenfunction of the Laplacian,
the distribution $D^h$ must be as well. But this only depends on the position of the twisted super conjugacy class $a=\text{exp}(\psi t)$ (here $t$ is fixed under conjugation by $X$), thus the eigenfunctions $E_h(\psi)$ satisfy the second order differential equation
\begin{equation}
 \Delta E_h(\psi) \ = \ h(h-1/2) E_h(\psi) \, .
 \label{ELap}
\end{equation}
Hence, we get
\begin{equation}
 \int d^2ud^2\lambda \ |vX^{-1}v^\ddagger|^{2h-1} \text{sgn}^\epsilon (ivX^{-1}v^\ddagger) \Phi_h(u,\lambda|k) \ = \ K^+_{h,\epsilon} E_h^+(\psi)+K^-_{h,\epsilon} E_h^-(\psi) \, ,
\end{equation}
where $E^{\pm}_h (\psi)$ are two independent solutions to \eqref{ELap}.

Our goal is to rewrite the delta-distribution $\delta(\psi-r)$ in terms of wave-functions. We have reduced this problem to finding the eigenfunctions $E_h^\pm(\psi)$, the coefficients $K^\pm_{h,\epsilon}$ and functions $f^\epsilon_h(r)$, such that
\begin{equation}
\delta(\psi-r) \ = \
 \int dh\, (K^+_{h,\epsilon} E_h^+(\psi)+K^-_{h,\epsilon} E_h^-(\psi))f^\epsilon_h(r)\, .
\end{equation}
This will be done case by case.

\subsubsection{Super AdS$_2$ branes}

We start with the AdS-type branes with gluing automorphism conjugation by $X$ \eqref{glueautomorphism}. We parameterize a group element in twisted super conjugacy class form, i.e. $g=X bX^{-1}ab^{-1}$) with
\begin{equation}
\begin{split}
a\ =\ \ \left(\begin{array}{ccc}1 & 0 & 0 \\
                             0 & \cosh \psi & i\sinh \psi \\
                             0 & -i\sinh\psi & \cosh\psi \end{array}\right)\, .
	\end{split}
\end{equation}
In these coordinates the Dirichlet boundary condition \eqref{dirichletads1} reads
\begin{equation}\label{adsbdy}
\str(X^{-1}g) \ = \ 1-2i\sinh\psi \ = \ 1+e^{ - \phi }(\theta\theta^\sharp+\gamma-\bar\gamma) \ = \ \text{constant}\, .
\end{equation}

The Laplacian modulo $b$ is
\begin{equation}
\begin{split}
\Delta_\psi \ &= \ \frac{1}{4}\del^2_\psi + \frac{1}{4}\frac{\sinh\psi}{\cosh\psi}\del_\psi+\frac{i}{4}\frac{1}{\cosh\psi}\del_\psi\\
&= \ \frac{1}{4}\del^2_\psi + \frac{1}{4}\frac{\sinh(\psi/2+i\pi/4)}{\cosh(\psi/2+i\pi/4)}\del_\psi\\
&= \ \frac{1}{4}\frac{1}{\cosh(\psi/2+i\pi/4)}(\del^2_\psi-\frac{1}{4})\cosh(\psi/2+i\pi/4)\, ,\\
\end{split}
\end{equation}
and its eigenfunctions with eigenvalue $h(h-1/2)$ are
\begin{equation}
 E_h^\pm(\psi) \ = \ \frac{e^{\pm(2h-1/2)\psi}}{\cosh(\psi/2+i\pi/4)} \, .
\end{equation}
The coefficients $K^\pm_{h,\epsilon}$ are fixed by the asymptotic behavior of the functions $\Phi_h$. We use the following result of \cite{PST}
\begin{equation}
 \frac{2h-1}{\pi}(e^{\phi}+e^{ - \phi} |u-\gamma|^2)^{-2h} \ \ \ \substack{\phi\,\rightarrow\, - \infty\\ \sim}\ \ \ \frac{2h-1}{\pi}|\gamma-u|^{-4h}e^{2h\phi}+\delta^2(\gamma-u)e^{(2 - 2h)\phi}\, .
\end{equation}
This implies the asymptotic behavior
\begin{equation}\label{asymptotic}
\begin{split}
 \Phi_h(u,\lambda|g) \ \ \ \substack{\phi\,\rightarrow\, - \infty\\ \sim}\ \ \ &|\gamma-u-\lambda\theta|^{-4h}e^{2h\phi}+\frac{\pi}{2h-1}\delta^2(\gamma-u-\lambda\theta)e^{(2 - 2h)\phi}+\\
  +&|\lambda-\theta|^2|\gamma-u|^{-4h-2}e^{(2h+1)\phi}+\pi|\lambda-\theta|^2\delta^2(\gamma-u)e^{(1 - 2h)\phi}\\
\substack{\phi\,\rightarrow\, - \infty\\ \sim}\ \ \ &|\gamma-u-\lambda\theta|^{-4h}e^{2h\phi}+\pi|\lambda-\theta|^2\delta^2(\gamma-u)e^{(1 - 2h)\phi}\, .\\
\end{split}
\end{equation}
Another result of \cite{PST} is the integral
\begin{equation}
\int_\C d^2u\ |u+\bar u|^a\sgn^\epsilon(u+\bar u)|u-\gamma|^{-2a-4} \ = \ \frac{\pi}{a+1}(-1)^\epsilon|\gamma+\bar\gamma|^{-a-2}\sgn^\epsilon(\gamma+\bar\gamma)\, .
\label{intform}
\end{equation}
This integral together with $2\sinh\psi \ \substack{\phi\,\rightarrow\, - \infty\\ \sim} \ e^{|\psi|}$ as well as \eqref{adsbdy} and \eqref{asymptotic} implies
\begin{equation}
\begin{split}
\frac{1}{1+2h} \int d^2u&d^2\lambda \ |vX^{-1}v^\ddagger|^{2h-1} \text{sgn}^\epsilon (ivX^{-1}v^\ddagger) \Phi_h(u,\lambda|k) \ \ \ \substack{\phi\,\rightarrow\, - \infty\\ \sim}\ \ \ \\
&\substack{\phi\,\rightarrow\, - \infty\\ \sim}\ \ \ \sgn^\epsilon(i(\bar\gamma-\gamma))\Bigl((i(\bar\gamma-\gamma-\theta\theta^\sharp))^{2h-1}+i(i(\bar\gamma-\gamma-\theta\theta^\sharp))^{-2h}\Bigr) \\
&\substack{\phi\,\rightarrow\, - \infty\\ \sim}\ \ \ e^{-\frac{1}{2}|\psi|+\frac{i\pi}{4}}
   \Bigl(e^{-|\psi|(-2h+\frac{1}{2})-\frac{i\pi}{4}}+(-1)^\epsilon e^{|\psi|(-2h+\frac{1}{2})+\frac{i\pi}{4}}\Bigr)\, .
\end{split}
\end{equation}
This allows us to rewrite
\begin{equation}
\begin{split}
e^{(\psi-r)(-2h+1/2)}+&e^{-(\psi-r)(-2h+1/2)} \ = \\
= \ &\kappa_h\int d^2ud^2\lambda \ |vX^{-1}v^\ddagger|^{2h-1}\Phi_h(u,\lambda|g)(d_0(r)-\text{sgn}( ivX^{-1}v^\ddagger)d_1(r))\\
\end{split}
\end{equation}
with
\begin{equation}
\begin{split}
d_0(r) \ &= \ \text{sgn}(r)\cosh(-r(-2h+1/2)-i\pi/4)\, ,\\
d_1(r) \ &= \ \text{sgn}(r)\sinh(-r(-2h+1/2)-i\pi/4)\ \text{and}\\
\kappa_h\ &= \ \frac{\cosh(\psi/2+i\pi/4)}{(h+1/2)}\, .\\
\end{split}
\end{equation}
Notice that $h=1/4-iP/2$ for non-negative real number $P$. Hence
we arrive at the following expression for the delta-distribution of the AdS-like brane
\begin{equation}
\begin{split}
\delta(\psi-r)  =  \int dP \kappa_h\int d^2ud^2\lambda \, |vX^{-1}v^\ddagger|^{2h-1}\Phi_h(u,\lambda|g)(d_0(r)-\text{sgn}( ivX^{-1}v^\ddagger)d_1(r)) \, .\\
\end{split}
\end{equation}
Recall that the dual wave function is $(\Phi_h(u,\lambda|g))^*=\Phi_{-h+1/2}(u,\lambda|g)$.

We can finally conclude that the bulk one-point function behaves as follows in the semiclassical limit%
\footnote{Here $k$ means the level of OSP(1$|$2) current algebra.}
\begin{equation}
\langle \Phi_h(u,\lambda|z)\rangle^{\text{AdS}}_{r,+} \ \ \ \substack{k\,\rightarrow\,\infty\\ \sim}\ \ \
|u-\bar u+\lambda\lambda^\sharp|^{-2h}e^{-\sgn(i(u-\bar u))r(-2h+1/2)}\, .
\end{equation}
There are two choices of lifting complex conjugation to Grassmann numbers, the super star $\sharp$ and the bar operation.
In the full field theory it is more convenient for us to work with the bar. The two operations are related as in \eqref{eq:barsuperstar}.
With $u=ix$ the semiclassical limit is
\begin{equation}\label{eq:semiclassicalADSplus}
\langle \Phi_h(x,\lambda|z)\rangle^{\text{AdS}}_{r,+} \ \ \ \substack{k\,\rightarrow\,\infty\\ \sim}\ \ \
|x+\bar x+\lambda\bar\lambda|^{-2h}e^{\sgn(x+\bar x)r(-2h+1/2)}\, .
\end{equation}
We can repeat the analysis for conjugation by $X^{-1}$ as gluing automorphism instead of $X$. In that case the semiclassical limit behaves as
\begin{equation}\label{eq:semiclassicalADSminus}
\langle \Phi_h(x,\lambda|z)\rangle^{\text{AdS}}_{r,-} \ \ \ \substack{k\,\rightarrow\,\infty\\ \sim}\ \ \
\sgn(x+\bar x)|x+\bar x-\lambda\bar\lambda|^{-2h}e^{\sgn(x+\bar x)r(-2h+1/2)}\, .
\end{equation}
Eventually, as in \eqref{ads1pt1} and \eqref{ads1pt2}, we will see that this is indeed the behavior of the bulk one-point functions in the semiclassical limit.

\subsubsection{Fuzzy supersphere branes}

The fuzzy-sphere like branes are described by the identity gluing automorphism.
We parameterize an group element as before ($bab^{-1}$) with
\begin{equation}
\begin{split}
a\ =\  \left(\begin{array}{ccc}1 & 0 & 0 \\
                             0 & e^\Lambda & 0 \\
                             0 & 0 & e^{-\Lambda} \end{array}\right)\, .
	\end{split}
\end{equation}
The branes are characterized by the equation
\begin{equation}\label{soherebdy}
\str(g) \ = \ e^{ - \phi } (\theta\bar\theta-\gamma\bar\gamma-1)+1+\theta\theta^\sharp-e^{\phi} \ = \ 1-2\cosh(\Lambda) \ = \ \text{constant} \, , 
\end{equation}
where 
$\Lambda$ is the radius of the spherical branes and hence we restrict it to be non-negative.

The Laplacian in these coordinates modulo $b$ is
\begin{equation}
\begin{split}
 \Delta_\Lambda \ &= \ \frac{1}{4}\del^2_\Lambda + \frac{1}{4}\frac{\cosh\Lambda}{\sinh\Lambda}\del_\Lambda-\frac{1}{4}\frac{1}{\sinh\Lambda}\del_\Lambda\\
&=\ \frac{1}{4}\del^2_\Lambda + \frac{1}{4}\frac{\sinh(\Lambda/2)}{\cosh(\Lambda/2)}\del_\Lambda\\
&=\ \frac{1}{4}\frac{1}{\cosh(\Lambda/2)}(\del^2_\Lambda-\frac{1}{4})\cosh(\Lambda/2) \, .
\end{split}
\end{equation}
Its eigenfunctions with eigenvalue $h(h-1/2)$ are
\begin{equation}
 E_h^\pm(\Lambda) \ = \ \frac{e^{\pm(-2h+1/2)\Lambda}}{\cosh(\Lambda/2)}
 \, .
\end{equation}
In this case a careful analysis of the asymptotics gives
\begin{equation}
\begin{split}
\cosh(\Lambda(2h-1/2)) \ = \
 \kappa_h\int d^2ud^2\lambda \ |vv^\ddagger|^{2h-1}\Phi_h(u,\lambda|g)\\
\end{split}
\end{equation}
with
\begin{equation}
\kappa_h\ = \ \frac{\cosh(\Lambda/2)}{(h+1/2)} \, .
\end{equation}
Since $\Lambda$ is non-negative, the delta-distribution for spherically branes is
\begin{equation}
\begin{split}
\delta(\Lambda-\Lambda_0)& \ = \ \int dP \kappa_h\int d^2ud^2\lambda \ |vv^\ddagger|^{2h-1}\Phi_h(u,\lambda|g)\cosh(\Lambda_0(2h-1/2))\, . \\
\end{split}
\end{equation}

We conclude that the bulk one-point function behaves as follows in the semiclassical limit
\begin{equation}\label{eq:semiclassicalsphere1}
\langle \Phi_h(u,\lambda|z)\rangle^{\text{sphere}}_{\Lambda_0,+} \ \ \ \substack{k\,\rightarrow\,\infty\\ \sim}\ \ \
|1+u\bar u-\lambda\lambda^\sharp|^{-2h}\cosh(\Lambda_0(2h-1/2))\,.
\end{equation}
The analogous result for spherical branes with gluing automorphism conjugation by $X^2$ instead of the identity map is
\begin{equation}\label{eq:semiclassicalsphere2}
\langle \Phi_h(u,\lambda|z)\rangle^{\text{sphere}}_{\Lambda_0,-} \ \ \ \substack{k\,\rightarrow\,\infty\\ \sim}\ \ \
|1+u\bar u+\lambda\lambda^\sharp|^{-2h}\sinh(\Lambda_0(2h-1/2))\,.
\end{equation}
In this case, the result in the full quantum theory will have a similar form but the parameter $\Lambda_0$ will be imaginary. This is analogous to the spherical branes in $H_3^+$ \cite{PST}.

\section{OSP(1$|$2) WZNW model}

Closed string sector of OSP(1$|$2) WZNW model has been studied
in \cite{HS2}, and in this section we review the results in a
different form with manifest OSP(1$|$2) symmetry.
First we introduce the OSP(1$|$2) current algebra, then we give some general arguments concerning correlation functions.
After this we restrict ourselves to concrete examples,
such as two, three and four point functions.
In particular, we give explicit forms of two and three-point functions.
Finally, we discuss several useful facts on degenerate representations since
a degenerate operator will be utilized to compute a specific
four point function.

In the SL(2) WZNW model, there are several bases for vertex operators,
such as $m$-basis, $x$-basis and $\mu$-basis, which are related by
Fourier transforms. Among them, $x$-basis is useful to keep
track of the SL(2) symmetry, and it might be interpreted as
a coordinate of boundary theory as in, e.g., \cite{BORT,KS}.
In our case we introduce one fermionic parameter $\xi$ along
with $x$, which may be interpreted as a super coordinate of
boundary theory. The super coordinate $(x,\xi)$ has been
used to describe ${\cal N} = (1,1)$ superconformal field theories
in two dimension, so we can utilize the methods developed
for them. As a good review see for example \cite{AZ}.

\subsection{OSP(1$|$2) current algebra}

Let us start from superalgebra OSP(1$|$2). The generators of
bosonic subalgebra SL(2) are denoted by $E^\pm, H$ and for
the other fermionic generators we use $F^\pm$.
The relations between these generators are
\begin{align}
&[ H , E^{\pm} ] = \pm E^{\pm} ~,
\qquad [ H , F^{\pm} ] =  \pm \tfrac{1}{2} F^{\pm} ~,
\qquad [ E^+ , E^- ] =  - 2 H ~,
\\
&[ E^{\pm} , F^{\mp} ] =  \pm F^{\pm} ~,
\qquad \{ F^+ , F^- \} = - \tfrac{1}{2} H ~,
\qquad \{ F^{\pm} , F^{\pm} \} =  \tfrac{1}{2} E^{\pm} ~. \nonumber
\end{align}
As mentioned above, here we use the representation labeled by
a commuting parameter $x$ and an anti-commuting parameter $\xi$.
With these parameters the generators can be written as
\begin{align}
 &{\cal D}^{E^+} =  \partial_x ~, \qquad
 {\cal D}^{H} = - x \partial_x - \tfrac12 \xi \partial_\xi - h ~, \qquad
 {\cal D}^{E^-} =  x^2 \partial_x + x \xi \partial_\xi + 2 x  h ~,
 \nonumber \\
 &{\cal D}^{F^+} = \tfrac12 ( \partial_\xi + \xi \partial_x )~,\qquad
 {\cal D}^{F^-} = \tfrac12 x (\partial_\xi + \xi \partial_x ) +  \xi h
 ~ . 
 \label{xandxi}
\end{align}
The second Casimir
\begin{align}
 \Delta =  H H - \frac12 (E^+ E^- + E^- E^+)
  + (F^+ F^- - F^- F^ + ) ~ 
  \label{casimir}
\end{align}
is $h(h-1/2)$ in the above representation.%
\footnote{Here $h$ is related to $j + 1/2 = - h$ with respect to $j$ in \cite{HS2}.}

An affine extension of the above superalgebra may be expressed by
currents. The bosonic subalgebra is generated by $J^{3} (z)$ and
$J^{\pm} (0)$, whose OPEs are
\begin{align}
 J^+(z) J^-(0) \sim  \frac{k}{z^2} - \frac{2J^3(0)}{z} ~ , \qquad
 J^3(z) J^{\pm}(0) \sim \pm \frac{J^\pm (0)}{z} ~ ,\qquad
 J^3(z)J^3(0)\sim -  \frac{k}{2z^2} ~ .
\end{align}
In addition to these bosonic generators, there are fermionic ones with
\begin{align}
 &J^3(z) j^{\pm}(0)\sim \pm\frac{j^\pm (0)}{2z} ~ , \qquad
 J^{\pm} (z) j^\mp (0) \sim  \pm \frac{j^\pm(0)}{z} ~ , \\  \nonumber
 &j^+(z)j^-(0) \sim  \frac{k}{2z^2} - \frac{J^3(0)}{2z} ~ , \qquad
 j^{\pm}(z) j^{\pm}(0) \sim  \frac{J^{\pm}(0)}{2z} ~ .
\end{align}
Here $k$ represents the level of current algebra.
The energy momentum tensor can be given by the Sugawara
construction as
\begin{align} \label{sugawara}
 T (z) &= \frac{1}{ k - 3/2 }
 \Bigl[ -  J^3 (z) J^3 (z)
 + \frac12 (J^+ (z) J^- (z) + J^- (z) J^+ (z) ) \\
  & \qquad \qquad \qquad
  + j^+ (z) j^- (z) - j^- (z) j^ + (z)  \Bigr]~, \nonumber
\end{align}
where normal ordering is assumed for the operators
at the same position. The central charge is
$c = 2 k / (2 k - 3)$.

With these currents primary fields can be defined as
\begin{align}
 J^A (z) \Phi_h (x, \xi | w) \sim \frac{- {\cal D}^A}{z - w}
 \Phi_h (x, \xi | w) ~, \label{pfinx}
\end{align}
where $J^H = J^3$, $J^{H^{\pm}} = J^\pm$, and $J^{F^\pm} = j^\pm$.
The eigenvalue of the energy
momentum tensor is given by $\Delta_h = - 2 b^2 h ( h - 1/2 )$
with $b^2 = 1/(2k - 3)$. 
The anti-holomorphic part can be defined similarly. It is related to complex conjugation as
\begin{equation}
 (J^3(z))^* \ = \ -\bar J^3(\bar z)\  , \ \
(J^\pm(z))^* \ = \ \bar J^\mp(\bar z)\  , \ \
(j^\pm(z))^* \ = \ \pm\bar j^\mp(\bar z)\  , \ \
\end{equation}
thus we may define, for instance
\begin{align}
 &\bar {\cal D}^{E^-} =  \partial_{\bar x} ~, \qquad
 \bar {\cal D}^{H} =  \bar x \partial_{\bar x}
+ \tfrac12 \bar \xi \partial_{\bar \xi } + h ~, \qquad
 \bar {\cal D}^{E^+} =  { \bar x }^2 \partial_{ \bar x}
 + \bar x \bar \xi \partial_{\bar \xi} + 2 \bar x  h ~,
 \nonumber \\
 &\bar {\cal D}^{F^-} = \tfrac12 ( \partial_{\bar \xi} + \bar \xi
 \partial_{\bar x } )~,\qquad
 \bar {\cal D}^{F^+} = - \tfrac12 \bar x (\partial_{\bar \xi} + \bar \xi \partial_{\bar x} ) -  \bar \xi h ~.
\end{align}

\subsection{Correlation functions}

Recall that the Lie superalgebra OSP(1$|$2)
appears as a subalgebra of super Virasoro algebra.
If $x$ is treated as a holomorphic coordinate
of some (space-time) conformal field theory, then $\xi$
can be thought of as a super-coordinate in a superfield
formalism as mentioned above.
Just like the form of correlation functions
in ${\cal N} = (1,1)$ superconformal field theories is
restricted due to the superconformal symmetry, the form of correlation
functions of OSP(1$|$2) WZNW model is also
fixed to some extent by the OSP(1$|$2) global symmetry.

Consider a $N$-point function
\begin{align}
Z_N = \left \langle
 \prod_{i=1}^N \Phi_{h_i} (x_i , \xi_i | z_i) \right \rangle ~,
\end{align}
where $z_i$ denote the positions of vertex operators inserted
on the worldsheet. The dependence on $z_i$ is fixed in part by
SL(2) subalgebra of Virasoro algebra as usual.
A novel point is the dependence on $x_i$ and $\xi_i$.
Since the superalgebra
OSP(1$|$2) has super-dimension $3|2$, we can fix three of $x_i$ and two
of $\xi_i$. In other words, the $N$-point function can be given by
a function of $N-3$ bosonic cross ratios and $N-2$ fermionic cross
ratios. Bosonic ones are given by
\begin{align}
 X_a = \frac{X_{12}X_{3a}}{X_{13}X_{2a}} ~ , \qquad
 X_{ij} = x_i - x_j + \xi_i \xi_j ~,
 \label{largex}
\end{align}
with $a = 4,5, \cdots , N$. On the other hand the fermionic
ones are
\begin{align}
 \eta_\alpha = (X_{12} X_{1 \alpha} X_{2 \alpha} )^{-1/2}
  (X_{2\alpha}\xi_1 + X_{ \alpha 1} \xi_2 + X_{12} \xi_\alpha
  -  \xi_1 \xi_2 \xi_\alpha )
\end{align}
with $\alpha=3,4,\cdots N$.
Utilizing the symmetry we can restrict the $N$-point function as
\begin{align}
\left \langle \prod_{i=1}^N \Phi_{h_i} (x_i , \xi_i | z_i) \right \rangle
 = \prod_{i < j} |z_{ij}|^{-2  \Delta_{ij}}
 |X_{ij}|^{- 2 \gamma_{ij}}
  F(X_a , \bar X_a ,  \eta_\alpha , \bar \eta_\alpha ; w_a, \bar w_a )
  ~, 
\end{align}
where
\begin{align}
 \sum_{j \neq i} \Delta_{ji} = 2 \Delta_i ~, \qquad
  \sum_{j \neq i} h_{ji} = 2 h_i ~, \qquad
  w_a = \frac{z_{12}z_{3a}}{z_{13}z_{2a}} ~.
\end{align}
Here $\Delta_i$ represents the conformal weight of $\Phi_{h_i}$.

In WZNW models we can construct the energy momentum
tensor in terms of currents by Sugawara construction as in \eqref{sugawara}.
This leads to Knizhnik-Zamolodchikov (KZ) equation which
the correlation functions satisfy.
In our case they are written as
\begin{align}
 [ \kappa \partial_{z_i} - \sum_{j \neq i} \frac{Q_{ij}}{z_i -z_j} ]
 Z_N = 0
 \label{KZ}
\end{align}
with $\kappa = - 1/4b^2$ and
\begin{align}
 Q_{ij} =  {\cal D}^{H}_i {\cal D}^{H}_j - \tfrac12 ({\cal D}^{E^+}_i {\cal D}^{E^-}_j + {\cal D}^{E^-}_i {\cal D}^{E^+}_j )
  + ({\cal D}^{F^+}_i {\cal D}^{F^-}_j - {\cal D}^{F^-}_i {\cal D}^{F^ +}_j ) ~ .
\end{align}
With the explicit form of \eqref{xandxi} we find
\begin{align}
  Q_{12} & = - ( \tfrac12 x_{12}^2 + \tfrac14 x_{12}\xi_1 \xi_2)
  \partial_{x_1 } \partial_{x_2} + (x_{12} h_2 + \tfrac12 h_2 \xi_1 \xi_2)
  \partial_{x_1}
  + (x_{21} h_1 + \tfrac12 h_1 \xi_2 \xi_1 )
  \partial_{x_2}
 \nonumber \\ & + \tfrac{h_2}{2} (\xi_1 - \xi_2) \partial_{\xi_1}
   + \tfrac{h_1}{2} (\xi_2 - \xi_1) \partial_{\xi_2}
    + (\tfrac12 x_{12} \xi_2 - \tfrac14 x_{12} \xi_1) \partial_{x_1}
    \partial_{\xi_2} \\ &+ ( \tfrac12 x_{21} \xi_1 - \tfrac14 x_{21} \xi_2)
    \partial_{\xi_1} \partial_{x_2}
 - \tfrac14 ( x_{12} + \xi_1 \xi_2) \partial_{\xi_1} \partial_{\xi_2}
  + h_1 h_2 ~ ,\nonumber
\end{align}
and similarly for other $Q_{ij}$.

\subsection{Two, Three and four point functions}

Two and three point functions of OSP(1$|$2) model have been obtained
explicitly
in \cite{HS2} by utilizing the relation to ${\cal N}=(1,1)$ super-Liouville
field theory.
Let us start from the two point function.
It can be fixed by symmetry and is given by
\begin{align}
 \label{2ptfn}
 & \left \langle \Phi_h (x, \xi | z)  \Phi_{h ' } (x ', \xi ' | z ')
 \right \rangle  \\ & \qquad =  \frac{1}{
|z - z '|^{ 4 \Delta_h} }
 \left [
  \pi \delta (h + h ' - \tfrac12) \delta^{(2)} (x - x') | \xi - \xi ' |^2 +
  \frac{\delta (h - h ') D(h)}{| x - x ' + \xi \xi '|^{4h}  } \right ]
   \nonumber
\end{align}
with some function $D(h)$.
The function was obtained in \cite{HS2} as
\begin{align}
 D(h) = \nu^{- 4 h + 1} \gamma (b^2 (k - 2 h - 1) )
 \label{2ptD}
\end{align}
with $\gamma (x) = \Gamma (x) / \Gamma (1-x)$.
Here $\nu$ is some parameter.

For the three point function,
 the form is restricted as
\begin{align}
Z_3 = \left \langle \prod_{i=1}^3 \Phi_{h_i} (x_i , \xi_i | z_i) \right \rangle
 = \prod_{i < j} |z_{ij}|^{- 2 \Delta_{ij}}
 |X_{ij}|^{- 2 \gamma_{ij}} ( C (h_1,h_2,h_3)
 + \tilde C ( h_1,h_2,h_3 ) \eta \bar \eta)
 \label{3pt}
\end{align}
with
\begin{align}
 \Delta_{12} =  \Delta_1 + \Delta_2 - \Delta_3 ~,
 \qquad
 \gamma_{12} = h_1 + h_2 - h_3 ~,
\end{align}
and so on. Notice that the three point function depends on one
fermionic cross ratio
\begin{align}
 \eta = (X_{12} X_{2 3} X_{1 3})^{-1/2}
  (X_{2 3}\xi_1 + X_{ 3 1} \xi_2 + X_{12} \xi_3
  - \xi_1 \xi_2 \xi_3 ) ~,
\end{align}
while the form of three point functions in a bosonic theory
can be fixed uniquely by symmetry up to an overall coefficient.
The information of three point function is therefore encoded
in the two functions $C (h_1,h_2,h_3) $ and
$\tilde C (h_1,h_2,h_3) $.

The explicit expressions are \cite{HS2} 
\begin{align} \label{C}
 &C (h_1,h_2,h_3)
    = (\nu b^{b^2})^{- 2 h + 2}
   \frac{\Upsilon ' _{\text{NS}}(0)
   \Upsilon_{\text{R}}(4 b h_1 - b )
     \Upsilon_{\text{R} } (4 b h_2 - b )
   \Upsilon_{\text{R}} (4 b h_3 - b )}
     {\Upsilon_{\text{R}}( 2 b h - b )
     \Upsilon_{\text{NS} } (2 b h_{12} )
     \Upsilon_{\text{NS}} (2 b h_{23} )
     \Upsilon_{\text{NS}} (2 b h_{31} ) } ~, \\
& \tilde C (h_1,h_2,h_3)
    = b^{-1 }(\nu b^{b^2})^{- 2 h + 2}
   \frac{\Upsilon ' _{\text{NS}}(0)
   \Upsilon_{\text{R}}(4 b h_1 - b )
     \Upsilon_{\text{R} } (4 b h_2 - b )
   \Upsilon_{\text{R}} (4 b h_3 - b )}
     {\Upsilon_{\text{NS}}( 2 b h - b )
     \Upsilon_{\text{R} } (2 b h_{12} )
     \Upsilon_{\text{R}} (2 b h_{23} )
     \Upsilon_{\text{R}} (2 b h_{31} ) }
     \nonumber
\end{align}
with $h=h_1 + h_2 + h_3$, $h_{12} = h_1 + h_2 - h_3$ and so on.
Here we have used the notation in \cite{FH}:
\begin{align}
 \Upsilon_{\text{NS}} (x) &= \Upsilon ( \tfrac{x}{2})
   \Upsilon (\tfrac{x+Q}{2}) ~,
 &\Upsilon_{\text{R}} (x) &= \Upsilon (\tfrac{x+b}{2})
   \Upsilon (\tfrac{x+b^{-1}}{2}) ~,
\end{align}
where the $\Upsilon$ function is
introduced in \cite{DO,ZZ} as follows
\begin{align}
 \ln \Upsilon (x) \ = \ \int_0^{\infty} \frac{dt}{t}
 \left[ e^{-2t} \left( \frac{Q}{2} - x\right)^2 -
   \frac{\sinh ^2 (\frac{Q}{2} - x) t}{\sinh bt \sinh \frac{t}{b} } \right] ~.
\end{align}
Under the shifts of their argument, they behave as
\begin{align}
 \Upsilon_{\text{NS}} (x+b) &= b^{-b x} \gamma ( \tfrac12 + \tfrac{bx}{2})
 \Upsilon_{\text{R}}(x) ~,
 &\Upsilon_{\text{R}} (x+b) &= b^{1-b x} \gamma (\tfrac{bx}{2})
 \Upsilon_{\text{NS}} (x)~, \\
  \Upsilon_{\text{NS}} (x+ \tfrac{1}{b})
  &= b^{\frac{x}{b}} \gamma (\tfrac12 + \tfrac{x}{2b})
 \Upsilon_{\text{R}} (x)~,
 &\Upsilon_{\text{R}} (x+ \tfrac{1}{b}) &= b^{-1+ \frac{x}{b} }
 \gamma (\tfrac{x}{2b})
 \Upsilon_{\text{NS}} (x) ~.
 \label{uprel}
\end{align}

{}From the above argument we can see that
the operator product expansion (OPE) is of the form
\begin{align}
 &\Phi_{h_1} (x_1 , \xi_1 | z_1) \Phi_{h_i} (x_2 , \xi_2 | z_2)
 =
  \int [ d h ] |z_{12}|^{2 \Delta - 2 \Delta_1 - 2 \Delta_2}
  |X_{12}|^{2 h - 2 h_1 - 2 h_2} \times \\
  & \nonumber \qquad \qquad
  \left( C (h ; h_1 ,h_2) [\Phi_{h} (x_2 , \xi_2 | z_2)]_\text{ee}
   + |X_{12}|^{-1}
 \tilde C (h ; h_1 ,h_2) [\Phi_{h} (x_2 , \xi_2 | z_2)]_\text{oo}
 \right) ~,
\end{align}
where $[ * ]_\text{ee}$ represents the terms with even number of
Grassmann odd variables and starts from $\Phi_h$.
One the other hand $[ * ]_\text{oo}$ represents the terms with odd number of Grassmann odd variables and starts from
$| \xi_1 - \xi _2 | ^2 \Phi_h$. The rest is fixed by the symmetry.
The explicit forms of the first few terms may be found in, e.g., \cite{AZ}.

In a similar way, the four point function can be written
in terms of one bosonic cross ratio and two fermionic
cross ratios.
The four point function takes the form of
\begin{align}
Z_4 = \left \langle \prod_{i=1}^4 \Phi_{h_i} (x_i , \xi_i | z_i) \right \rangle
 = \prod_{i < j} |z_{ij}|^{-2  \Delta_{ij}}
 |X_{ij}|^{- 2 \gamma_{ij}}
  F(X , \bar X ,  \chi_2 , \bar { \chi}_2 , \chi_3 , \bar \chi_3 ; z , \bar z ) ~,
  \label{fourpt}
\end{align}
and we set
\begin{align}
 &\Delta_{24} = - \Delta_1 + \Delta_2 + \Delta_3 + \Delta_4 ~,
 \qquad \Delta_{13} = 2 \Delta_3 ~, \nonumber \\
 & \Delta_{12} = \Delta_1 + \Delta_2 - \Delta_3 - \Delta_4 ~ , \qquad
 \Delta_{14} = \Delta_1 + \Delta_4 - \Delta_2 - \Delta_3 ~,  \\
& \gamma_{24} = - h_1 + h_2 + h_3 + h_4 ~,
 \qquad \gamma_{13} = 2 h_3 ~, \nonumber \\
 &\gamma_{12} = h_1 + h_2 - h_3 - h_4 ~ , \qquad
 \gamma_{14} = h_1 + h_4 - h_2 - h_3 ~.
\end{align}
Addition to the worldsheet cross ratio $z = w_4$,
the bosonic cross ratio is $X = X_4$ in \eqref{largex} as
\begin{align}
 X = \frac{X_{12}X_{34}}{X_{13}X_{24}}
 = \frac{(x_1 - x_2 + \xi_1 \xi_2 ) (x_3 - x_4 + \xi_3 \xi_4 )}{(x_1 - x_3 + \xi_1 \xi_3 ) ( x_2 - x_4 + \xi_2 \xi_4 ) }~,
 \label{crossX}
\end{align}
and the fermionic ones used here are
\begin{align} \label{eta31}
 &\chi_3 = (X_{12} X_{2 4} X_{1 4})^{-1/2}
  (X_{2 4}\xi_1 + X_{ 4 1} \xi_2 + X_{12} \xi_4
  - \xi_1 \xi_2 \xi_4 ) ~,  \\
 &\chi_2 = - \sqrt{X} (X_{13} X_{1 4} X_{34})^{-1/2}
  (X_{3 4}\xi_1 + X_{ 4 1} \xi_3 + X_{13} \xi_4
  - \xi_1 \xi_3 \xi_4 ) ~.\nonumber
\end{align}
We may fix the parameters as
$x_1 \to \infty$, $x_2 \to 1$, $x_3 \to x$, $x_4 \to 0$
by setting $\xi_1 \to x_1 \eta$, $\xi_2 \to 0$, $\xi_3 \to \xi$,
$\xi_4 \to 0$. Then we can express the four point function
with the three parameters $(x,\eta,\xi)$ in a quite simple way.
Notice that $(X,\chi_3 ,\chi_2)$ map to $(x,\eta , \xi - x \eta)$.
In general it is quite difficult to find out the explicit expression of
the four point function.
However, if one of the operators is in a degenerate
representation, then we may be able to have the explicit form
by solving the Knizhnik-Zamolodchikov equation \eqref{KZ}.
In fact we will utilize four point function with a degenerate
operator
$\Phi_{k/2}$ to obtain one point functions of bulk
operator on a disk. The expression of the four point function
is given in appendix \ref{sec4pt}.

\subsection{Degenerate representation}
\label{degenerate}

A state belongs to a degenerate representation,
when a descendant constructed by the action of currents becomes
a null state. If we require the null state to vanish, then we
have some equations which correlation functions have to satisfy.
Along with the KZ equation we may be able to obtain
explicit expressions for correlation functions.
According to \cite{KW} (see also \cite{ERS} for instance),
there are degenerate representations with
\begin{align}
 - 4 h_{r,s} + 1 = r +  \frac{s}{b^2}  ~.
 \label{labeldeg}
\end{align}
The integers $r,s$ satisfy $r + s = 2 \mathbb{Z} + 1$,
and $r > 0 , s \geq 0$ or $r < 0 , s < 0 $.

We can easily understand the case of $s = 0$. In
this case the representation of zero modes with $r > 0 $
is $r$-dimensional, and in particular
\begin{align}
 j_0^+ (J_0^+)^{(r-1)/2} | h_{r,0} \rangle = 0 ~,
\end{align}
which yields the constraint
\begin{align}
 ( \partial_\xi + \xi \partial_x ) \partial_x^{(r-1)/2}
  \Phi_h (x, \xi ) = 0 ~.
\label{deg3}
\end{align}
It looks useful to utilize the simplest one, $\Phi_{-1/2}$
with $(r,s) = (3,0)$. However the OPE is of the form as
$\Phi_h \Phi_{-1/2} \sim {\cal C}[ \Phi_{h - 1/2} ] _\text{ee}
+ {\cal C}[ \Phi_{h } ] _\text{oo} +
{\cal C}[ \Phi_{h + 1/2} ] _\text{ee} $, and related to this
fact we have to solve the third order differential equations.
Similar computations have been done for ${\cal N}=(1,1)$
super-Liouville field theory in \cite{BBNZ,Belavin}.

In the analysis of SL(2) WZNW model, degenerate
operators with  $r > 0 , s \geq 0$ are used to compute
four point functions, see e.g. \cite{TeschnerH}.
However, there are another type of  degenerate operators with $r<0,s<0$,
and it might make the analysis simpler to utilize them.
One line of degenerate operators is with $(r,s) = (- 2 n,-1)$,
which implies $h_{- 2 n , - 1}= ( k-1 + n ) / 2$.
This type of operators have been discussed in \cite{su2}
for the SU(2) WZNW model (see also \cite{PS} for
applications to the SL(2) WZNW model).
Denoting the corresponding state as
$| ( k-1 + n ) / 2 \rangle$, we can find
a null state as
\begin{align}
| \theta \rangle =  (J_{-1}^-)^n | \tfrac{k -1 + n}{2} \rangle ~.
\end{align}
Later we will use the case with $n=1$ since it is the simplest one.
In the bosonic case, the OPE is simply given by
$\Phi_h \Phi_{k/2} \sim {\cal C} [ \Phi_{k/2 - h } ]$, however in the super case
the corresponding OPE is complicated enough like
$\Phi_h \Phi_{k/2} \sim {\cal C}[ \Phi_{k/2 - h } ] _\text{ee} +
{\cal C}[ \Phi_{k/2 - h - 1/2} ] _\text{oo} $.
More detailed analysis can be found in appendix \ref{sec4pt}.

\section{Boundary OSP(1$|$2) model}
\label{sec:BOSP}

In this section we study boundary OSP(1$|$2) WZNW model 
focusing on bulk one point functions on a disk.
In the presence of boundary, we have to assign boundary conditions in a proper way as discussed in section 2.
In terms of currents the gluing conditions at the boundary $z = \bar z$ are
\begin{align}
 J^{\pm} = \bar J^{\mp} = (J^{\pm})^* ~, \qquad
 J^3 = -\bar J^3 =  (J^3)^* ~ , \qquad j^{\pm} = \pm\epsilon \bar j^{\mp} =  \epsilon(j^{\pm})^*
 \label{bcads2}
\end{align}
with $\epsilon = \pm$. These boundary
conditions correspond to branes similar to the AdS$_2$-branes
in the SL(2) WZNW model.
We can assign another type of boundary conditions as
\begin{align}
 J^{\pm} = \bar J^{\pm} = (J^{\mp})^* ~, \qquad
 J^3 = \bar J^3 = -(J^3)^*~ , \qquad j^{\pm} = \epsilon j^{\pm} = \pm\epsilon(j^{\mp})^* ~.
 \label{bcfuzzy}
\end{align}
This case corresponds to branes
similar to the fuzzy sphere branes in the SL(2) WZNW model.
Things are quite analogous to the SL(2) WZNW model with
boundary, which are analyzed, for example, in \cite{PST,LOP}.

Overlaps of closed strings to these branes can be read off from
one point functions of bulk operators on a disk with the
above boundary conditions.
For the super AdS brane, one point function is restricted to
have the form locally
\begin{align}
 \left \langle \Phi_h (x , \xi | z) \right \rangle =
  \frac{U^{\sigma} (h;\epsilon) }{|z - \bar z|^{2 \Delta_h }|x + \bar x + \epsilon \xi \bar \xi|^{2 h } }
  \label{ads2}
\end{align}
from the boundary condition.
The above form has a singularity at $x + \bar x  = 0$,
thus
the coefficient $U^{\sigma} (h)$ could depend on the sign $\sigma = {\rm \, sgn} (x + \bar x ) $.
We remark here that the same singularity appears in the SL(2) WZNW model, see \cite{PST,LOP}.
This function will be obtained below  by solving constraint equations coming
from the crossing symmetry.
For the fuzzy supersphere brane, one point function is restricted as
\begin{align}
 \left \langle \Phi_h (x , \xi | z) \right \rangle =
  \frac{U (h;\epsilon) }{|z - \bar z|^{2 \Delta_h }| 1 + x  \bar x + \epsilon \xi \bar \xi|^{2 h } } ~.
  \label{fuzzy}
\end{align}
In this case we do not have any singularity on the complex $x$-plane.
The coefficient $U(h;\epsilon)$ will be again obtained by
solving constraints from the crossing symmetry.

\subsection{Super AdS$_2$ branes}

We would like to compute bulk one point function on a disk
with boundary condition corresponding to super AdS brane.
Since the boundary condition \eqref{bcads2} restricts
the form of one point function as \eqref{ads2}, we just
need to know the function $U^\sigma (h ; \epsilon)$.
In order to compute the function, we obtain two kinds of
constraints which the one point function has to satisfy.
One comes from the reflection relation which connects
$\Phi_h$ to $\Phi_{1/2 - h}$. The other originates from
the crossing symmetry of four point function. Solutions
to these constraints are found and consistency checks
are performed in other sections.

First let us study the reflection relation.
The information of reflection relation
can be read off from the coefficient $D(h) $ \eqref{2ptD}
in the two point function \eqref{2ptfn} as in the cases of
Liouville field theory and SL(2) WZNW model.
It might be convenient to define
\begin{align}
 V_h (x|z) = \Phi_h (x , 0 | z)  ~,
 \qquad
 W_h (x  | z) = \partial_{\bar \xi}
 \partial_\xi \Phi_h (x , \xi | z)   ~,
\end{align}
then the two point functions are
\begin{align}
& \left\langle V_h (x | z ) V_{h '} (x ' | z') \right \rangle
    = \frac{ \delta (h - h ') D(h) }{ | z - z ' |^{4 \Delta_h}| x - x ' |^{4h} } ~,
    \\
 &\left\langle W_h (x | z ) W_{h '} (x ' | z ') \right \rangle
    = - \frac{ (2h)^2 \delta (h - h ') D(h) }{| z - z ' |^{4 \Delta_h} | x - x ' |^{4h + 2} }
     ~,
\end{align}
and moreover
\begin{align}
 &\left\langle W_{\frac12 - h} (x | z ) V_{h '} (x ' | z ) \right \rangle
    = \frac{ \pi \delta (h - h ') \delta^{(2)} (x - x ') }{ | z - z ' |^{4 \Delta_h} } ~,
    \\
 &\left\langle V_{\frac12 - h} (x | z ) W_{h '} (x ' | z ') \right \rangle
    =\frac{ \pi\delta (h - h ') \delta^{(2)} (x - x ') }{ | z - z ' |^{4 \Delta_h} }  ~.
    \nonumber
\end{align}
{}From the above forms we can see that there should be a relation
between $V_h$ and $W_{\frac12 - h}$ or $W_h$ and $V_{\frac12 - h}$.
Explicitly, the reflection relations are given by
\begin{align}
 &V_h ( x |z ) = \frac{D(h)}{\pi} \int d^2 x ' |x - x '|^{- 4 h}
 W_{\frac12 - h} (x' | z )~, \\
 &W_h ( x | z ) = - (2h)^2 \frac{D(h)}{\pi} \int d^2 x ' |x - x '|^{- 4 h - 2}
 V_{\frac12 - h} (x' | z )~.
 \nonumber
\end{align}
Since the vertex operators satisfy the above reflection relations,
the one point function \eqref{ads2} should respect these relations.
This yields a constraint on the function $U^\sigma (h ; \epsilon)$
\begin{align}
 U^{ - \sigma } (h ; \epsilon )
 = \epsilon D(h) U^\sigma (\tfrac12 - h ; \epsilon ) ~,
 \label{1ptconst}
\end{align}
which can be obtained as in \cite{PST,LOP}
with the help of \eqref{intform}.
Notice that the condition \eqref{1ptconst} relates the
function $U^\sigma (h ; \epsilon )$ to
the function with the opposite sign
$U^{ - \sigma } (\frac12 -  h ; \epsilon )$.

Next we move to the crossing symmetry.
We study the following two point function on a disk as
\begin{align}
 Z_2^o =\left \langle \Phi_{\frac{k}{2}} (x_1 , \xi_1 | z_1 )  \Phi_h (x_2 , \xi_2 | z_2 ) \right \rangle
 \label{o2pt}
\end{align}
with the operator $\Phi_{k/2}$
belonging to a degenerate representation.
In the limit of $z_1 \to z_2$, we can use the operator product
expansion of $\Phi_{k/2}$ and $\Phi_h$, thus we obtain a sum of one point functions. On the other hand, in the limit
of ${\rm Im}\, z_2 \to 0$, the two point function can be expanded
in terms of boundary operators, and it reduces to a one point
function if we pick up the contribution with identity boundary
operator. Comparing these two expressions, we obtain a constraint
equation for the coefficient $U^\sigma (h;\epsilon)$.

In the limit of  $z_1 \to z_2$, the bulk operators can be
expanded as
\begin{align}
\Phi_{ \frac{k}{2}} (x_1 , \xi_1 | z_1) \Phi_{h} (x_2 , \xi_2 | z_2)
  &= | z_{12} |^{2 \Delta_{\frac{k}{2} - h} - 2 \Delta_{\frac{k}{2}} - 2 \Delta_{h} } |X_{12}|^{- 4 h} C (h) [ \Phi_{\frac{k}{2} - h} ]_\text{ee}
 \label{oped2}
  \\ \nonumber
 & + | z_{12} |^{2 \Delta_{\frac{k}{2} - h - \frac12} - 2 \Delta_{\frac{k}{2}} - 2 \Delta_{h} } |X_{12}|^{- 4 h - 2} \tilde C (h) [ \Phi_{\frac{k}{2} - h - \frac12 } ]_\text{oo} ~,
\end{align}
as mentioned before. The coefficients are computed in appendix
\ref{candtc} as
\begin{align}
 C(h) = D(h) ~, \qquad \tilde C(h) = \frac{D(h)}{b^2 \nu  \gamma ( b^2 (k - 2h - 1)) \gamma (2 b^2 h)} ~.
\end{align}
Here $D(h)$ was given in \eqref{2ptD}.
With the help of this operator product expansion, the
two point function can be written as%
\footnote{
We have two choices of sign in the coefficient $U^{\pm \sigma}$,
and here we adopt the minus sign. Notice that one
reflection relation should be applied as in \eqref{1ptconst}
when we compute the OPE in \eqref{oped2}.
Similar computation can be done for
the SL(2) WZNW model, and the same sign should be used to
reproduce the known result.
}
\begin{align}
\label{schannel}
 Z_2^o &= | z_1 - \bar z_2 |^{- 4 \Delta_h}
         | z_1 - \bar z_1 |^{2 \Delta_h - 2 \Delta_{\frac{k}{2}}}
         | x_1 + \bar x_2 + \epsilon \xi_1 \bar \xi_2 |^{- 4 h}
         | x_1 + \bar x_1 + \epsilon \xi_1 \bar \xi_1 |^{2 h - k}
         \\ &\times \left( C( h)
        U^{- \sigma} (\tfrac{k}{2} - h;\epsilon)
   {\cal F}^S_1 (X , \chi_2 ,\chi_3 ; z )
         + \epsilon \tilde C( h)
        U^{- \sigma} (\tfrac{k}{2} - h - \tfrac12 ;\epsilon) {\cal F}^S_2  (X, \chi_2 , \chi_3 ; z)
        \right) \nonumber ~.
\end{align}
The functions ${\cal F}^S_1$ and ${\cal F}^S_2$ are defined in \eqref{fsi},%
\footnote{We should replace $x(1-\eta \xi )$ and $\eta \xi$ by
the cross ratios $X$ and $\chi_3 \chi_2$, respectively.}
and behave as in \eqref{fs1} and \eqref{fs2}, respectively.
Two point function of bulk operators on a disk can be mapped to four point function of chiral operators on a plane by the standard mirror trick.
Because of the boundary conditions \eqref{bcads2} the parameters are $(h_3,x_3,\xi_3,z_3)
 = (h, - \bar x_2 , \epsilon \xi_2,\bar z_2)$,
$(h_4,x_4,\xi_4,z_4)
 = (k/2, - \bar x_1 , \epsilon \xi_1,\bar z_1)$
along with $h_1 = k/2$ and $h_2 = h$.
In this set up, the cross ratios are, for example,
\begin{align}
 z = \frac{ | z_1 - z_2 | ^2 }{ | z_1 - \bar z_2 |^2} ~,
  \qquad
X = \frac{ | x_1 - x_2 + \xi_1 \xi_2 | ^2 }
  { | x_1 + \bar x_2 + \epsilon \xi_1 \bar \xi_2 |^2} ~.
\end{align}
The fermionic cross rations $\chi_2,\chi_3$ can be
written in a similar way.

On the other hand, when we take the limit of ${\rm Im} \, z_2 \to 0$,
the two point function can be expanded in terms of
boundary operators as intermediate states.
The degenerate operator $\Phi_{k/2}$ can be expanded
by boundary operators near ${\rm Im} \, z_1 \sim 0$,
and among them there is the identity operator.
Therefore, if we pick up the contribution with the
identity operator as an intermediate state, then we have
\begin{align}
 Z_2^o \sim | z_1 - \bar z_1 |^{- 2 \Delta_{\frac{k}{2}}}
           | x_1 + \bar x_1 + \epsilon \xi_1 \bar \xi_1 |^{- k}
           A^{\sigma}(\tfrac{k}{2};\epsilon)
           \left \langle \Phi_h (x_2 , \xi_2 | z_2 ) \right \rangle ~.
           \label{tchannel}
\end{align}
For rational conformal field theories, factors like
$A^\sigma (k/2 ; \epsilon)$ are proportional to one point functions as
$\langle \Phi_{k/2} \rangle$. However, for non-rational
conformal field theories, such as, OSP(1$|$2) model, this
is not always true. Therefore, here we treat $A^\sigma (k/2 ; \epsilon)$
as parameters. See \cite{PST,LOP} for more detail.

We can relate these two expressions by utilizing
channel duality from $s$-channel to $t$-channel.
With the help of channel duality formula in \eqref{stdual},
we can rewrite the expression \eqref{schannel} in a
suitable way to expand around $\text{Im}\, z_2 \to 0$.
Then, compared with the expression \eqref{tchannel},
we have the constraint equation as%
\footnote{The conformal blocks \eqref{fsi} actually have a singularity
at $z \sim x$ as discussed in \cite{Ponsot}. On a sphere it might not
lead to any problem, since we can avoid the singularity by analytic continuation,
which yields a phase $\exp(- \pi i \gamma_{24})$ in \eqref{stdual}.
Anyway, only products of holomorphic and anti-holomorphic expressions enter.
However, on a disk, this singularity has a significant meaning, since both $z$ and $x$
take real values, and hence we cannot connect the regions
$z < x$ and $z > x$ by analytic continuation  \cite{HR}.
Since $0 < x < 1$ in our case, we have to cross the singularity
when we compare the expressions with $z \sim 0$ and with $z \sim 1$.
In this sense the constraint coming from the crossing symmetry is weaker
than in the bulk case as discussed in \cite{HR}.
Here we assume that  phases like $\exp(- \pi i \gamma_{24})$ do not appear,
and this assumption  may be confirmed
by the comparison with the results obtained in next section.
 }
\begin{align}
 C(h) U^{ - \sigma} (\tfrac{k}{2} - h;\epsilon) F^{ST}_{11}
  + \epsilon \tilde C(h)
  U^{ - \sigma} (\tfrac{k}{2} - h - \tfrac12 ;\epsilon) F^{ST}_{21}
  = U^\sigma (h;\epsilon) A^\sigma (\tfrac{k}{2};\epsilon) ~.
  \label{1ptconst2}
\end{align}
In order to solve this equation, it is useful to set
\begin{align}
 U^\sigma (h;\epsilon) = \nu^{- 2 h + \frac{1}{2} }
 \Gamma (\tfrac12 - b^2 (2h - \tfrac12)) E^\sigma (h;\epsilon) ~,
\end{align}
then the constraint coming from the reflection relation
\eqref{1ptconst} can be written as
\begin{align}
 E^{- \sigma} (h ; \epsilon ) = \epsilon E^\sigma (\tfrac{1}{2} - h; \epsilon) ~.
\end{align}
Moreover, the other constraint equation \eqref{1ptconst2}
reduces to
\begin{align}
 E^{- \sigma} ( \tfrac{k}{2} - h ; \epsilon)
  - \epsilon E^{- \sigma} ( \tfrac{k}{2} - h - \tfrac12; \epsilon)
  = \nu^{k - 1} \frac{\Gamma (1 + b^2 (1 - k))}{\Gamma (- b^2)}
 A^\sigma (\tfrac{k}{2};\epsilon) E^{ \sigma} ( h ; \epsilon)~.
\end{align}
A solution satisfying the both constraints
\eqref{1ptconst} and \eqref{1ptconst2} is given by
\begin{align}
 E^\sigma (h ; + 1) =   A_b
  e^{ - \sigma (2h - \frac12 ) r } ~, \qquad
   E^\sigma (h ; - 1) =  \sigma A_b
  e^{ - \sigma (2h - \frac12 ) r }
\end{align}
with
\begin{align}
A^\sigma (\tfrac{k}{2};\epsilon)
  = \nu^{1-k}
  ( \epsilon e^{ \sigma k r } - e^{\sigma (k - 1)r} )   \frac{\Gamma (- b^2)}{\Gamma (1 + b^2 (1 - k))} ~.
 \end{align}
 Here $A_b$ is a constant.

In this way, we obtain the one point functions of bulk operator
on a disk as
\begin{align}
 &\left \langle \Phi_h (x , \xi | z) \right \rangle_{r; + 1} =
 A_b  \frac{\nu^{- 2 h + \frac{1}{2} }
 \Gamma (\tfrac12 - b^2 (2h - \tfrac12))
  e^{ - \sigma (2h - \frac12 ) r } }{|z - \bar z|^{2 \Delta_h }|x + \bar x +  \xi \bar \xi|^{2 h } } ~, \label{ads1pt1} \\
 &\left \langle \Phi_h (x , \xi | z) \right \rangle_{r; - 1} =
 \sigma A_b  \frac{ \nu^{- 2 h + \frac{1}{2} }
 \Gamma (\tfrac12 - b^2 (2h - \tfrac12))
  e^{ - \sigma (2h - \frac12 ) r } }{|z - \bar z|^{2 \Delta_h }|x + \bar x - \xi \bar \xi|^{2 h } }  \label{ads1pt2}
\end{align}
for $\epsilon = + 1$ and  $\epsilon = - 1$, respectively.
The parameter $r$ is related to the boundary condition and
it represents the position of brane as seen before.
Solutions to the constraints are not unique, but we can see
the above choice is the proper one by checking it in several ways.
We already see that the semiclassical limit ($k\rightarrow\infty$ and thus $b\rightarrow 0$)  
of these one-point functions agrees with our semi-classical analysis \eqref{eq:semiclassicalADSplus} and \eqref{eq:semiclassicalADSminus}.

\subsection{Fuzzy supersphere branes}

As mentioned above, we can take another type of boundary condition
as in \eqref{bcfuzzy}, and in this case the one point function
is restricted to have the form of \eqref{fuzzy} due to the symmetry.
In order to obtain the coefficient $U(h;\epsilon)$, we again
utilize the constraint, particularly from the crossing symmetry.
For this purpose we examine the two point function \eqref{o2pt}
now with the different boundary condition \eqref{bcfuzzy}.
The two point function of bulk operators can be mapped to
four point function of chiral operator as before, but the
parameters are $(h_3,x_3,\xi_3,z_3)
 = (h, - 1/\bar x_2 , \epsilon \bar \xi_2/ \bar x_2 ,\bar z_2)$,
$(h_4,x_4,\xi_4,z_4)
 = (k/2, - /\bar x_1 , \epsilon \bar \xi_1 / \bar x_1,\bar z_1)$
along with $h_1 = k/2$ and $h_2 = h$.
In particular, the bosonic cross ratio is  given by
\begin{align}
 X = - \frac{ | x_1 - x_2 + \xi_1 \xi_2 | ^2 }
  { |1 + x_1 \bar x_2 + \epsilon \xi_1 \bar \xi_2 |^2} ~.
  \label{minus}
\end{align}
As before, we first express the two point function in the
limit of $z_1 \to z_2$ as in \eqref{schannel}.
Then, we study the limit of $\text{Im} \, z_2 \to 0$
as in \eqref{tchannel}. However, here we assume that
the parameter $A(\frac{k}{2}; \epsilon)$ is the same
as the one point function 
$A(\frac{k}{2}; \epsilon) = U(\frac{k}{2} ; \epsilon) $.
With the help of channel duality formula \eqref{stdual}, we can rewrite
the expression for  $z_1 \to z_2$ in terms of those for
$\text{Im} \, z_2 \to 0$. Comparing to the other expression
we can obtain the constraint equation as for super AdS$_2$ brane.

The constraint equation corresponding to \eqref{1ptconst2}
can be found as
\begin{align}
 C(h) U (\tfrac{k}{2} -
 h;\epsilon) F^{ST}_{11}
  - \epsilon \tilde C(h)
  U (\tfrac{k}{2} - h - \tfrac12 ;\epsilon) F^{ST}_{21}
  =  U (h;\epsilon) U (\tfrac{k}{2};\epsilon) ~.
\end{align}
As before it is convenient to set
\begin{align}
 U (h;\epsilon) = \nu^{- 2 h + 1 }
 \Gamma (\tfrac12 - b^2 (2h - \tfrac12)) E (h;\epsilon) ~,
\end{align}
then the above constraint equation reduces to
\begin{align}
 E( \tfrac{k}{2} - h ; \epsilon)
  +  \epsilon E ( \tfrac{k}{2} - h - \tfrac12; \epsilon)
  = \Gamma (b^2 ( k - 2 ))
 E (\tfrac{k}{2};\epsilon) E ( h ; \epsilon)~.
\end{align}
For $\epsilon = + 1$, we can find a solution
\begin{align}
 E (h) = \frac{\sin (s (2 h - 1/2))}
 {\sin (- s/2) \Gamma (b^2 ( k - 2 ))} ~,
 \label{sol1}
\end{align}
where $s = 2 \pi b^2 n$ with a positive integer $n$.
On the other hand we have for $\epsilon = - 1$
\begin{align}
 E (h) =  \frac{\cos (s (2 h - 1/2))}
 {\cos (- s/2) \Gamma (b^2 ( k - 2 ))} ~,
\end{align}
where $s = 2 \pi b^2 (n + 1/2)$ with a non-negative integer $n$.
Thus the one point functions are summarized as
\begin{align}
 &\left \langle \Phi_h (x , \xi | z) \right \rangle_{s; + 1} =
  \frac{\nu^{- 2 h + 1 }
 \Gamma (\tfrac12 - b^2 (2h - \tfrac12))
   }{|z - \bar z|^{2 \Delta_h }|x + \bar x +  \xi \bar \xi|^{2 h } } \frac{\sin (s (2 h - 1/2))}
 {\sin (- s/2) \Gamma (b^2 ( k - 2 ))}~, \label{fuzzy1pt1} \\
 &\left \langle \Phi_h (x , \xi | z) \right \rangle_{s; - 1} =
  \frac{ \nu^{- 2 h + 1 }
 \Gamma (\tfrac12 - b^2 (2h - \tfrac12))
  }{|z - \bar z|^{2 \Delta_h }|x + \bar x - \xi \bar \xi|^{2 h } }  \frac{\cos (s (2 h - 1/2))}
 {\cos (- s/2) \Gamma (b^2 ( k - 2 ))} \label{fuzzy1pt2} 
\end{align}
for $\epsilon = + 1$ and  $\epsilon = - 1$, respectively.
These solutions are not unique either, and we check them by
examining the Cardy condition in section 6.
In this case the semiclassical analysis with \eqref{eq:semiclassicalsphere1} and \eqref{eq:semiclassicalsphere2} differs slightly from our results, i.e. the branes' position labeled by $s$ is purely imaginary. This is not unexpected since the same already happened for spherical branes in $H_3^+$ model \cite{PST}.

\section{Relation to boundary super-Liouville theory}
\label{sec:bsLt}

In the previous section, we have computed one point functions
of bulk operators with two different boundary conditions by utilizing
the crossing symmetry. In this section, we reproduce the disk
amplitudes for super AdS$_2$ branes
by making use of the relation to super-Liouville
field theory with boundary. In fact, the structure constants
of bulk OSP(1$|$2) WZNW model have been computed by utilizing
the relation to super-Liouville field theory in \cite{HS2}.
The relation is a generalization of the one by Ribault and
Teschner in \cite{RT}, which relates sphere amplitudes of
the SL(2) WZNW model (or the $H_3^+$ model) and those
of Liouville field theory.
The extension of Ribault-Teschner relation to disk amplitudes
has been done in \cite{HR}, see also \cite{FR}.

\subsection{Boundary super-Liouville theory}

Let us first review ${\cal N}=1$ super-Liouville field theory
with boundary by closely following \cite{FH}.
The theory consists of a bosonic field $\varphi$ and its super
partner $\psi$, whose action is given by
\begin{align}
   S^{L}
   \ = \ \frac{1}{2\pi} \int d^2 z \left[  \partial \varphi \bar
   \partial \varphi
   + \frac{Q_\varphi}{4} \sqrt g {\cal R} \varphi
  + \psi \bar \partial \psi + \bar \psi \partial \bar \psi \right] +
  2 i \mu_L b^2 \int d^2 z \psi \bar \psi e^{ b \varphi}  ~,
\end{align}
where ${\cal R}$ represents the curvature of the worldsheet.
Here the background charge is related to the parameter $b$
by $Q_\varphi = b + 1/b$, and the central charge is $c = 3/2 + 3 Q_\varphi^2$.
In the NSNS-sector, primary fields are
$V_{\alpha} = e^{\alpha \varphi}$ with conformal weight
$\Delta_\alpha ^L = \alpha (Q - \alpha)/2$.
In the RR-sector, they are defined with
chiral spin fields $\sigma^{\pm}$ and  $\bar \sigma^{\pm}$,
which have operator products with the fermions as
\begin{align}
 \psi (z) \sigma^{\pm} (0) &\sim \frac{\sigma^{\mp}(0)}{\sqrt2 z^{\frac{1}{2}}} ~,
& \bar \sigma^{\pm} (\bar z) \bar \psi (0)
  &\sim \frac{i \bar \sigma^{\mp}(0)}{\sqrt2 \bar z^{\frac{1}{2}}} ~.
\end{align}
With non-chiral products $\sigma^{ a \bar a} = \sigma^a
\bar \sigma^{\bar a}$ ($a, \bar a  = \pm $),
we define primary fields in the RR-sector as
$\Theta^{a \bar a}_\alpha =
\sigma^{a \bar a} e^{\alpha \varphi}$.
Their conformal weights are
$\tilde \Delta_\alpha ^L = \alpha (Q - \alpha)/2 + 1/16$.

In the presence of boundary we can add boundary terms as
\begin{align}
 S^L_B = \int dt \left[ \frac{1}{4 \pi} Q_\varphi \sqrt{g} {\cal K} \varphi + \frac14 \Theta \partial_t \Theta +
  \mu_B b \Theta \psi e^{b \varphi /2} \right] ~.
  \label{bLaction}
\end{align}
Here we denote the curvature of boundary as ${\cal K}$,
and $\Theta$ is a Grassmann odd field living only at the boundary.
In particular, free field correlator is given by
$\langle \Theta (t_1 ) \Theta (t_2) \rangle = \text{sgn} (t_1 - t_2)$.
As boundary conditions we assign
\begin{align}
 T(z) = \bar T(\bar z) ~ , \qquad T_F (z) = \zeta \bar T_F (\bar z )
 \label{bczeta}
\end{align}
at $z = \bar z$, where $T$ is the energy momentum tensor and $T_F$ is
its superpartner.
The boundary operators are defined as
\begin{align}
 B_\beta (t) = e^{\beta \varphi /2 } (t ) =  e^{\beta \varphi_L} (t)~,
 \qquad \Theta^\pm_\beta (t) = \sigma^\pm e^{\beta \varphi /2} (t)
 = \sigma^\pm e^{\beta \varphi_L} (t) ~,
\end{align}
which are inserted at the boundary of worldsheet. Here
$\varphi_L$ is the holomorphic part of the original field
$\varphi$.

As we will see below,  bulk one point functions on a disk
in the OSP(1$|$2) WZNW model are related to those in
super-Liouville field theory.
These one point functions in super-Liouville field theory are
of the form 
\begin{align}
 \langle V_\alpha (z) \rangle_{u_\zeta} =
  \frac{U_+ (\alpha ; u_\zeta) }{| z - \bar z |^{ 2 \Delta_\alpha ^L }} ~,
  \qquad
   \langle \Theta ^{a a}_\alpha (z) \rangle_{u_\zeta}
  =  \frac{U_- (\alpha , a ; u_\zeta )}{ | z - \bar z |^{ 2 \tilde  \Delta_\alpha ^L }} ~.
  \label{1pt_L}
\end{align}
Here we have assigned the boundary condition \eqref{bczeta} with
parameter $\zeta$. The subscript $u_\pm$ represents the boundary
condition and is related to the parameter $\mu_B$ in
\eqref{bLaction} as
\begin{align}
 \mu_B =
\left ( \tfrac{2 \mu _L}{\cos (\pi b^2 /2)}\right )^{\frac12} \sinh (\pi u_+ b)
 ~, \qquad
 \mu_B =
\left ( \tfrac{2 \mu _L}{\cos (\pi b^2 /2)}\right )^{\frac12} \cosh (\pi u_- b)
~. \label{mu2u}
\end{align}
The coefficients are obtained in \cite{FH} as
\begin{align}
 \label{FHNS}
 &U_+ = - 2^{- \frac{1}{2}} \pi^{- \frac12}
 (\mu_L \pi \gamma (\tfrac{bQ_\varphi}{2} ))^{\frac{2 \alpha - Q_\varphi}{2b}}
  ( \alpha - \tfrac{Q\varphi}{2} )
  \Gamma (b (\alpha - \tfrac{Q_\varphi}{2} ) )
 \Gamma (\tfrac{1}{b} ( \alpha - \tfrac{Q_\varphi}{2}) ) \hat U _+ ~, \\
 &\hat U _+  (\alpha  ; u_\pm ) = \cosh (\pi (2 \alpha - Q_\varphi ) u_\pm )
 \nonumber
\end{align}
for the NSNS-sector and
\begin{align}
\label{FHR}
 &U_- = 2^{-\frac12} \pi ^{-\frac12}
 (\mu _L \pi \gamma (\tfrac{b Q_\varphi}{2}))^{\frac{ 2 \alpha - Q_\varphi }{2b}}
 \Gamma (\tfrac12 + b(\alpha - \tfrac{Q_\varphi}{2}) )
 \Gamma (\tfrac12 + \tfrac{1}{b}(\alpha - \tfrac{Q_\varphi}{2}) ) \hat U_-
 ~, \\
 &\hat U _ - (\alpha , a ; u_+ )=
 a \sinh (\pi (2 \alpha - Q_\varphi) u_+ ) ~,
 \qquad
 \hat U _ - (\alpha , a ; u_- )=
 \cosh (\pi (2 \alpha - Q_\varphi) u_- )
\nonumber
\end{align}
for the RR-sector.

\subsection{Boundary OSP(1$|$2) model}

The action of WZNW model associated with supergroup can be obtained
as in the case with bosonic group. First adopt a specific parametrization
of the (super) group element, and then insert it in the general expression of
WZNW action. Introducing auxiliary fields, we may express the action
of the OSP(1$|$2) WZNW model as \cite{HS2}
\begin{align}
S^O = \frac{1}{\pi} \int d^2 z & \left[ \frac12 \partial \phi
 \bar \partial \phi + \frac{b}{8} \sqrt{g} {\cal R} \phi
 + \beta \bar \partial \gamma + \bar \beta \partial \bar \gamma
 + p \bar \partial \theta + \bar p \partial \bar \theta  \right]
 \\ & + i \lambda
 \int d ^2z  (p + \beta \theta )
 (\bar p - \bar \beta \bar \theta) e^{b \phi} ~.\nonumber
\end{align}
Namely, we have a free boson $\phi$ with background charge
 $Q_\phi=b = 1/\sqrt{2k-3}$, and
$(\beta , \gamma)$-system with the conformal weights $(1,0)$.
In addition to these bosonic fields, there are free fermions
$(p,\theta)$ with the conformal weights $(1,0)$.
The central charge of the system is $c = 1 + 3 b^2 = \frac{2k}{2k-3}$
as desired.
The interaction terms can be treated perturbatively.

In order to define vertex operators, it is convenient to
bosonize the fermionic fields as
\begin{align}
 \theta = \exp ( i Y_L ) ~, \qquad p = \exp ( - i Y_L) ~, \qquad
 \bar \theta = \exp ( i Y_R ) ~, \qquad \bar p = \exp ( - i Y_R) ~,
\end{align}
and then define $Y = Y_L + Y_R$.
Using the new field the vertex operators are
\begin{align}
V_h^s  (\mu | z) = |\mu |^{-2h+1+s}
  e^{is Y} e^{\mu \gamma - \bar \mu \bar \gamma }
  e^{ - 2b(h - \frac12 ) \phi}
  \label{mubasisex}
\end{align}
with $s=0,1$ in the RR-sector and $s=1/2$ in the NSNS-sector.
In the RR-sector and the NSNS-sector, the conformal weights are
$\Delta_h = - 2 b^2 h (h -  1/2)$ and
$\tilde \Delta_h = - 2 b^2 h (h -  1/2) - 1/8$, respectively.
For the vertex operators with $s = 0,1$,we can change the
basis by the Fourier transform as
\begin{align}
 V_h^{\pm} (\mu | z ) = \frac{1}{\pi}
  | \mu |^{-2h+2} \int d^2 x \int d \bar \xi d \xi
  e^{\mu x - \bar \mu \bar x } (1 \pm | \mu |^{-1} \xi \bar \xi )
  \Phi_h (x,\xi | z) ~.
\end{align}
Here 
\begin{align}
 V_h^\pm (\mu | z) = |\mu |^{-2h+1} ( | \mu | \bar \theta \theta \pm 1 )
  e^{\mu \gamma - \bar \mu \bar \gamma }
  e^{ - 2b(h - \frac12 ) \phi}
  \label{vhpm}
\end{align}
are the orthogonal basis as discussed in \cite{HS2}
and $\Phi_h (x,\xi | z) $ is defined in \eqref{pfinx}.

Now we would like to include a boundary to the worldsheet, and
hence to find out proper boundary terms.
In the $H_3^+$ model, the corresponding boundary terms are
constructed in \cite{FR}, and in our case they are discussed in
appendix \ref{app:bdyaction} (see also appendix \ref{app:b2pt}).
For super AdS$_2$ branes they are given by
\begin{align}
 S_B^O =
  \int du \left[ \frac{1}{2 \pi}\frac{b}{8} \sqrt{g} {\cal K }\phi
 +   \frac{1}{4} \Theta  \partial_u \Theta +
 \lambda_B  e^{b \phi /2} \Theta
 (\beta  \theta + p ) \right] \nonumber
\end{align}
with $\beta = - \bar \beta $ and $p = \zeta \bar p$ at the boundary.
In terms of bosonization, the boundary condition for fermions are
mapped to $Y_L = Y_R + i \ln \zeta$.
The parameter $\lambda_B$ plays the same role as
$\mu_B$ in super-Liouville field theory.
The boundary operators may be defined as
\begin{align}
 B^\tau_l (\upsilon | u)= |\upsilon|^{- l +\frac12 + \frac{\tau}{2}} e^{i \frac{\tau}{2} Y}
  e^{\frac12 (\upsilon \gamma - \upsilon \bar \gamma)}
  e^{ - b(l - \frac12 ) \phi}
\end{align}
with $\tau = 0,1/2,1$.
Next task is to rewrite
the following generic correlation function
\begin{align}
 \Omega = \left \langle \prod_{i=1}^n V_{h_i}^{s_i} (\mu_i | z_i )
 \prod_{a = 1}^m B_{l_a}^{\tau_a} (\upsilon_a  | u_a ) \right \rangle
 \label{ospcorr}
\end{align}
on a disk in terms of that of super-Liouville field theory.

\subsection{Relation between the two theories}

In \cite{HS} it was shown that the relation between $H_3^+$ model
and Liouville theory in \cite{RT} can be rederived in the path integral
formulation. Utilizing this method the relation has been extended to that for
sphere amplitudes of OSP(1$|$2) model and super-Liouville theory in \cite{HS2}.
In this subsection we extend the result to the boundary case.
In the path integral formulation the correlation function \eqref{ospcorr}
is written as
\begin{align}
 \Omega = \int d \phi d^2 \gamma d^2 \beta d^2 p d^2 \theta
 e^{- S^O - S^O_B}
  \prod_{i=1}^n V_{h_i}^{s_i} (\mu_i | z_i )
 \prod_{a = 1}^m B_{l_a}^{\tau_a} (\upsilon_a  | u_a ) ~.
 \label{ospcorrpath}
\end{align}
Following \cite{HS2} we first integrate over the fields $\beta,\gamma$,
and then perform field redefinitions to reduce to super-Liouville field
theory. The analysis is quite analogous to the bulk case, so here
we explain it only briefly.

We start from the observation that integration over the zero mode of
$\gamma - \bar \gamma $ leads to the momentum conservation
\begin{align}
 \sum_{i=1}^n ( \mu_i + \bar \mu_i ) + \sum_{a=1}^m \upsilon_a = 0 ~.
\end{align}
Then we move to the non-zero modes.
Since $\gamma $ appears only linearly in the exponent
of the path integral expression \eqref{ospcorrpath}, we can integrate
them out, which yields delta-functionals of $\beta$.
After integration over $\beta$, the field $\beta (z)$ is replaced
by a function ${\cal B} (z)$ defined by
\begin{align}
 &{\cal B} (z) = - \sum_{i=1}^n \frac{\mu_i}{z - z_i}
  - \sum_{i=1}^n  \frac{\bar \mu_i}{z - \bar z_i}
   - \sum_{a=1}^m \frac{\upsilon_a}{z - u_a} ~.
\end{align}
In the same way, $\bar \beta (z)$ is replaced
by $ -  \bar {\cal B} (\bar z)$.
As in \cite{FH} it is essential to define
$y_{i'} , \bar y_{i'} , t_{a '}$ by
\begin{align}
{\cal B} (z) = u \frac{\prod_{{i'}=1}^{n'} (z - y_{i'})(z - \bar y_{i'})
 \prod_{a' = 1}^{m ' }(z - t_{a'})}
 {\prod_{{i}=1}^{n} (z - z_{i})(z - \bar z_{i})
 \prod_{a = 1}^m(z - u_{a})}
\end{align}
with $2 n' + m' = 2 n + m -2$. Up to the permutation of $y's$ or $t's$,
the above equation defines a map from the old variables $\mu,\bar\mu,\upsilon$
to the new variables $u, y, \bar y , t$.

Since the field $\beta(z)$ is now replaced by ${\cal B} (z)$,
the coefficients in the action are some functions depending on $z$.
In order to remove the coordinate dependence, we perform the shifts as
\begin{align}
 \varphi := \phi + \frac{1}{2b} | {\cal B}|^2 ~, \qquad
 Y_L ' := Y_L - \frac{i}{2} \ln {\cal B} ~. \qquad 
  Y_R ' := Y_R - \frac{i}{2} \ln \bar {\cal B} ~. 
  \label{yprime}
\end{align}
If we fermionize the new boson $Y' = Y'_L + Y'_R$ by
\begin{align}
 \psi \pm i \chi = \sqrt{2} \exp (\pm i Y_L ') ~, \qquad
 \bar \psi \pm i \bar \chi = \sqrt{2} \exp (\pm i Y_R ') ~,
\end{align}
then we find that the action becomes the one of super-Liouville field theory
plus the one of free fermion $(\chi,\bar \chi)$
\begin{align}
   S^{F}
   \ = \ \frac{1}{2\pi} \int d^2 z \left[
   \chi \bar \partial \chi + \bar \chi \partial \bar \chi \right]  ~.
\end{align}
The boundary conditions for fermions are mapped to
\begin{align}
 \psi = \zeta \text{sgn} \, \cal{B} \bar \psi ~, \qquad
 \chi = \zeta \text{sgn} \, \cal{B} \bar \chi ~.
\end{align}
This is because $Y'_L$ and $Y'_R$ are defined in \eqref{yprime},
and ${\cal B} = \bar {\cal B} =  | {\cal B} | $ or 
$e^{- \pi i }{\cal B}= e^{\pi i}  \bar {\cal B} =  | {\cal B} | $
at the boundary.
The relation between parameters of boundary term  are given by
\begin{align}
\sqrt{2 \, \text{sgn} \, \cal B } \lambda_B =  \mu_B b  ~.
 \label{relcou}
\end{align}

During the change of variables, we receive extra contributions from
kinetic terms. One is the insertion of extra fields
$V^{1/2}_{-1/2b}(y_{i'})$ and $B^{1/2}_{-1/2b} (t_{a'})$ with
\begin{align}
V^s_\alpha = e^{i s Y'} e^{\alpha \varphi} ~, \qquad
B^\tau_\beta =  e^{i \tau Y'/2} e^{\beta \varphi /2 } ~,
\end{align}
and another is an overall factor $|u| |\Xi|^{^{\frac{1}{4b^2}- \frac{1}{4}} }$ with
\begin{align}
  \Xi &=  \prod_{i < j} |z_{ij} |^2 \prod_{i , j} ( z_i - \bar z_j )
  \prod_{i , a} | z_i - u_a |^2 \prod_{a < b} u_{ab}
    \prod_{i '< j'} |y_{i'j'} |^2 \prod_{i' , j'} ( y_{i'} - \bar y_{j'} )
  \prod_{i' , a'} | y_{i'} - t_{a'} |^2 \prod_{a' < b'} t_{a'b'}
  \nonumber \\
  &\times \prod_{i , i'} |z_i - y_{i'}|^{-2}
  \prod_{i,i'}|z_i - \bar y_{i'}|^{-2}
   \prod_{i, a'} |z_i - t_{a'}|^{-2}
   \prod_{a , i'} | u_a - y_{i'} |^{-2}
   \prod_{a, a'} (u_a - t_{a'})^{-1} ~.
\end{align}
In the end we arrive at the expression as
\begin{align}
  \label{RT}
 \Omega &= \delta ( \sum_{i} (\mu_i + \bar \mu_i ) + \sum_a \nu_a ))
 |u|  |\Xi|^{^{\frac{1}{4b^2}- \frac{1}{4}} }
  \times \\ \nonumber &\times
 \left \langle \prod_{i=1}^n
 V^{s_i-\frac{1}{2}}_{\alpha_i} (z_i)
  \prod_{a=1}^m B^{\tau_a -\frac{1}{2}}_{\beta_a}  (u_a)
  \prod_{i ' =1 }^{n '} V^{\frac{1}{2}}_{-\frac{1}{2b}} (y_{i'})
  \prod_{a ' = 1}^{m '} B^{\frac{1}{2}}_{-\frac{1}{2b}} (t_{a'})
  \right \rangle ~,
\end{align}
where
\begin{align}
 \alpha_i = - 2b(h_i - \tfrac{1}{2}) + \tfrac{1}{2b} ~, \qquad
 \beta_a = - 2b (l_a - \tfrac12 ) + \tfrac{1}{2b} ~.
 \label{mshift}
\end{align}
The momenta receive extra contributions as in \eqref{mshift},
and  the background charge is shifted from $Q_\phi = b$ to
$Q_\varphi = b  + 1/b$.
The right hand side should be computed in super-Liouville
field theory with a pair of free fermions $(\chi , \bar \chi)$.
Thanks to the condition $2 n' + m ' = 2 n + m - 2 $, we do not
need to include extra fields for one
point functions of bulk operator on a disk and two point functions
of boundary operators. In the next subsection we compute one
point functions of bulk operator on a disk in the OSP(1$|$2)
WZNW model utilizing the relation \cite{RT}.%
\footnote{
If we want to study more complicated cases,
then we need to develop the map between correlators with free
boson $Y'$ and with free fermions $\chi,\psi$ on a worldsheet with boundary.
These cases will not be treated in this paper,
but it might be useful to consult with the work \cite{hori},
which deals with similar problems.
}

\subsection{One point function of bulk operators}

According to \eqref{RT}, one point function of bulk operator
on a disk can be written as
\begin{align}
 \langle V^{s}_h (\mu | z ) \rangle = \delta (\mu + \bar \mu )
 | \mu | | z - \bar  z |^{1 + \frac{1}{4 b^2} - \frac{1}{4}}
 \left \langle V^{s - \frac12}_{ - 2 b (j-\frac{1}{2}) + \frac{1}{2b}} (z) \right \rangle ~,
 \label{1ptfors}
\end{align}
where the right hand side should be computed in the super
Liouville field theory with a pair of free fermions.
Let us first restrict the label to $s=0,1$.
In these cases, the right hand side involves one point function of
$e^{\pm iY'/2}$, which should be rewritten in terms of fermions $\psi,\chi$.
If we define order and disorder operators of Ising model by $\Sigma^\pm$,
then we have a map as
\begin{align}
\sqrt{2} \cos Y '/2 , \sqrt{2} \sin Y '/2  \leftrightarrow   ( \Sigma^+ )^2 ,
   ( \Sigma^- )^2 ~.
\end{align}
With the description of Ising model in terms of free fermion we
moreover have%
\footnote{The choice of $\mp$ may be related to the definition of energy operator ${\cal E} = \pm i \psi \bar \psi$. Moreover, we have a duality transformation as 
${\cal E} \leftrightarrow - {\cal E}$ and $\Sigma^+ \leftrightarrow \Sigma^-$. 
For the bulk case as in \cite{HS2}, it is merely convention. However, in the presence of boundary, it should matter and the
choice should be related to the boundary condition for fermion $\psi$. }
\begin{align}
 \Sigma ^+ = \frac{1}{\sqrt{2}} (\sigma^{++} \mp \sigma^{--}) ~,
 \qquad
 \Sigma ^- = \frac{1}{\sqrt{2 i}} (\sigma^{-+} \mp \sigma^{+-}) ~.
 \label{order}
\end{align}
With these maps we can rewrite the expression of \eqref{1ptfors} in terms of
one point function in the RR-sector of Liouville field theory.
Defining $\Sigma^{\pm}_\alpha = \Sigma^{\pm}_\psi e^{\alpha \varphi}$
and $\Sigma^{\pm}_\chi$, we therefore have
\begin{align}
 \langle V^{\pm}_h (\mu | z ) \rangle \propto \delta (\mu + \bar \mu )
 | \mu | | z - \bar  z |^{1 + \frac{1}{4 b^2} - \frac{1}{4}}
 \left \langle \Sigma^{+}_{ - 2 b (h - \frac{1}{2}) + \frac{1}{2b}} (z) \right \rangle
 \langle \Sigma^{+}_\chi (z) \rangle ~,
\end{align}
where $V^{\pm}_h (\mu | z )$ was defined in \eqref{vhpm}.%
\footnote{Once we fix the bosonization rule, only one sign
of $\langle V^{\pm}_h (\mu | z ) \rangle$ is non-vanishing.}
Notice that one point functions with $\Sigma^-$ vanish.
Furthermore, the one point function of free fermion $\chi$ is of the form as
$\langle \Sigma^+_\chi (z) \rangle = U_\chi^+|z - \bar z|^{ - 1/8}$
with some constant $U_\chi^+$.
The explicit expression of \eqref{FHR} then leads to
\begin{align}
 \langle V^\pm _h (\mu | z ) \rangle =\frac{ \delta (\mu + \bar \mu )
 | \mu | U_h}{ | z - \bar  z |^{ 2 \Delta_h}}
\end{align}
with
\begin{align} \nonumber
 &U_h = A_0
 (\mu _L \pi \gamma (\tfrac{b^2 + 1}{2}))^{ - 2h + \frac12}
 \Gamma (\tfrac12 -  b^2 (2 h  - \tfrac{1}{2}) )
 \Gamma ( - 2 h + 1) \sinh ( 2 \pi b u_+ ( 2 h  - \tfrac{1}{2})) ~, \\
 &U_h = A_0
 (\mu _L \pi \gamma (\tfrac{b^2 + 1}{2}))^{ - 2 h + \frac12}
 \Gamma (\tfrac12 - b^2 (2  h - \tfrac{1}{2}) )
 \Gamma ( - 2 h + 1) \cosh (2  \pi b u_- (2 h - \tfrac{1}{2}))
 \nonumber
\end{align}
for $\zeta \, \text{sgn} \, {\cal B} = + 1$ and 
$\zeta \, \text{sgn} \, {\cal B} = - 1$, respectively.
Here $A_0$ is some constant independent of $h$.
The above results are consistent with the previous ones
\eqref{ads1pt1} and \eqref{ads1pt2} with the identifications of
$\nu = \mu _L \pi \gamma (\tfrac{b^2 + 1}{2})$ and
$\zeta = - \epsilon$.
Moreover, the parameters for boundary condition should be related as
 $2 \pi b u_+ = - r - \pi i /2 \,  \, {\rm sgn} \, {\rm Im} \, \mu  $ and  $2 \pi b u_- = - r - \pi i /2 \, {\rm sgn}\,{\rm Im} \, \mu  $,
which correspond to  \eqref{relcou} through \eqref{mu2u}.

One advantage of this approach is that it is easy to extend to the
NSNS-sector contrary to the method with the crossing symmetry.
Using the relation \eqref{RT} (or \eqref{1ptfors} with $s=1/2$)
one point function of this operator can be written
in terms of super-Liouville field theory as
\begin{align}
 \langle V^{\frac12}_h (\mu | z ) \rangle = \delta (\mu + \bar \mu )
 | \mu | | z - \bar  z |^{1 + \frac{1}{4 b^2} - \frac{1}{4}}
\left \langle V^0_{ - 2 b (h - \frac{1}{2}) + \frac{1}{2b}} (z) \right \rangle ~.
\end{align}
The above correlator does not include spin operators, and hence
we just need to use the expression of \eqref{FHNS} in super-Liouville
field theory. The result is given by
\begin{align}
 \langle V^{\frac12} _h (\mu | z ) \rangle = \frac{ \delta (\mu + \bar \mu )
 | \mu | \tilde U_h}{  | z - \bar  z |^{ 2 \tilde \Delta_h}}
 \label{1ptR}
\end{align}
with
\begin{align}
 \tilde U_h =  \tilde A_0
 (\mu_L \pi \gamma (\tfrac{1 + b^2}{2} ))^{ - 2 h + \frac{1}{2}}
  \Gamma (1  - b^2 ( 2 h - \tfrac{1}{2}) )
 \Gamma ( - 2 h + \tfrac{1}{2} )  \cosh (2 \pi b u_\pm ( 2 h  - \tfrac{1}{2}) )  ~.
\end{align}
Here $\tilde A_0$ is a constant independent of $h$,
and the identification is used as $2 \pi b u_\pm = - r - \pi i /2
 \, {\rm sgn} \, {\rm Im} \, \mu  $.

\section{Annulus amplitude and Cardy condition}

In section \ref{sec:BOSP}, we have computed one point functions
of bulk operator by solving the constraint coming from the channel duality.
Addition to this constraint, there is another constraint for one point
function, which arises from the open-closed duality.
{}From the information of bulk one point functions, we can
construct boundary states describing branes. Then, we can compute the
overlap between boundary states by considering the exchange of closed
strings. On the other hand, we can obtain the same amplitude in the
language of open string as a partition function with proper boundary
conditions. They are related by modular transformation of the annulus
amplitude, which yields a strong consistency condition to the one point functions of closed strings. This condition is called the Cardy condition.
In rational conformal field theories the Cardy condition almost fixes
boundary states, but in non-rational theories like our model it
is not that strong. Nonetheless this condition still gives useful
information to boundary states as we will see in this section.

\subsection{Boundary states for super AdS brane in the RR-sector}
\label{sec:aaNS}

Boundary states have information of how closed strings couple to branes.
For a while we focus on super AdS branes and closed strings in the RR-sector.
In particular, the coupling to primary states is given by
\begin{align}
{}_\text{B} \langle r ; \epsilon| h; x , \xi \rangle \equiv
 \left\langle \Phi_h (x,\xi | \tfrac{i}{2}) \right\rangle _{r; \epsilon } ~.
\end{align}
Here $r$ and $\epsilon$ represent the boundary conditions as before.
The coupling to higher modes is fixed by the boundary condition
\eqref{bcads2}. Later we change the basis from $x$-basis to $m$-basis
by utilizing a Fourier transformation%
\footnote{Orthogonal basis consists of linear combinations, such as 
$\Phi^{h,\pm}_{m,\bar m} = 
\frac{1}{\sqrt2} ( \Phi^{h,0,0}_{m,\bar m}  \pm \Phi^{h,1,1}_{m , \bar m})$.}
\begin{align}
 \Phi^{h,a, \bar a}_{m , \bar m} (z)
  = \int \frac{d ^2 x }{| x|^2} d \bar \xi d \xi
   x^{h - m + \frac{a}{2}} \bar x ^{h - \bar m + \frac {\bar a }{2}}
    \xi^{1 - a} \bar \xi ^{1 - \bar a}
   \Phi_h (x , \xi | z) ~.
   \label{xtom}
\end{align}
Here we assume that $m = (n +i p)/2$, $\bar m = (n - ip)/2$
with $n \in \mathbb{Z}$ and $p \in \mathbb{R}$.
Following \cite{PST}, the overlaps are computed as
\begin{align}
{}_\text{B} \langle r ; \epsilon| h; n , p , a, \bar a \rangle
 = 2 \pi \delta (p) \delta_{a, \bar a}
\cdot 2 \pi A_b \nu^{- 2h + \frac12}
 \Gamma (\tfrac{1}{2} - b^2 (2 h - \tfrac{1}{2}) )
 A(h,n,s| r , \epsilon) ~,
\end{align}
with
\begin{align}
 &A(h,n,0|r, +1)
 = d_n^h (\pi_n^0 \cosh (2h-\tfrac12) r - \pi_n^1 \sinh (2h-\tfrac12) r) ~, \\
 &A(h,n,1|r, +1)
 = - 2 h d_n^{h+ \frac12 }
 (\pi_n^0 \cosh (2h-\tfrac12) r - \pi_n^1 \sinh (2h-\tfrac12) r) ~, \\
 &A(h,n,0|r, - 1)
 = d_n^h (\pi_n^1 \cosh (2h-\tfrac12) r - \pi_n^0 \sinh (2h-\tfrac12) r) ~, \\
 &A(h,n,1|r,  - 1)
 = 2 h d_n^{h+ \frac12 }
 (\pi_n^1 \cosh (2h-\tfrac12) r - \pi_n^0 \sinh (2h-\tfrac12) r) ~.
\end{align}
The functions $d_n^h$ and $\pi_n^\delta$ $(\delta = 0,1)$
are defined as
\begin{align}
 d_n^h = \frac{\Gamma ( - 2 h + 1)}
 { \Gamma (1 - h + \frac{n}{2}) \Gamma (1 - h - \frac{n}{2})} ~, \qquad
 \pi^ \delta_{2 m} = 1 - \delta ~ , \qquad
 \pi^\delta_{2 m + 1}  = \delta
\end{align}
for $m \in \mathbb{Z}$.

Using the boundary states for super AdS branes,
the annulus amplitude can be written as
\begin{align}
 Z_{r; \epsilon}^\text{R}
 = {}_\text{B} \langle r  ; \epsilon  | (-1)^F
\tilde q^{\frac12 H^\text{R} }
 | r ; \epsilon \rangle_\text{B}
  = \int _\mathbb{ S } \frac{d h}{2 \pi} \int_\mathbb{C} d ^2 x
 \int d \bar \xi d  \xi  \chi^h _{\text{R} , - }(\tilde q )
 {}_\text{B} \langle r ; \epsilon | h; x , \xi \rangle
 \langle h; x , \xi|   r  ; \epsilon \rangle_\text{B} ~.
\end{align}
Here the Hamiltonian is
$H^\text{R} =L_0 + \tilde L_0 - c/12 $, and only 
the states in the RR-sector are summed over.
The worldsheet modulus is given by  $\tilde \tau = -1/\tau$
and $\tilde q = \exp (2 \pi i \tilde \tau)$.
We define $F = \frac12 (F_c + \bar F_c)$, where
the numbers of Grassmann odd fields are counted by $F_c$ and $\bar F_c$ for
holomorphic and anti-holomorphic parts, respectively.
The integral domain $\mathbb{S}$ is restricted to the physical line as
$h = 1/4 + i P$ with $P \in \mathbb{R}$.
Since there are three bosons and two fermions with periodic boundary 
condition and without any singular vectors, the character is given by $\chi^h _{\text{R} , - }(\tilde q ) = \tilde q ^{2 b^2 P^2} \eta ^{-1 }
(\tilde q)$.

The overlap is actually a divergent quantity basically due to
the infinite volume of super AdS branes, so we have to
adopt a regularization scheme.
In order to see this it is convenient to change the basis as
\begin{align}
Z_{r; \epsilon}^\text{R}
  &= \int _\mathbb{ S } \frac{d h}{2 \pi}
    \frac{1}{2 \pi} \sum_{n \in \mathbb{Z}}
    \int _\mathbb{R}  \frac{dp}{2 \pi} \chi^h _{\text{R} , - }(\tilde q)
\Bigl[
 {}_\text{B} \langle r ; \epsilon | h; n , p , 1 , 1 \rangle
 \langle h; n , p , 0 , 0  | r ; \epsilon \rangle_\text{B}
  \nonumber \\
 & \qquad \qquad \qquad
 -  {}_\text{B} \langle r ; \epsilon | h; n , p , 0 , 0 \rangle
 \langle h; n , p , 1 , 1  | r ; \epsilon \rangle_\text{B}
\Bigr] ~.
\label{overlaps}
\end{align}
First of all we replace a
delta-function as $2 \pi \delta_T (p)
 = \frac{2}{p} \sin T p$ as in \cite{PST}.
A difficult part is on the summation over $n$. Because we
have to sum over bosonic and fermionic contributions, naively
they are canceled with each other. Therefore, we pick up
effectively one bosonic contribution among them.%
\footnote{There should be a better regularization than this ad hoc way,
but here we adopt this anyway.
This regularization may be justified by the character of degenerate
representation with $(r,s) = (2 J + 1 , 0)$.
The representation is of $2J+1$-dimensional
and Grassmann odd and even states appear alternatively.
Therefore, the character is given by
$\pm \chi^h _{\text{R} , - }(\tilde q ) = \tilde q ^{2 b^2 P^2} \eta ^{-1 }
(\tilde q)$ with $P = - i r/4$, and an analitic continuation would lead to
our character.
}
With this regularization, the overlap can be computed as
\begin{align}
Z_{r; \epsilon}^\text{R} =  \epsilon  4 \pi T | A_b | ^2
 \int_{- \infty}^\infty \frac{ d P }{ 4 \pi }
 \chi^h_{\text{R} , - } (\tilde q) \frac{\cos 4 P r  }
 { \cosh 2 \pi P \cosh 2 b^2 \pi P } ~.
 \label{NSoverlap}
\end{align}
Here we have used the fact that $\cos ^2 \pi h + \cos ^2 \pi (\frac12 - h) = 1$.

Performing the open-closed duality, we may map the overlap between
boundary states into the partition function of open string.
Utilizing the formula of modular transformation
\begin{align}
 \chi^h _{\text{R} , - }( \tilde q) = 2 b \int_{- \infty}^\infty  d P '
\chi^{h'} _{\text{R} , - }(q ) \exp (8 \pi i b^2 P ' P) ~,
\label{NS-S}
\end{align}
we can rewrite the overlap \eqref{NSoverlap} as
\begin{align}
Z_{r; \epsilon}^\text{R}
 =   T \int_{0}^\infty d P '
 N (P' | r ; \epsilon) \chi^{h '} _{\text{R} , - }( q)  ~.
 \label{NSamp}
\end{align}
For the following analysis it might be convenient to
express the density of state as
\begin{align}
 N (P ' | r ; \epsilon) = \frac{ \epsilon}{ 2 \pi}
\frac{\partial}{\partial P '}
 \int_0^\infty \frac{dt}{t}
 \frac{ \left( \cosh \frac{Qt}{4}
- \cosh \frac{(b- b^{-1})t}{4} \right ) \left ( \sin t b (P ' + \frac{r}{2 \pi b^2} ) +
 \sin t b ( P ' - \frac{r}{2 \pi b^2} ) \right)}{\sinh tb/2 \sinh t/2b }
\end{align}
where we set $| A_b |^2 =  2 b$.

\subsection{Annulus amplitude from open strings}
\label{sec:aaopen}

Before going into the detail of the Cardy condition, we first
study the partition function of open string stretched between
the super AdS$_2$ brane with the label $(r,\epsilon)$.
For the calculation of the partition function, we sum up the
spectrum of open strings with the density of states $\rho ( P| r; \epsilon)$.
The spectral density
is related to the reflection amplitude $R( P|r; \epsilon)$ as
\begin{align}
 \rho (P | r; \epsilon) \sim \frac{L}{\pi}
  + \frac{1}{2 \pi i } \frac{\partial}{\partial P} \ln R( P|r; \epsilon)
   ~. \label{rhoPre}
\end{align}
Here $L$ is a cut-off scale, and the wave-function is assumed not
to go into the region $L < \phi$ due to the exponential potential.
For the detail, see, for example, appendix B in \cite{PST}.

The reflection relation can be read off from the two point functions
of boundary operators, which are computed in appendix \ref{app:b2pt}.
Boundary operators may be constructed by  $\Psi_l^{\rho , \rho '}
(t , \eta | u) $
as in the bulk case. The representation of zero modes of OSP(1$|$2) current algebra is parameterized by $l,t, \eta$, and the position on the boundary of worldsheet is denoted by $u$. The extra labels
$\rho, \rho '$ represent boundary conditions across the inserted point $u$,
which are related to $r,r'$ by $\rho = r/(4 \pi b^2) - i / (8 b^2) $.
Let us denote ${\cal V}_l^{\rho , \rho '}
(t | u) = \Psi_l^{\rho , \rho '} (t , 0 | u) $, then the two point
functions
\begin{align}
 & \left \langle {\cal V}^{\rho , \rho ' }_{l_1} (1  | 1)
{\cal V}^{\rho ' , \rho}_{l_2} (0  | 0) \right \rangle
 = \delta (l_1 - l_2 ) d(l_1 ; \rho , \rho ' ) ~, \\
 & \left  \langle p {\cal V}^{\rho , \rho ' }_{l_1} (1  | 1)
 p {\cal V}^{\rho ' , \rho }_{l_2} (0  | 0) \right \rangle
 = \delta (l_1 - l_2 ) d ' (l_1 ; \rho , \rho ' ) 
\end{align}
are computed with the function $d(l_1 ; \rho , \rho ' )$
in \eqref{d-} or \eqref{d+} and  $d ' (l_1 ; \rho , \rho ' )$
in \eqref{dprime-} or \eqref{dprime+}. 

Following the above
general argument we can compute the spectral density
from the two point functions.
Let us start from the primary ${\cal V}_l^{\rho , \rho '} $.
Since the identification is $\epsilon = - \zeta $ as discussed 
in section \ref{sec:bsLt}, we have
\begin{align}
 &\rho (P|r) \sim - \frac{1}{2 \pi i} \frac{\partial}{\partial P'} \ln {\bf S }_\text{NS} (2b(l+2i\rho))
 {\bf S} _\text{NS} (2b(l -2i\rho))  \label{dosrho}
\end{align}
 for $\epsilon = + 1$ and 
\begin{align} 
 &\tilde \rho (P|r) \sim - \frac{1}{2 \pi i} \frac{\partial}{\partial P'} \ln { \bf S}_\text{R} (2b(l+2i\rho))
 {\bf S}_\text{R} (2b(l -2i\rho)) \label{dosrhot}
\end{align}
for $\epsilon = - 1$.
Here we picked up the terms depending on the parameter $r$
(or $\rho$). We also set $l= 1/4 + i P$.
The spectral density for the descendant 
$p {\cal V}_l^{\rho , \rho '} $ 
can be obtained in the same way,
and  given by $\tilde \rho (P|r)$ in \eqref{dosrhot}
for $\epsilon = + 1$ and $ \rho (P|r)$ in \eqref{dosrho}
for $\epsilon = - 1$.

Here we remark that the spectral densities are different for
${\cal V}_l^{\rho , \rho '} $ and $p {\cal V}_l^{\rho , \rho '}$, even though
in principle they could be related with each other by OSP(1$|$2) current algebra.
A similar situation arises in ${\cal N}=1$
super-Liouville theory, where the reflection relation and
worldsheet supersymmetry do not commute \cite{FH}.
Since the first type operator is Grassmann even and the second
type operator is Grassmann odd, we may construct the following
partition function as
\begin{align}
 Z^o_{\text{R}} (r;\epsilon) = \text{Tr}_\text{R} \, \tfrac{1 + (-1)^F}{2} q^{H_o}
  + \text{Tr}_\text{R} \, \tfrac{1 - (-1)^F}{2} q^{H_o} ~.
\end{align}
Here $H_o = L_0 - c/24$ is the Hamiltonian for the open strings
and we sum over states in the R-sector.%
\footnote{Similar quantity can be defined in the NS-sector,
but we do not deal with it in this paper.}
The fermion number is counted by $F$.
For the first term we use the spectral density of the primary
${\cal V}_l^{\rho , \rho '} $  and for the second term
we use the one of the descendant $p {\cal V}_l^{\rho , \rho '} $.

The above partition function is a divergent quantity, so we
need to regularize it. There are two type of divergence.
One type comes from the limit of $L \to \infty$ in \eqref{rhoPre},
and it is useful to consider the relative partition function
$Z^o_\text{R} (r;\epsilon) - Z^o_\text{R} (r_*;\epsilon) $ with
reference boundary condition $r_*$. Another is due to
the sum over the eigen-value $m$ of $J_0^3$. For
the contribution from $\text{Tr}_\text{R} (-1) ^F q^{H_o}$, naively
bosonic and fermionic contributions cancel out. Thus we pick up
one type of contribution as before.
The character is given by
$\chi^h _{\text{R} , - }( q ) = \tilde q ^{2 b^2 P^2} \eta ^{-1 }
(q)$ as well.
For the contribution from $\text{Tr}_\text{R} q^{H_o}$, the summation
over $m$ simply diverges. The situation is the same as in the
case of SL(2) WZNW model \cite{PST}, so we adopt the same regularization.
Namely, we just sum up in the range $- \kappa < m \leq \kappa$.
The character for this case is given by
\begin{align}
 \chi^{h}_{\text{R} , +} (q)
 =  \frac12 q^{2 b^2 {P}^2 } \frac{\vartheta_2 (0 | \tau) }{ \eta ^{4 } ( \tau)}
\end{align}
with
\begin{align}
\vartheta_2 ( y | \tau )
  &= \sum_{n = - \infty}^\infty  q^{(n - \frac12 )^2 /2} z^{n - \frac12 }
  \\
  &= 2 e^{\pi i \tau / 4} \cos \pi y \prod_{m=1}^\infty ( 1 - q^m) ( 1 + z q^{m} )
  (1 + z^{-1} q^{m}) ~ .
\nonumber
\end{align}
Here we set $z = \exp (2 \pi i y)$.

\subsection{Boundary states for super AdS branes in the NSNS-sector}

It is easy to observe that the overlap computed in subsection
\ref{sec:aaNS} cannot be mapped to the one in subsection
\ref{sec:aaopen} by the modular transformation.
The reason is that the character $\chi^h _{\text{R} , - }( q )$
is related to the same character by the transformation
$\tilde \tau =  -1/\tau\to \tau$ as in \eqref{NS-S}.
In order to resolve this problem,
we include the NSNS-sector in the spectrum of closed strings exchanged
between the branes.
The character for the sector is given by
\begin{align}
\chi^h_{\text{NS},-} (\tilde q ) = \tilde q ^{2 b^2 P^2}
 \frac{\vartheta_4 (0 , \tilde \tau) }{ \eta ^{4 } (\tilde \tau)} ~.
\end{align}
Here we have used theta function defined as
\begin{align}
 \vartheta_4 ( y | \tau )
  = \sum_{n = - \infty}^\infty ( - 1 )^n q^{n^2 /2} z^n
  = \prod_{m=1}^\infty ( 1 - q^m) ( 1 - z q^{m - \frac12} )
  (1 - z ^{-1} q^{m - \frac12}) ~ ,
\end{align}
whose modular transformation is
\begin{align}
 \vartheta_4 ( y/\tau  , - 1/\tau )
  = (- i \tau)^{1/2} \exp (\pi i y^2 / \tau ) \vartheta_2 (y , \tau) ~.
\end{align}
This character is related to the other character $\chi^h _{\text{R} ,+}( q )$
in the partition function of open string as
\begin{align}
 \chi^h_{\text{NS},-} (\tilde q ) = 4 b \frac{i}{\tau}
  \int_{- \infty}^\infty d P '
 \chi^{h'}_{\text{R} , +} (q) \exp ( 8 \pi i b^2 P P')~.
 \label{modularR}
\end{align}

With the help of the relation between OSP(1$|$2) WZNW model and
${\cal N} = 1 $ super-Liouville field theory, we have obtained
the one point function of closed string in the NSNS-sector as in
\eqref{1ptR}.
We can map from the $\mu$-basis to $x$-basis as
\begin{align}
 V_h^{\frac12} (\mu | z)
  = \frac{1}{\pi} | \mu |^{- 2 h + 2}
  \int _\mathbb{C} d^2 x e^{\mu x - \bar \mu \bar x}
  \Phi^{\frac{1}{2}}_h (x | z) ~,
\end{align}
and in the $x$-basis the one point function can be written as
\begin{align}
 \left \langle \Phi^{\frac12}_h (x | z) \right \rangle
  = \frac{\tilde U^\sigma (h;r)}{|x + \bar x|^{2 h + \frac12}
  | z - \bar z | ^ {2 \tilde \Delta_h } } ~,
\end{align}
 where
\begin{align}
 \tilde  U^{\sigma} (h; r )
  = \tilde A _b b^{-1} \nu^{ - 2 h + \frac12 }
  \Gamma (1 - b^2 (2h - \tfrac12) ) e^{- \sigma (2h-\frac12) r} ~.
\end{align}
Notice that the coefficient depends on the sign
$\sigma = \text{sgn} (x + \bar x)$.
The overall  factor $\tilde A_b$ will be fixed later
by using the Cardy condition. 
As before we define the boundary state as
\begin{align}
{}_\text{B} \langle r ; \epsilon| h; x , \tfrac12 \rangle \equiv
 \left\langle \Phi^{\frac12}_h (x | \tfrac{i}{2}) \right\rangle _{r; \epsilon } ~.
\end{align}
We again change the basis from $x$-basis to $m$-basis by utilizing
the formula 
\begin{align}
 \Phi^{h,\frac{1}{2}}_{m , \bar m} (z)
  = \int \frac{d ^2 x }{| x|^2} 
   x^{h - m + \frac{1}{4}} \bar x ^{h - \bar m + \frac {1 }{4}}
   \Phi_h^{\frac{1}{2}} (x  | z) ~,
   \label{xtomns}
\end{align}
then we have
\begin{align}
{}_\text{B} \langle r ; \epsilon| h; n , p , \tfrac12 \rangle
 = 2 \pi \delta (p)
\cdot 2 \pi \tilde A_b b^{-1} \nu^{- 2h + \frac12}
 \Gamma (1 - b^2 (2 h - \tfrac{1}{2}) )
 A(h,n| r ) 
\end{align}
with
\begin{align}
 &A(h,n|r)
 = d_n^{h + \frac14 } (\pi_n^0 \cosh (2h-\tfrac12) r - \pi_n^1 \sinh (2h-\tfrac12) r) ~.
\end{align}

Let us compute the following annulus amplitude as
\begin{align}
 Z_{r}^\text{NS}
 = {}_\text{B} \langle r  ; \epsilon  | (-1)^F
\tilde q^{\frac12 H^\text{NS} }
 | r ; \epsilon \rangle_\text{B}
  = \int _\mathbb{ S } \frac{d h}{2 \pi} \int_\mathbb{C} d ^2 x
   \chi^h_{\text{NS} , -} (\tilde q )
 {}_\text{B} \langle r ; \epsilon | h; x , \tfrac12 \rangle
 \langle h; x , \tfrac12 |   r  ; \epsilon \rangle_\text{B} ~,
\end{align}
where we consider the Hamiltonian for the NSNS-sector.
The overlap with closed strings in the NSNS-sector
diverges but the type of divergence is the same as
the one for SL(2) WZNW model. Therefore we can adopt the same
regularization as in \cite{PST}. In the $m$-basis the overlap
can be regularized as
\begin{align}
Z_{r}^\text{NS}
  &= \int _\delta ^ \infty  \frac{d P}{ \pi}
    \frac{1}{2 \pi} \sum_{n = - \lambda + 1}^\lambda
    \int _\mathbb{R}  \frac{dp}{2 \pi} \chi^h_{\text{NS} , -} (\tilde q)
 {}_\text{B} \langle r ; \epsilon | h; n , p , \tfrac12 \rangle
 \langle h; n , p , \tfrac12 | r ; \epsilon \rangle_\text{B}
\nonumber \\
 & =  \lambda 4 \pi T | \tilde A_b |^2  \int _\delta ^ \infty  \frac{d P}{ \pi}
 \frac{\cosh ^2 \pi P \cos ^2 2 P r + \sinh ^2 \pi P \sin ^2 2 P r}
 { \sinh 2\pi b^2  P  \sinh 2 \pi P }\chi^h_{\text{NS} , -} (\tilde q) ~.
\end{align}
Here the range of $n$ is set as $- \lambda < n \leq \lambda $
and the delta-function is replace as  $2 \pi \delta_T (p)
 = \frac{2}{p} \sin T p$.
We also set the cut-off $\delta$ for the integration over $P$, as
the integral diverges in the limit of $\delta \to 0$.
In order to obtain the finite part we may consider a difference as
\begin{align}
 Z_{r}^\text{NS} - Z_{r_*}^\text{NS}
 =   \lambda 4 \pi T  | \tilde A_b |^2  \int _0 ^ \infty  \frac{d P}{ \pi}
 \frac{\cos ^2 2 P r - \cos ^2 2 P r_* }
 { \sinh 2\pi b^2  P  \sinh 2 \pi P }\chi^h_{\text{NS} , -} (\tilde q)
\end{align}
with a reference boundary parameter $r_*$.
After the modular transformation with \eqref{modularR} we find
\begin{align}
 Z_{r}^\text{NS} - Z_{r_*}^\text{NS}
 =  4 \kappa   T
 \int_{0}^\infty d P '
 ( \tilde N (P' | r ) -  \tilde N (P' | r_* )  )
\chi^{h '}_{\text{R} , + } ( q) 
\label{Ramp}
\end{align}
with
\begin{align}
 \tilde N (P ' | r ) = \frac{1}{ 2 \pi }
\frac{\partial}{\partial P '}
 \int_0^\infty \frac{dt}{t}
 \frac{ \left( \cosh \frac{Qt}{4}
+ \cosh \frac{(b- b^{-1})t}{4} \right ) \left ( \sin t b (P ' + \frac{r}{2 \pi b^2} ) +
 \sin t b ( P ' - \frac{r}{2 \pi b^2} ) \right)}{\sinh tb/2 \sinh t/2b }  ~.
\end{align}
Here we have set $|\tilde A_b|^2 = 4 b$. The relation between $\lambda$
and $\kappa$ is given by $ - i \tau \kappa =  \lambda $ as discussed in
\cite{PST}.

\subsection{Open-closed duality}

As pointed out in the previous subsection, we  should sum over
the NSNS-sector of closed strings as well as the RR-sector.
Now the amplitude is given by
\begin{align}
Z_{r;\epsilon} &=
 {}_\text{B} \langle r  ; \epsilon  | (-1)^F
\tilde q^{\frac12 H^\text{R} }
 | r ; \epsilon \rangle_\text{B}
 + {}_\text{B} \langle r  ; \epsilon  | (-1)^F
\tilde q^{\frac12 H^\text{NS} }
 | r ; \epsilon \rangle_\text{B}
 =   Z_{r;\epsilon}^\text{R} +
   Z_{r}^\text{NS} ~.
 \label{bbamp}
\end{align}
With the application of modular transformation, the above amplitude
can be written in terms of open strings in the R-sector.
In fact, using \eqref{NSamp} and \eqref{Ramp}, we find
\begin{align}
 \nonumber
 & Z_{r; \epsilon } - Z_{r_*; \epsilon }  = 2 T \int_{- \infty}^{\infty} d P'
 ( \rho (P ' | r) - \rho (P ' | r_*) )
  (4 \kappa \chi^{h'}_{\text{R},+} (q) + \epsilon \chi^{h'}_{\text{R}, -} (q) )\\
  & \qquad \qquad \qquad \qquad \qquad \qquad
  +  ( \tilde \rho (P ' | r) - \tilde \rho (P ' | r_*) )
  ( 4 \kappa \chi^{h'}_{\text{R},+} (q)  - \epsilon \chi^{h'}_{\text{R}, -} (q)  ) ~,
\end{align}
where the spectral densities are given by \eqref{dosrho} and \eqref{dosrhot}.
{}From this we can conclude that the amplitude \eqref{bbamp} 
for super AdS branes is consistent
with the partition function of open strings in the R-sector.

\subsection{Fuzzy supersphere brane}

The couplings of closed strings in the NSNS-sector to fuzzy supersphere branes 
are obtained in \eqref{fuzzy1pt1} and \eqref{fuzzy1pt2}
by utilizing the factorization constraint.
Here we would like to check the Cardy condition for this type of branes.
We define the boundary states  by
\begin{align}
{}_\text{B} \langle s ; \epsilon| h; x , \xi \rangle \equiv
 \left\langle \Phi_h (x,\xi | \tfrac{i}{2}) \right\rangle _{s; \epsilon } ~,
\end{align}
then the annulus amplitude can be written as
\begin{align}
 {}_\text{B} \langle s ' ; \epsilon ' |(-1)^F \tilde
 q^{\frac12 H^\text{R} }
 | s ; \epsilon \rangle_\text{B}
  = \int _\mathbb{ S } \frac{d h}{2 \pi} \int_\mathbb{C} d ^2 x
 \int d \bar \xi d  \xi  \chi^h_{\text{R},-} (\tilde q )
 {}_\text{B} \langle s ' ; \epsilon '| h; x , \xi \rangle
 \langle h; x , \xi|   s  ; \epsilon \rangle_\text{B} ~.
\end{align}
First we study the case with $\epsilon = \epsilon ' = + 1$.
Using the explicit from of one point function \eqref{fuzzy1pt1},
the overlap can be evaluated as
\begin{align}
 {}_\text{B} \langle s ' ; +1| (-1)^F \tilde q^{\frac12 H^\text{R} }
 | s ; + 1  \rangle_\text{B} \propto
  \int d P \frac{\sinh 2 s' P \sinh 2 s P}{ \cosh 2 \pi b^2 P}
  \chi^h_{\text{R},-} (\tilde q )~,
\end{align}
where we have set
$s = 2 \pi b^2 n$, $s ' = 2 \pi b^2 m$ with positive integer $n,m$.
Using the relation
\begin{align}
 \frac{\sinh 4 \pi b^2 m P \sinh 4 \pi b^2 n P}{ \cosh 2 \pi b^2 P}
  = \sum_{l=0}^{2 \, \text{min} \, (n,m) - 1 } (-1)^l
   \cosh 2 \pi b^2 (2 n + 2 m - 2l - 1) P ~,
\end{align}
we may find
\begin{align}
 Z (q |s, s ' ; + 1 )
 = \sum_{J = | n - m | }^{ n + m - 1}
  ( - 1 )^{ n + m - J - 1 } \chi^{-J/2}_{\text{R},-} ( q) ~.
\end{align}
Here the representation with $h = - J/2$ is of $2J + 1$-dimensional,
and the character is given by $\pm \chi^{-J/2}_{\text{R},-} ( q)$.
Next we choose $\epsilon = \epsilon ' = - 1$.
Then the overlap becomes
\begin{align}
 {}_\text{B} \langle s ' ; - 1| (-1)^F \tilde q^{\frac12 H^\text{R} }
 | s ; - 1  \rangle_\text{B} \propto
  \int d P \frac{\cosh 2 s' P \cosh 2 s P}{ \cosh 2 \pi b^2 P}
  \chi^h_{\text{R},-} (\tilde q )~,
\end{align}
where we have set
$s = \pi b^2 ( 2 n + 1 )$, $s ' = \pi b^2 ( 2 m + 1 )$
with non-negative integer $n,m$.
Now the relation
\begin{align}
 \frac{\cosh 2 \pi b^2 ( 2 m + 1 ) P \cosh 2 \pi b^2 ( 2 n + 1 ) P}
 { \cosh 2 \pi b^2 P}
  = \sum_{l=0}^{2 \, \text{min} \, (n,m)  } (-1)^l
   \cosh 2 \pi b^2 (2 n + 2 m - 2l + 1) P 
\end{align}
leads to
\begin{align}
 Z (q |s, s ' ; - 1 )
 = \sum_{J = | n - m | }^{ n + m }
  ( - 1 )^{ n + m - J} \chi^{-J/2}_{\text{R},-} ( q) ~.
\end{align}
In this way, we have shown that the couplings of RR-states to the 
fuzzy supersphere branes are consistent with the partition function
of open strings.

Let us consider the couplings of closed strings in the NSNS-sector as well.
For the super AdS branes, we obtained the coupling of NSNS-states 
with the help of the relation to ${\cal N}=1$ super
Liouville field theory as in \eqref{1ptR}. 
However, this approach is not applicable to the
super spherical branes as far as we know. Fortunately, the Cardy condition is
actually strong enough to guess the coupling of NSNS-states almost uniquely
at least in this case.
Suppose that the one point function is given by
\begin{align}
 \left \langle \Phi^{\frac12}_h (x | z) \right \rangle_{s;\epsilon}
  = \frac{\tilde U (h;s,\epsilon)}{|x + \bar x|^{2 h + \frac12}
  | z - \bar z | ^ {2 \tilde \Delta_h } }
\end{align}
with
\begin{align}
 \tilde U (h;s,\epsilon) \propto \nu^{- 2 h + 1 }
 \Gamma ( 1 - b^2 (2h - \tfrac12))  \sin ( s (2 h - \tfrac12) )
\end{align}
up to an overall factor.
Moreover we set $s = \pi b^2 n$, $s ' = \pi b^2 m$
with non-negative integer $n,m$.
With this ansatz the overlap is
\begin{align}
 {}_\text{B} \langle s ' ; \epsilon |
 (-1)^F \tilde q^{\frac12 H^\text{NS} }
 | s ; \epsilon  \rangle_\text{B} \propto
  \int d P P \frac{\sinh 2 s' P \sinh 2 s P}{ \sinh 2 \pi b^2 P}
   \chi^h_{\text{NS},-} (\tilde q )~,
\end{align}
which becomes after the modular transformation
\begin{align}
 Z (q |s, s ' ; \epsilon )
 = \sum_{J = | n - m | /2 }^{ ( n + m )/2 - 1 }
  \chi^{-J/2}_{\text{R},+} ( q)
  \label{RFS}
\end{align}
if we choose a proper normalization.
Here we have used the relation
\begin{align}
 \frac{\sinh 2 \pi b^2 m P \sinh 2 \pi b^2 n P}
 { \sinh 2 \pi b^2 P}
  = \sum_{l=0}^{ \, \text{min} \, (n,m)  - 1 }
   \sinh 2 \pi b^2 (n +  m - 2l - 1) P ~.
\end{align}
The expression \eqref{RFS} has a meaning as the partition function of open strings only if $m+n \in 2 \mathbb{Z}$.
Referring to the coupling of RR-states, it is natural to guess that
$m,n \in 2 \mathbb{Z}$ for $\epsilon = +1$ and
$m,n \in 2 \mathbb{Z} + 1 $ for $\epsilon = -1$.

\section{Conclusion and discussions}

In this article, we have analyzed branes in the OSP(1$|$2) WZNW model. We have used two different methods. The first and common method is to determine correlation functions
in the boundary theory via factorization constraints. We used this to determine bulk one-point functions and boundary two-point functions.
The results can be checked by strong consistency conditions. These are the Cardy condition and agreement with the semiclassical limit. For non-compact branes worldsheet
duality is subtle and involves a spectral density. We were able to treat this issue as in the case of the bosonic $H_3^+$ model \cite{PST}.
The second method is to establish a correspondence to boundary $\mathcal{N}=1$ super-Liouville theory and using the known results from this model, e.g. \cite{FH}. This correspondence is a generalization of
bulk correspondence between the OSP(1$|$2) model and super-Liouville theory
in \cite{HS2}, see also the bosonic counterpart \cite{RT}.

A boundary CFT is completely specified by bulk one-point functions, bulk-boundary two-point functions and boundary three-point functions.
Thus some work is still left in solving the boundary OSP(1$|$2) WZNW model. For the super AdS$_2$ branes the correspondence to
$\mathcal{N}=1$ super-Liouville theory seems to be a good way to address this issue.
Another open problem is the treatment of NS-sectors. The fermions in WZNW models naturally have Ramond-boundary conditions. So far, we have only
studied the NS-sector through the correspondence to super-Liouville theory.
A more direct way might be relevant. In \cite{CR2010} GL(N$|$N) models with NS-boundary conditions for the fermions are discussed.

There are several interesting applications and generalizations of our work.
First of all, it would be interesting to generalize our analysis to WZNW models on other supergroups. A candidate is the OSP(2$|$2) WZNW model.
This model has several types of branes; spherical ones and AdS-type branes with either Dirichlet or Neumann boundary conditions in the U(1) direction. For the AdS-type branes it seems possible to establish a correspondence to boundary $\mathcal{N}=2$  super-Liouville theory.
The appropriate boundary action of the OSP(2$|$2) WZNW model is suggested using a
 relation between supergroup WZNW models and $\mathcal{N}=(2,2)$ superconformal theories \cite{CR2010}.
Such a correspondence would provide bulk one-point functions from \cite{Hosomichi:2004ph}.

The correspondence between the $H_3^+$ model and bosonic Liouville theory has been
 used to show a strong-weak duality (the Fateev-Zamolodchikov-Zamolodchikov duality \cite{FZZ}) between the cigar CFT and sine-Liouville theory \cite{Hikida:2008pe}.
Using the boundary action of \cite{FR} we could extend this duality to worldsheets with boundary.
Furthermore, we would like to understand correspondences and dualities involving supergroups, such as of the OSP-type and their cosets.
There are also $\mathcal{N}=(2,2)$ super conformal field theories on these
spaces \cite{GHT,Creutzig:2009fh}. To these cases mirror symmetry should be applied similarly to the case of the fermionic Euclidean black hole \cite{Hori:2001ax} .

\subsection*{Acknowledgement}

We are very grateful to Vincent Bouchard, Gaston Giribet,
David Ridout, Peter Roenne, and Volker Schomerus.
The work of YH is supported in part by JSPS Research Fellowship.

\appendix

\section{Some super analysis}\label{app:complex}

We list some useful formulae.
There are two ways to extend complex conjugation to the ring of Grassmann numbers. We denote them by a bar and by the super star $\sharp$.
The ordinary bar-operation is defined by
\begin{equation}
\overline{c\theta}\ = \bar c\bar\theta\, ,\qquad \bar{\bar\theta}\ = \ \theta \, ,\qquad \overline{\theta_1\theta_2}\ = \ \bar\theta_2\bar\theta_1
\end{equation}
for a complex number $c$ and Grassmann odd elements $\theta,\theta_1$ and $\theta_2$.
The superstar is
\begin{equation}
(c\theta)^\sharp\ = \bar c\theta^\sharp\, ,\qquad \theta^{\sharp\sharp}\ = \ -\theta \, ,\qquad (\theta_1\theta_2)^\sharp\ = \ \theta_1^\sharp\theta_2^\sharp\, .
\end{equation}
They are related as follows; define the map $\alpha$ as $\alpha(x)=1$ if $x$ is Grassmann even and $\alpha(\theta)=i$ if $\theta$ is Grassmann odd.
Then it is straightforward to check, that
\begin{equation}\label{eq:barsuperstar}
X^\sharp={\alpha(\overline X)}
\end{equation}
 for any Grassmann number $X$.

The analog of transposition for super matrices is the super transpose $st$
\begin{equation}\label{eq:supertranspose}
	\begin{split}
		\left(\begin{array}{cc}A & B \\ C & D\\ \end{array}\right)^{st}=\left(\begin{array}{cc}A^t & -C^t \\ B^t & D^t\\\end{array}\right)\, .
	\end{split}
\end{equation}
Hermitean conjugation is defined as
\begin{equation}\label{eq:hermiteanconjugation}
\ddagger\ = \ st\circ \sharp\, .
\end{equation}
It is important that this map is of order two (while the supertranspose and the superstar are each of order four).

\section{Four point function with a degenerate operator}
\label{sec4pt}

Generically it is quite difficult to obtain explicit expressions of
four point function. The situation might change if
one of the vertex
operators belongs to a degenerate representation. In this appendix we
study the case with $h_1 = k/2$.

\subsection{Null equation from the degenerate operator}

As discussed in subsection \ref{degenerate}
the state $| k/2 \rangle$ is degenerate since a descendant
$|\theta \rangle = J_{-1}^- | k/2 \rangle $ is null.
Therefore, we can set the operator corresponding
to $|\theta \rangle$ to be zero, and this yields differential equation which
correlation functions with the degenerate operator should satisfy.
In the language of operator the null condition can be written as
\begin{align}
 J_{-1}^- (x , \xi ) \Phi_{\frac{k}{2}} ( x , \xi | z ) = 0 ~.
 \label{nullop}
\end{align}
Here $J_{-1}^- (x , \xi ) $ is given by the mode expansion of the
following operator as
\begin{align}
 J^- (x , \xi | z) &=
 e^{ - 2 \xi j^+_0} e^{ - x J^+_0} J^- (z)
  e^{  x J^+_0} e^{  2 \xi j^+_0} \\
   &= J^-(z) + 2 x J^3 (z) + x^2  J^+ (z)
    - 2 \xi j^- (z) + 2 x \xi j^+ (z) ~. \nonumber
\end{align}
We consider correlation functions with the insertion of
\eqref{nullop}. With the help of Ward identity
\begin{align}
 & \left \langle
 J^- (x , \xi | z) \prod_{i=1}^N \Phi_{h_i}(x_i , \xi_i | z_i )
  \right  \rangle  \\ \nonumber
  & = - \sum_{i=1}^N \left [ (x - x_i + \tfrac12 \xi \xi_i)^2 \partial_{x_i}
  + (x - x_i ) (\xi - \xi_i ) \partial_{\xi_i} -
 2 (x - x_i + \xi \xi_i) h_i
 \right ]
 \left \langle  \prod_{i=1}^N \Phi_{h_i}(x_i , \xi_i | z_i )
  \right \rangle
\end{align}
we can see that the correlation function with $ \Phi_{k/2}$
should satisfy the following differential equations
\begin{align}
\left[ \sum_{i=1}^N  \frac{1}{z - z_i }\left \{
  (x - x_i + \tfrac12 \xi \xi_i)^2 \partial_{x_i}
  + (x - x_i ) (\xi - \xi_i ) \partial_{\xi_i} -
 2 (x - x_i + \xi \xi_i) h_i \right \}
 \right ]  \nonumber \\
 \times
 \left \langle  \Phi_{\frac{k}{2}}(x , \xi | z )
 \prod_{i=1}^N \Phi_{h_i}(x_i , \xi_i | z_i )
  \right \rangle = 0 ~,
\end{align}
which is called as the null equation.

Let us focus on four point function.
In this case we can set as
 $(z_1 , z_2 , z_3 , z_4 ) = (z_\infty , 1 , z , 0)$,
  $(x_1 , x_2 , x_3 , x_4 ) = (x_\infty , 1 , x , 0)$,
  and $(\xi_1 , \xi_2 , \xi_3 , \xi_4 ) =
 (x_\infty \eta , 1 , \xi , 0)$, and finally take the limit
of $z_\infty,x_\infty \to \infty$.
Picking up the term proportional to $z_\infty^{-1}x_\infty^2$,
we obtain a differential equation
\begin{align}
  {\cal D}_0
 \left \langle  \Phi_{\frac{k}{2}}(x_1 , \xi_1 | z_1 )
 \prod_{i=2}^4 \Phi_{h_i}(x_i , \xi_i | z_i )
  \right \rangle  = 0
  \label{calD0}
\end{align}
with
\begin{align}
 {\cal D}_0 \equiv
 z \partial_{x_2} + (1 + \eta \xi)
 \partial_{x_3} + (z + 1) \partial_{x_4}
  + \eta (z \partial_{\xi_2} + \partial_{\xi_3} + (z+1) \partial_{\xi_4} )
  ~.
\end{align}
This equation largely restricts the form of correlator.
Recall that the four point function is of the form \eqref{fourpt}.
Expanding by the fermionic parameters,
the holomorphic part may be written as
\begin{align}
 \Omega = X_{12}^{- \gamma_{12}} X_{13}^{- \gamma_{13}}
 X_{14}^{- \gamma_{14}}
  X_{24}^{-\gamma_{24} } (A(X,z) + B(X,z) \chi_3 \chi_2
 + C(X,z) \chi_3 + D(X,z) \chi_2  ) ~.
  \label{omega2}
\end{align}
Here $X_{ij}$ are defined in \eqref{largex} and the cross ratios
$X,\chi_3,\chi_2$ are in \eqref{crossX} and \eqref{eta31}.
Replacing the correlator by its holomorphic part $\Omega$ \eqref{omega2}
and setting $(x_i,\xi_i,z_i)$ as the fixed values at the final point,
we find
\begin{align}
&(x - z ) A' + \gamma_{24} A + \eta \xi ((x -z ) B '
 + (\gamma_{24}+1) B - x A ') \\  &+ (x-z) \eta C '
  + (\gamma_{24} + \tfrac12) \eta C
 + (x-z) (\xi - x \eta) D '
  + (\gamma_{24} + \tfrac12) (\xi - x\eta) D
  = 0 ~. \nonumber
\end{align}
Here $'$ represents the derivative with respect to $x$.
Expanding this equation by $\eta,\xi$ the above equation
yields four independent differential equations.
Solutions can be find as
\begin{align}
 &A(X,z) = a(z) (X - z)^{- \gamma_{24}} ~, \qquad
 B(X,z) = b(z) (X - z)^{- \gamma_{24} - 1} ~,
 \nonumber \\
  &C(X,z) =c (z) (X - z)^{- \gamma_{24} - \frac12} ~ , \qquad
  D(X,z) =d (z) (X - z)^{- \gamma_{24} - \frac12} ~ . \qquad
\end{align}
Here we should notice that $A(X) \to A(x) - x \eta \xi A ' (x)$.
In this way, the null equation fixes the $(X , \chi_3 , \chi_2)$-dependence completely, and now the problem is to obtain the explicit
form of the functions $a(z),b(z),c(z),d(z)$ of $z$.

\subsection{Solutions to the KZ equation}

It is known that correlators of WZNW models satisfy the KZ equation,
and in our case it is written as \eqref{KZ}. Here we fix the
functions $a(z),b(z),c(z),d(z)$ by utilizing the KZ equation.
Since the KZ operator itself is bosonic, it does not mix
the bosonic and fermionic parts of conformal blocks.
First we focus on the bosonic part with $a(z), b(z)$ and later
we study the fermionic part with $c(z), d(z)$.
Notice that only the bosonic part appears in the two point function
on a disk as seen in section \ref{sec:BOSP}.

Differential equations for $a(z)$ and $b(z)$ can be obtained by
inserting the holomorphic part of correlator $\eqref{omega2}$
into the KZ equation \eqref{KZ} with $N=4$.
After a tedious but straightforward calculation we find
\begin{align}
 \kappa z (z-1) \partial_z a(z) &=
  [ \alpha _1 (z - 1) + \beta _1 z ] a (z) - \tfrac14 b(z) ~, \\
 \kappa z (z-1) \partial_z  b(z) &= z \gamma_{24}
  [ \tfrac14 (\gamma_{12} - 1) + \tfrac{h_3}{2}] a (z) +
  [ \alpha ' _1 (z - 1) + \beta '_1 z ] b (z) ~,
 \end{align}
where we have compared the terms with
$(x - z) ^{- \gamma_{24}}$ and
 $\eta \xi (x - z) ^{- \gamma_{24}}$.
The coefficients are given by
\begin{align}
\alpha_1 &= \tfrac12 \gamma_{24} ( \gamma_{24} + 1 )
 - \gamma_{24} ( h_3 + h_4 + \tfrac14  ) + h_3 h_4 ~, \\
\beta _1 &= \tfrac12 \gamma_{24} ( \gamma_{24} + 1 ) -
 \gamma_{24} ( h_3 + h_2 ) + h_3 h_2 ~, \\
\alpha ' _1 &= \tfrac12 ( \gamma_{24} + 1 ) ( \gamma_{24} + 2 )
 - ( \gamma_{24} + 1 ) ( h_3 + h_4 + \tfrac34)
 + \tfrac{ h_3 }{2} + \tfrac{ h_4 }{2} + h_3 h_4 ~, \\
\beta ' _1 &= \tfrac12 ( \gamma_{24} + 1 )( \gamma_{24} + 2 )
  - (  \gamma_{24} + 1 ) ( h_3 + h_2 + \tfrac34 ) + \tfrac{h_3}{2} + \tfrac{h_2}{2} + h_3 h_2 - \tfrac14 \gamma_{24} ~.
\end{align}
Since they are two independent first order differential equations,
we have two independent solutions to them.

With the two solutions, we can write down the conformal
block as a linear combination of two independent functions.
We set the form as
\begin{align}
{\cal F}^S_{i} &=
  z^{\frac{1}{2 \kappa} [ (h_1 - h_2) (h_1 - h_2 - \frac12)
   - h_3 (h_3 - \frac12) - h_4 (h_4 - \frac12) ]}
    (1 - z)^{\frac{1}{2 \kappa} [(h_1 - h_4 - \frac12) (h_1 - h_4 - 1)
   - h_2 (h_2 - \frac12) - h_3 (h_3 - \frac12)] }
   \nonumber \\ & \qquad \times \left(
f^S_{a,i} (z) (x ( 1 - \eta \xi ) - z)^{- \gamma_{24}}
 + f^S_{b,i} (z) \eta \xi (x - z )^{- \gamma_{24} - 1}
 \right)
 \label{fsi}
\end{align}
with $i=1,2$, which behave as
\begin{align}
 {\cal F}^S_{1} & =  z^{\frac{1}{2 \kappa} [ (h_1 - h_2) (h_1 - h_2 - \frac12)
   - h_3 (h_3 - \frac12) - h_4 (h_4 - \frac12) ]}
  x ^{- \gamma_{24}} + \cdots ~, \label{fs1} \\
  {\cal F}^S_{2} & =  z^{\frac{1}{2 \kappa} [ (h_1 - h_2 - \frac12) (h_1 - h_2 - 1)
   - h_3 (h_3 - \frac12) - h_4 (h_4 - \frac12) ]}
\eta \xi  x^{- \gamma_{24} - 1 } + \cdots \label{fs2} ~.
\end{align}
The overall normalization is our convention.
The functions $f^S_{a,i} (z) $ and $f^S_{b,i} (z) $ are found to be
\begin{align}
 \label{smallfsi}
 f^S_{a,1} &= F(1 + \tfrac{ -1 + h_1 - h_2 + h_3 + h_4}{4 \kappa}
  , - \tfrac{ -1 + h_1 + h_2 + h_3 - h_4}{4 \kappa} ,
   1 - \tfrac{1 - 2 h_1 + 2 h_2}{4 \kappa} ; z ) ~, \\
 f^S_{b,1} &= - \tfrac{(- 1 + h_1 + h_2 + h_3 - h_4 ) (h_1 - h_2 - h_3 - h_4)}{ 4 \kappa - 1 + 2 h _1 - 2 h_2 } \times \nonumber \\
 & \qquad  \times
 z F(1 + \tfrac{ -1 + h_1 - h_2 + h_3 + h_4}{4 \kappa} ,
1 - \tfrac{ -1 + h_1 + h_2 + h_3 - h_4}{4 \kappa}  ,
   2 - \tfrac{1 - 2 h_1 + 2 h_2}{4 \kappa} ; z ) ~, \nonumber \\
 f^S_{a,2} &= -
  \tfrac{1}{1 - 2 h_1 + 2 h_2 } z^{\frac{1 - 2 h_1 + 2 h_2 }{4 \kappa} }
  F(\tfrac{2 - 3 h_1 + h_2 - h_3 + h_4}{4 \kappa} ,
1 + \tfrac{ - h_1 + h_2 + h_3 + h_4}{4 \kappa}  ,
   1 + \tfrac{1 - 2 h_1 + 2 h_2}{4 \kappa} ; z ) ~, \nonumber \\
 f^S_{b,1} &= z^{\frac{1 - 2 h_1 + 2 h_2 }{4 \kappa} }
F(\tfrac{2 - 3 h_1 + h_2 - h_3 + h_4}{4 \kappa} ,
\tfrac{ - h_1 + h_2 + h_3 + h_4}{4 \kappa}  ,
   \tfrac{1 - 2 h_1 + 2 h_2}{4 \kappa} ; z ) ~. \nonumber
\end{align}
Here $F(a,b,c;z)$ denotes  the hypergeometric function.

Above expressions are given in terms of $z$, thus
they are suitable for the region with $z \sim 0$ or $z_3 \sim z_4$.
We may say that they are in the $s$-channel expression.
In the $t$-channel expression with $z \sim 1$ or $z_3 \sim z_2$, it is suitable to
express in terms of $1-z$ as well as $1-x$.
These expressions are simply given with ${\cal F}^S_{i}$ by exchanging
$z,x$ and $1-z,1-x$, and moreover $h_2$ and $h_4$.
We denote them by ${\cal F}^T_{i}$ with $i=1,2$.
These two expression are related by
\begin{align}
 {\cal F}^S_{1} = e^{- \pi i \gamma_{24}}
 (F^{ST}_{11} {\cal F}^T_{1} + F^{ST}_{12} {\cal F}^T_{2} )~, \qquad
  {\cal F}^S_{2} = e^{- \pi i \gamma_{24}}
 (F^{ST}_{21} {\cal F}^T_{1} + F^{ST}_{22} {\cal F}^T_{2}) 
  \label{stdual}
\end{align}
with
\begin{align}
 F^{ST}_{11} &= \frac{\Gamma (1 - \frac{1 - 2 h_1 + 2 h_2}{4 \kappa})
\Gamma (\frac{1 - 2 h_1 + 2 h_4}{4 \kappa}  ) }
 {\Gamma (1 - \frac{1 - h_1 + h_2 - h_3 - h_4}{4 \kappa} )
 \Gamma (\frac{1 - h_1 - h_2 - h_3 + h_4}{4 \kappa} )} ~, \\
 F^{ST}_{12} &= - \frac{ 4 \kappa \Gamma (1 - \frac{1 - 2 h_1 + 2 h_2}{4 \kappa})
\Gamma ( 1 + \frac{ - 1 + 2 h_1 - 2 h_4}{4 \kappa}  ) }
 {\Gamma ( \frac{ h_1 - h_2 - h_3 - h_4}{4 \kappa} )
 \Gamma (1 - \frac{2 - 3 h_1 + h_2 - h_3 + h_4}{4 \kappa} )} ~, \\
 F^{ST}_{21} &= - \frac{\Gamma ( \frac{1 - 2 h_1 + 2 h_2}{4 \kappa})
\Gamma (\frac{1 - 2 h_1 + 2 h_4}{4 \kappa}  ) }
 { 4 \kappa \Gamma ( \frac{2 - 3 h_1 + h_2 - h_3 + h_4}{4 \kappa} )
 \Gamma (1 + \frac{ - h_1 + h_2 + h_3 + h_4}{4 \kappa} )} ~, \\
 F^{ST}_{22} &= \frac{ \Gamma ( \frac{1 - 2 h_1 + 2 h_2}{4 \kappa})
\Gamma (1 + \frac{ - 1 + 2 h_1 - 2 h_4}{4 \kappa}  ) }
 {\Gamma ( 1 + \frac{ -1 +  h_1 + h_2 + h_3 - h_4}{4 \kappa} )
 \Gamma (\frac{1 -  h_1 + h_2 - h_3 - h_4}{4 \kappa} )} ~.
\end{align}

Let us move to the fermionic part with $c(z), d(z)$.
As mentioned above, there are no such contributions to
the disk amplitude, so we study them here only for completeness.
We hope to report on more systematic analysis of closed strings
in the OSP(1$|$2) WZNW model somewhere else.
In the same way as before, we can show that the KZ equation
leads to two independent first order differential
equations for  $c(z)$ and $d(z)$ as
\begin{align}
 \kappa z (z-1) \partial_z c(z) &=
  [ \alpha _2 (z - 1) + \beta _2 z ] c (z)
 + \tfrac14 (\gamma_{12} - \tfrac12 ) z d(z) ~, \\
 \kappa z (z-1) \partial_z  d(z) &=
  - [ \tfrac14 (\gamma_{24} + \tfrac12) - \tfrac{h_3}{2}] c (z) +
  [ \alpha ' _2 (z - 1) + \beta ' _2 z ] d (z) ~.
 \end{align}
Here the coefficients are given by
\begin{align}
\alpha _2 &= \tfrac12 (\gamma_{24} + \tfrac12 )
 ( \gamma_{24} + \tfrac32 )
 - ( \gamma_{24} + \tfrac12 ) ( h_3 + h_4 + \tfrac14  ) + h_3 h_4 ~, \\
\beta _2 &= \tfrac12 (\gamma_{24} + \tfrac12 )
 ( \gamma_{24} + \tfrac32  )
 - ( \gamma_{24} + \tfrac12 )( h_3 + h_2  + \tfrac12) + \tfrac{h_3}{2} + h_3 h_2 ~, \\
\alpha ' _2 &= \tfrac12 (\gamma_{24} + \tfrac12 )
 ( \gamma_{24} + \tfrac32 )
 - ( \gamma_{24} + \tfrac12 ) ( h_3 + h_4 + \tfrac34)
+ \tfrac{h_3}{2} + \tfrac{h_4}{2} + h_3 h_4 ~, \\
\beta ' _2 &= \tfrac12 (\gamma_{24} + \tfrac12 )
 ( \gamma_{24} + \tfrac32 )
 - ( \gamma_{24} + \tfrac12 )( h_3 + h_2 + \tfrac12 )+ \tfrac{h_2}{2} + h_3 h_2 ~.
\end{align}
With two independent solutions to the above equations,
we can write down the conformal block as 
\begin{align}
{\cal G}^S_{i} &=
  z^{\frac{1}{2 \kappa} [ (h_1 - h_2) (h_1 - h_2 - \frac12)
   - h_3 (h_3 - \frac12) - h_4 (h_4 - \frac12) ]}
    (1 - z)^{\frac{1}{2 \kappa} [(h_1 - h_4 ) (h_1 - h_4 - \frac12 )
   - h_2 (h_2 - \frac12) - h_3 (h_3 - \frac12)] }
   \nonumber \\ & \qquad \times \left(
f^S_{c,i} (z) \eta (x - z )^{- \gamma_{24} - \frac12 }
 + f^S_{d,i} (z) ( \xi - x \eta ) (x - z )^{- \gamma_{24} - \frac12}
 \right)
\end{align}
with $i=1,2$, which behave like
\begin{align}
 {\cal G}^S_{1} & =   z^{\frac{1}{2 \kappa} [ (h_1 - h_2 - \frac12) (h_1 - h_2 - 1)
   - h_3 (h_3 - \frac12) - h_4 (h_4 - \frac12)  ]}
 \eta x ^{- \gamma_{24} - \frac12 } + \cdots ~, \\
  {\cal G}^S_{2} & = z^{\frac{1}{2 \kappa} [ (h_1 - h_2) (h_1 - h_2 - \frac12)
    - h_3 (h_3 - \frac12)- h_4 (h_4 - \frac12) ]}
(\xi - x \eta ) x^{- \gamma_{24} - \frac12  } + \cdots ~.
\end{align}
The functions are
\begin{align}
 f^S_{c,1} &= z^{\frac{1 - 2 h_1 + 2 h_2 }{4 \kappa} }
 F(\tfrac{ \frac12 - h_1 + h_2 + h_3 - h_4}{4 \kappa}
  ,  \tfrac{ - \frac12 + h_1 + h_2 - h_3 - h_4}{4 \kappa} ,
   \tfrac{1 - 2 h_1 + 2 h_2}{4 \kappa} ; z ) ~, \\
 f^S_{d,1} &= \tfrac{ \frac12 - h_1 + h_2 - h_3 + h_4 }
 { 1 - 2 h _1 + 2 h_2 }  z^{\frac{1 - 2 h_1 + 2 h_2 }{4 \kappa} }
 F(1 - \tfrac{ \frac12 - h_1 - h_2 + h_3 + h_4}{4 \kappa} ,
 \tfrac{ \frac12 - h_1 + h_2 + h_3 - h_4}{4 \kappa}  ,
   1 + \tfrac{1 - 2 h_1 + 2 h_2}{4 \kappa} ; z ) ~, \nonumber \\
 f^S_{c,2} &=
  \tfrac{ - \frac12 + h_1 + h_2 - h_3 - h_4}
 {4 \kappa - 1 + 2 h_1 - 2 h_2 } z
F( 1 - \tfrac{\frac12 -  h_1 + h_2 - h_3 + h_4}{4 \kappa} ,
 1 - \tfrac{\frac32  - 3 h_1 + h_2 + h_3 + h_4}{4 \kappa}  ,
  2 -  \tfrac{1 - 2 h_1 + 2 h_2}{4 \kappa} ; z )  ~, \nonumber \\
 f^S_{d,1} &=
F( - \tfrac{\frac12 -  h_1 + h_2 - h_3 + h_4}{4 \kappa} ,
 1 - \tfrac{\frac32  - 3 h_1 + h_2 + h_3 + h_4}{4 \kappa}  ,
  1 -  \tfrac{1 - 2 h_1 + 2 h_2}{4 \kappa} ; z ) ~. \nonumber
\end{align}
In the $t$-channel expression, functions are given by replacing $h_2 \leftrightarrow h_4$ and $(z,x) \leftrightarrow (1-z,1-x)$.
The relation between two expressions is
\begin{align}
 {\cal G}^S_{1} = e^{- \pi i (\gamma_{24} + \frac12 )}
 ( G^{ST}_{11} {\cal G}^T_{1} + G^{ST}_{12} {\cal G}^T_{2} )~, \qquad
  {\cal G}^S_{2} =  e^{- \pi i (\gamma_{24} + \frac12 )}
 ( G^{ST}_{21} {\cal G}^T_{1} + G^{ST}_{22} {\cal G}^T_{2} )~,
\end{align}
with
\begin{align}
 G^{ST}_{11} &= - \frac{\Gamma ( \frac{1 - 2 h_1 + 2 h_2}{4 \kappa})
\Gamma ( 1 + \frac{ - 1 +2 h_1 - 2 h_4}{4 \kappa}  ) }
 {\Gamma (1 - \frac{\frac12 - h_1 - h_2 + h_3 + h_4}{4 \kappa} )
 \Gamma (\frac{\frac12 - h_1 + h_2 + h_3 - h_4}{4 \kappa} )} ~, \\
 G^{ST}_{12} &=  \frac{ \Gamma (\frac{1 - 2 h_1 + 2 h_2}{4 \kappa})
\Gamma ( \frac{1 - 2 h_1 + 2 h_4}{4 \kappa}  ) }
 {\Gamma ( \frac{\frac{3}{2} - 3 h_1 + h_2 + h_3 + h_4}{4 \kappa} )
 \Gamma (\frac{\frac12 -  h_1 + h_2 - h_3 + h_4}{4 \kappa} )} ~, \\
 G^{ST}_{21} &= \frac{\Gamma (1 - \frac{1 - 2 h_1 + 2 h_2}{4 \kappa})
\Gamma (1 + \frac{- 1 + 2 h_1 - 2 h_4}{4 \kappa}  ) }
 { \Gamma ( 1 - \frac{\frac12  - h_1 + h_2 - h_3 + h_4}{4 \kappa} )
 \Gamma (1-\frac{\frac32 - 3 h_1 + h_2 + h_3 + h_4}{4 \kappa} )} ~, \\
 G^{ST}_{22} &= \frac{ \Gamma (1 - \frac{1 - 2 h_1 + 2 h_2}{4 \kappa})
\Gamma (\frac{  1 - 2 h_1 + 2 h_4}{4 \kappa}  ) }
 {\Gamma ( 1 - \frac{ \frac12 -  h_1 + h_2 + h_3 - h_4}{4 \kappa} )
 \Gamma (\frac{\frac12 -  h_1 - h_2 + h_3 + h_4}{4 \kappa} )} ~.
\end{align}

Now that we have the complete set of conformal blocks, the four point
function of bulk operators can be written down explicitly.
Recall that the operator product expansion involving $\Phi_{k/2}$
has the simple form as in \eqref{oped2}.
Therefore, the four point function $g_4 (z, x , \eta ,\xi)$ can be
expanded with the above conformal blocks as
\begin{align}
 g_4 (z,x,\eta ,\xi ) &= C(h_2 ) C(\tfrac{k}{2} - h_2 , h_3 , h_4)
 |{\cal F}_1^S |^2 -  \tilde  C(h_2 ) \tilde C(\tfrac{k}{2} - h_2 - \tfrac{1}{2}, h_3 , h_4) |{\cal F}_2^S |^2
  \\ \nonumber
 & + \tilde C(h_2 ) C(\tfrac{k}{2} - h_2 - \tfrac{1}{2} , h_3 , h_4)
 |{\cal G}_1^S |^2 +  C(h_2 ) \tilde C(\tfrac{k}{2} - h_2 , h_3 , h_4) |{\cal G}_2^S |^2 ~.
\end{align}
After the modular transformation, the above expression is transfered
into the $t$-channel. The expression is single-valued around $z \sim 1$
only if the three point functions satisfy the constraints
\begin{align}
  \frac{C(h_2) C(\frac{k}{2} - h_2 , h_3 , k_4)}{\tilde C(h_2) \tilde C(\frac{k}{2} - h_2 - \frac12 , h_3 , k_4)} =  \frac{F_{21}^{ST} F_{22}^{ST}}{F_{11}^{ST} F_{12}^{ST}} ~, \qquad
    \frac{\tilde C(h_2) C(\frac{k}{2} - h_2 - \frac12, h_3 , k_4)}{ C(h_2) \tilde C(\frac{k}{2} - h_2  , h_3 , k_4)} = - \frac{G_{21}^{ST} G_{22}^{ST}}{G_{11}^{ST} G_{12}^{ST}} ~.
    \label{const43}
\end{align}
These conditions are not enough to fix uniquely the three point functions 
as \eqref{C}, and we need to make use of another four point function with
a different degenerate operator, such as, $\Phi_{k-1}$. In this case
the operator product expansion is of the form as $\Phi_h \Phi_{k-1} \sim
 {\cal C} [ \Phi_{k-1-h}]_{\text{ee}} + {\cal C} [ \Phi_{h+\frac12}]_{\text{oo}} + {\cal C} [ \Phi_{h} ]_{\text{ee}} $.

\subsection{Basic three point functions of bulk operator}
\label{candtc}

The coefficients $C(h)$ and $\tilde C(h)$ in \eqref{oped2} are
actually important quantities to compute the one point function
of bulk operator on a disk.
Here we obtain them with the information of \eqref{const43}.
These can be computed with the explicit form of three point
function \eqref{C}, but it might be instructive to find them
in a different route. It would be also important if we want to
obtain the three point functions without referring super-Liouville
field theory.

Before using \eqref{const43} we would like to obtain
constraint equations for $C(h)$ and $\tilde C(h)$ from three point functions as analyzed in \cite{GK} for the SL(2) WZNW model.
We consider the correlator
\begin{align}
 \left \langle \Phi_{\frac{k}{2}} (x,\xi)
 \Phi_{h} (y_1 ,\eta_1) \Phi_{\frac{k}{2} - h } (y_2 ,\eta_2)
  \right \rangle ~.
\end{align}
Utilizing the OPE formula \eqref{oped2} we can expand around
$x \sim y_1$ and $x \sim y_2$.
Comparing the two expressions we can find the relation
\begin{align}
 C(\tfrac{k}{2} , h ,\tfrac{k}{2} - h  )
  = C(h) D( \tfrac{k}{2} - h )
  = C ( \tfrac{k}{2} - h ) D(h)  ~.
   \label{ccon1}
\end{align}
Here $D(h)$ is given in \eqref{2ptD}, which appears in the
two point function \eqref{2ptfn}.
In the same way from
\begin{align}
 \left \langle \Phi_{\frac{k}{2}} (x,\xi)
 \Phi_{h} (y_1 ,\eta_1) \Phi_{\frac{k}{2} - h - \frac12} (y_2 ,\eta_2)
  \right \rangle ~,
\end{align}
we would obtain
\begin{align}
 \tilde C(\tfrac{k}{2} , h ,\tfrac{k}{2} - h - \tfrac12 )
  = \tilde C(h) D( \tfrac{k}{2} - h - \tfrac12)
  = \tilde C ( \tfrac{k}{2} - h - \tfrac12 ) D(h)  ~.
  \label{ccon2}
\end{align}

Now is the time to use \eqref{const43}.
Setting $h_2 = h_3 = h , h_1 = h_4 = k/2$, we have
\begin{align}
 \frac{C(h) C(\frac{k}{2} - h,h,\frac{k}{2})}
 { \tilde C(h) \tilde C(\frac{k}{2} - h - \frac12 ,h,\frac{k}{2})}
  = b^4 \left( \gamma ( b^2 (k - 2h - 1)) \right)^2
   \gamma (2 b^2 h ) \gamma (2 b^2 (h - \tfrac12 )) ~.
\end{align}
With the first equations of \eqref{ccon1} and \eqref{ccon2},
this equation leads to
\begin{align}
 \frac{C(h)^2}{\tilde C(h)^2}
  = b^4 \nu^2 \left( \gamma ( b^2 (k - 2h - 1)) \gamma (2 b^2 h)\right)^2 ~.
\end{align}
Combining with the second equations of \eqref{ccon1} and \eqref{ccon2},
we may have
\begin{align}
 C(h) = D(h) ~, \qquad \tilde C(h) = \frac{D(h)}{b^2 \nu  \gamma ( b^2 (k - 2h - 1)) \gamma (2 b^2 h)}
\end{align}
up to a common constant factor. We can show that
the explicit form of three point functions \eqref{C}
also leads to the same expression.

\section{Super AdS$_2$ branes}\label{app:bdyaction}

In this appendix we find the boundary action of the AdS$_2$-like branes in the OSP(1$|$2) WZNW model.

\subsection{The bulk action}

Let us start by stating the bulk action. We follow \cite{HS2} with notation as
\begin{align}\label{bulkaction}
  S_{\text{bulk}}\ = \ \frac{k}{\pi}\int d^2z\ \Bigl[\del\phi\bar\del\phi+e^{-2\phi}(\del\bar\gamma-\bar\theta\del\bar\theta)(\bar\del\gamma-\theta\bar\del\theta)+2e^{-\phi}\bar\del\theta\del\bar\theta\Bigr] \, .
\end{align}
In the free field realization this becomes
\begin{equation}
 \begin{split}
  S_{\text{bulk}}\ &= \ S_0+S_{\text{int}}\, ,\\
S_0\ &= \ \frac{1}{\pi}\int d^2z\ \Bigl[\frac{1}{2}\del\phi\bar\del\phi+\frac{b}{8}\sqrt{g}\mathcal R\phi+\beta\bar\del\gamma+\bar\beta\del\bar\gamma+p\bar\del\theta+\bar p\del\bar\theta\Bigr] \, , \\
S_{\text{int}}\ &= \ -\frac{1}{\pi}\int d^2z\ \Bigl[\frac{1}{k}\beta\bar\beta e^{2b\phi}+\frac{1}{2k}(p+\beta\theta)(\bar p+\bar\beta\bar\theta)e^{b\phi}\Bigr]\, .\\
 \end{split}
\end{equation}
The bulk equation of motion for the auxiliary fields $p$ and $\beta$ are
\begin{equation}\label{eomaux}
\begin{split}
\beta \ &= \ ke^{-2b\phi}(\del\bar\gamma-\bar\theta\del\bar\theta)\, ,\qquad p+\beta\theta\ = \ -2ke^{-b\phi}\del\bar\theta \, , \\
\bar\beta \ &= \ ke^{-2b\phi}(\bar\del\gamma-\theta\bar\del\theta)\, ,\qquad \bar p+\bar\beta\bar\theta\ = \ 2ke^{-b\phi}\bar\del\theta\, . \\
\end{split}
\end{equation}

\subsection{The boundary action}

We use the same parametrization as \cite{HS2}. Further let $t^a$ be the generators of OSP(1$|$2), and denote  the current by
\begin{equation}
	\bar{J}(\bar{z}) \ = kg^{-1}\bar{\del}g \ = \ \sum_a t^a\bar{J}^a \ .
\end{equation}
Then we get
\begin{equation}
	\begin{split}
		\bar{J}^H(\bar{z}) \ &= \
		2k\bar{\del}\phi + 2ke^{-2\phi}\bar{\gamma}\bar{\del}\gamma-
		2k\bar{\gamma}\theta e^{-2\phi}\bar{\del}\theta +\bar{\theta}e^{-\phi}\bar{\del}\theta \, , \\
		\bar{J}^{E^+}(\bar{z}) \ &= \ 	ke^{-2\phi}\bar{\del}\gamma-ke^{-2\phi}\theta\bar{\del}\theta \, , \\
		\bar{J}^{E^-}(\bar{z}) \ &= \ 	
		k\bar{\theta}\bar{\del}\bar{\theta}-ke^{-2\phi}\bar{\gamma}^2\bar{\del}\gamma-2k\bar{\gamma}\bar{\del}\phi+k\bar{\del}\bar{\gamma}
		+ke^{-2\phi}\bar{\gamma}^2\theta\bar{\del}\theta+2ke^{-\phi}\bar{\gamma}\bar{\theta}\bar{\del}\theta \, , \\
		\bar{J}^{F^+}(\bar{z}) \ &= \ 	-2ke^{-2\phi}\bar{\del}\gamma\bar{\theta}-2ke^{-2\phi}\theta\bar{\theta}\bar{\del}\theta
		+2ke^{-\phi}\bar{\del}\theta \, , \\
                \bar{J}^{F^-}(\bar{z}) \ &= \ 	
		2k\bar{\del}\bar{\theta}-2k\bar{\theta}\bar{\del}\phi-2ke^{-2\phi}\bar{\theta}\bar{\gamma}\bar{\del}\gamma
		+2ke^{-2\phi}\bar{\gamma}\bar{\theta}\theta\bar{\del}\theta+2ke^{-\phi}\bar{\gamma}\bar{\del}\theta  \, , \\
	\end{split}
\end{equation}
and similarly for $J$.
We introduce a short-hand notation
\begin{equation}\label{shorthand}
\begin{split}
\beta' \ &= \ ke^{-2b\phi}(\del\bar\gamma-\bar\theta\del\bar\theta)\, ,\qquad p'+\beta'\theta\ = \ -2ke^{-b\phi}\del\bar\theta \, , \\
\bar\beta' \ &= \ ke^{-2b\phi}(\bar\del\gamma-\theta\bar\del\theta)\, ,\qquad \bar p'+\bar\beta'\bar\theta\ = \ 2ke^{-b\phi}\bar\del\theta\, . \\
\end{split}
\end{equation}
Then the gluing conditions for the gluing map given by conjugation with $X^\epsilon$ \eqref{glueautomorphism} read
\begin{equation}
	\begin{split}
		& \bar{\gamma}-\gamma \ = \ ce^\phi -\epsilon\theta\bar{\theta} \, , \qquad
		\beta' \ = \ \bar\beta' \, , \qquad
		p' \ = \ \epsilon\bar p' \, , \\		
 & 4k(\del-\bar{\del})\phi -4\beta' ce^\phi \ = \ (\theta+\epsilon\bar{\theta})(p'-\epsilon\bar p'+\beta'(\theta-\epsilon\bar\theta)) \, , \\
	&	\epsilon4k\bar{\del}\bar{\theta}+4k\del\theta \ = \ 2k(\theta+\epsilon\bar{\theta})(\del+\bar{\del})\phi +ce^\phi(p'-\epsilon\bar p'+\beta'(\theta-\epsilon\bar\theta))\, . \\
	\end{split}
\end{equation}
Using the bulk equations of motion \eqref{eomaux} the gluing conditions have the above form
in the free field formalism just with $\beta',\bar\beta',p',\bar p'$ replaced by $\beta,\bar\beta,p,\bar p$.

The position of the brane is parameterized by the number $c$.
If $c\neq 2$, then we propose the following action
\begin{equation}
	S \ = \ S_{\text{bulk}} \ + \ \frac{k}{2\pi}\frac{1}{2-c} \int du \ e^{-\phi}(\theta+\epsilon\bar{\theta})(\del+\bar{\del})(\theta+\epsilon\bar{\theta}) \, ,
\end{equation}
where $S_{\text{bulk}}$ is \eqref{bulkaction}. Variation of the action under the Dirichlet constraint $\bar{\gamma}-\gamma \ = \ ce^\phi -\epsilon\theta\bar{\theta}$ at the boundary vanishes, if the above gluing conditions hold and the usual bulk EOMs.

In the free field realization this becomes
\begin{equation}
 \begin{split}
  S \ = \ S_{\text{bulk}}+S_{\text{bdy}} \, , 
 \end{split}
\end{equation}
where the interaction term of the bulk action is as before, but the free term gets partially integrated
\begin{equation}
S_0\ = \ \frac{1}{2\pi}\int d^2z\ \Bigl[\frac{1}{2}\del\phi\bar\del\phi+\frac{b}{8}\sqrt{g}\mathcal R\phi-\gamma\bar\del\beta-\bar\gamma\del\bar\beta+\theta\bar\del p+\bar\theta\del\bar p\Bigr]\, .\\
\end{equation}
The boundary term is
\begin{equation}
 \begin{split}
  S_{\text{bdy}} \ &= \ S_{0,\text{bdy}}+S_{\text{int,bdy}} \, , \\
S_{0,\text{bdy}}\ &=\ \frac{1}{2\pi}\frac{k}{2-c}\int du\ \Bigl[\Theta(\del+\bar\del)\Theta\Bigr]+\frac{1}{2\pi}\int du \frac{b}{8}\sqrt{g}\mathcal K \phi \, , \\
S_{\text{int,bdy}}\ &=\ \frac{1}{2\pi}\int du\ \Bigl[ce^{b\phi}\beta+e^{b\phi/2}\Theta(\beta\frac{1}{2}(\theta-\epsilon\bar\theta)+p)\Bigr] \, .\\
 \end{split}
\end{equation}
Here, we introduced the additional fermionic boundary degree of freedom $\Theta$. Furthermore, we imposed the Dirichlet conditions
\begin{equation}
\beta \ = \ \bar\beta \qquad\text{and}\qquad p \ = \ \epsilon\bar p
\end{equation}
at the boundary. Integrating the auxiliary fields $\beta, \bar\beta, p,\bar p$ and $\Theta$ gives the old action. Moreover the variation of the boundary action vanishes provided our choice of gluing conditions for the currents holds.
Note also the extra boundary linear dilaton term coming with the geodesic curvature $\mathcal K$ normal to the boundary.

Finding a similar formalism for the fuzzy-spherical branes seems to be very difficult. One of the reasons is that in the first order form one has boundary conditions of the form $p=\bar\del\bar\theta$ which were considered in \cite{ghost}.

\section{Two point functions of boundary operator}
\label{app:b2pt}

In the presence of boundary, we can insert boundary operators and
consider their correlation functions.
We assign boundary conditions corresponding to the super AdS$_2$ branes,
then the boundary operators can be defined by $\Psi_l^{\rho , \rho '}
(t , \eta | u) = {\cal V}_l^{\rho , \rho '}
(t  | u) +  \eta   {\cal W}_l^{\rho , \rho '} (t| u)$
as explained in subsection \ref{sec:aaopen}.
In this appendix, we compute the following two types of two point functions.
The first type is the two point function of primary operators
\begin{align}
 & \left \langle {\cal V}^{\rho , \rho '}_{l_1} (1  | 1)
{\cal V}^{\rho ' , \rho}_{l_2} (0  | 0) \right \rangle
 = \delta (l_1 - l_2 ) d(l_1 ; \rho , \rho ')~,
 \label{first2}
\end{align}
where we set $\eta_i = 0$ just for simplicity.
The second type is the two point function of descendants,
and its form is assumed to be
\begin{align}
 & \left \langle p {\cal V}^{\rho , \rho ' }_{l_1} (1  | 1)
p {\cal V}^{\rho ' , \rho}_{l_2} (0 | 0) \right \rangle
 =\delta (l_1 - l_2 ) d ' (l_1 ; \rho , \rho ' ) ~.
 \label{second2}
\end{align}
The function $d ' (l_1 ; \rho , \rho ' )$ could be different
from the one for primaries.
In order to compute $d(l_1 ; \rho , \rho ')$ and
$d ' (l_1 ; \rho , \rho ' )$, we make use of  a degenerate operator
${\cal V}^{\rho ,\rho '}_{-1/2} (t | u)$.
Contrary to the case of bulk operator, the boundary operator
with $l = -1/2$ is not always degenerate and the situation depends
on the boundary conditions $\rho , \rho '$.
Here we assume that the operator is degenerate when
$\rho ' = \rho \pm i/2$, and we will see it is indeed the case.
See \cite{FZZb,FH} for more details.

\subsection{First type of two point function}

First we study the two point function of primary operators
\eqref{first2}. We obtain a constraint equation for the
coefficient $d(l_1 ; \rho , \rho ')$ by utilizing
the three point function with the insertion of
 a degenerate boundary operator
${\cal V}_{ - 1/2}^{\rho , \rho '} $.
The operator product expansion involving the operator is given as
\begin{align}
 &{\cal V}^{\rho '' , \rho ' }_{- \frac12} (t_1 | u_1)
{\cal V}^{\rho ' , \rho }_{l} (t_2 | u_2)
 = | u_{12} |^{2 b^2 l } c_+ (l)
 [ {\cal V}_{l - \frac12} ^{\rho '' , \rho }
 (t_2  | u_2)]_\text{e} \\ & \qquad
 + |u_{12}|^{  b^2 } \tilde c_0 (l) [ {\cal W}_{l }^{\rho '' , \rho }
 (t_2  | u_2)]_\text{o}
  + |u_{12}|^{  b^2 - 2 b^2 l } c_- (l)
  t_{12}   [ {\cal V}_{l + \frac12}^{\rho '' , \rho }
(t_2  | u_2) ]_\text{e} ~.
 \nonumber
\end{align}
Taking different limits of the three point function
$ \langle {\cal V}_{ - \frac12}^{\rho '' , \rho '} {\cal V}_{l}
 ^{\rho ' , \rho }  {\cal V}_{l + \frac12}
 ^{\rho  , \rho  '' } \rangle$, we have a constraint equation
\begin{align}
  c_- (l) d(l + \tfrac12 ; \rho '' , \rho) = c_+ (l + \tfrac12)
   d (l ; \rho ' , \rho) ~.
   \label{b2con}
\end{align}
We can obtain the explicit form of $d (l ; \rho ' , \rho)$
by solving this equation, however in order to do so we need to
know $c_- (l)$ and $c_+(l)$ in the above
operator product expansion.

In the following we compute $c_- (l), c_+(l)$ by utilizing a free
field realization of OSP(1$|$2) model developed in 
appendix \ref{app:bdyaction}.
In the free field formulation the interaction terms may be given by%
\footnote{Here we change the notation as $\bar \gamma \to - \bar \gamma $
and $\bar \beta \to - \bar \beta$ from the previous one.}
\begin{align}
 S_\text{int} = i \lambda
 \int d ^2z \left[
 (p + \beta \theta )
 (\bar p - \bar \beta \bar \theta) e^{b \phi} \right] +
  \lambda_B   \int d u
 \left[ e^{b \phi /2} \Theta
 (\beta  \theta + p ) \right] ~ .
  \label{freeint}
\end{align}
Here we omit the terms with $\beta \bar \beta e^{2 b \phi}$
and $\beta e^{ b \phi } $ since they can be generated from the
other terms.
Treating the interaction terms perturbatively,
the boundary conditions are set as
\begin{align}
 \beta = - \bar \beta ~, \qquad
 \gamma = - \bar \gamma ~ , \qquad
 p = \zeta \bar p  ~ , \qquad
 \theta = \zeta \bar \theta ~ .
\end{align}
With this type of free field realization, vertex operators
are usually given in the $m$-basis.
However, in our case, the expression in the $x$-basis is suitable,
which was actually discussed in \cite{HOS} for the SL(2) WZNW model.
Following their arguments we use the leading part as
\begin{align}
 {\cal V}_l (t  | u) =
| \gamma - t  |^{- 2 l} e^{ b l \phi} ~.
\end{align}
The overall factor is just our convention.
This convention is equivalent to set $c_+ (l) = 1$ since we have
\begin{align}
 {\cal V}_{- \frac12} (t_1  | u_1)
 {\cal V}_{ l} (t_2  | u_2)
  \sim | u_{12} |^{2 b^2 l } | \gamma - t_2  |^{- 2 l + 1}
  e^{ b (l - 1/2 ) \phi} (u_2 ) + \cdots
\end{align}
for $t_1 \to t_2 , u_1 \to u_2$.
See section 4 of \cite{GK} for similar calculations.

Similarly we compute $c_- (l)$ using the
operator product expansions of free fields
\begin{align}
 \beta (z) \gamma (w ) \sim - \frac{1}{z - w} ~, \qquad
 p (z) \theta (w) \sim \frac{1}{z - w} ~, \qquad
 \phi (z , \bar z ) \phi (w , \bar w) \sim - \ln |z - w | ^2 ~ .
\end{align}
Anti-holomorphic part is treated by utilizing the mirror trick
and operator product expansions are given in a similar manner.
In particular, the scalar field $\phi (u)$ inserted at the
boundary has the OPE relation as
$\phi (u_1) \phi (u_2) \sim - 4 \ln | u_{12}|$.
For the calculation of $c_- (l)$ we have to include
the effect of the interaction terms. Since there are both
bulk and boundary interaction terms, we separate the function
as $c_- (l) = c^{(v)}_- (l) + c^{(b)}_- (l)$, where
$c^{(v)}_- (l)$ and $c^{(b)}_- (l)$ denote
the contributions with bulk and boundary interactions, respectively.

When the bulk interaction term contributes,
we have to evaluate
\begin{align}
 - i \lambda \int d^2 z (p + \beta \theta)
  (\bar p - \bar \beta \bar \theta ) e^{b \phi } (z)
   | \gamma - t_1 | e^{ - \frac12 b \phi } (u_1 )
   | \gamma - t_2 |^{- 2 l} e^{ b l \phi } (u_2 ) ~,
\end{align}
where the integration is over the upper half plane $\text{Im}\, z \geq 0$.
Applying the Wick contraction of free fields, it reduces to
\begin{align}
 2 l \lambda \zeta | u_{12} |^{2 b^2 l } t_{12} 
 \int d^2 z \frac{| u_1 - z |^{2 b^2}}
 {|z - u_2 |^{4 b^2 l} | z - \bar z |^{b^2 + 1}}
  \left( \frac{1}{z - u_2 } + \frac{1}{\bar z - u_2} \right)
  |\gamma - t_2|^{- 2 l - 1} e^{ b (l + \frac12) \phi}~,
\end{align}
if we pick up the terms proportional to $t_{12}$.
Non-trivial contribution arises from the  
$(\beta , \gamma)$-system, which can be computed as
\begin{align}
 \beta (z) | \gamma (u_1) - t_2 - t_{12} |
  | \gamma (u_2 ) - t_2|^{- 2 l} \sim
  \frac{ - 2l}{z - u_2}  t_{12} 
  | \gamma (u_2 ) - t_2|^{- 2 l - 1}
\end{align}
for the term we want.
In the following we may set $u_1 = 1, u_2 = 0$ for simplicity.
{}From the above expression we can see that
 the bulk contribution $c^{(v)}_-$ is written in a integral form as
\begin{align}
 c_-^{(v)} =  2 l \lambda \zeta
 \int d^2 z \frac{| 1 - z |^{2 b^2}}
 {|z |^{4 b^2 l} | z - \bar z |^{b^2 + 1}}
  \left( \frac{1}{z} + \frac{1}{\bar z} \right) ~ .
  \label{bulkc}
\end{align}

When the boundary interaction terms contribute,
we have to compute
\begin{align}
 \sum_{i,j = 1}^3
  \frac{\lambda_B^i \lambda_B^j }{2}
 \int_{ {\cal C}_i} d x_1 \int_{ {\cal C}_j} dx_2
 \Theta (p + \beta \theta) e^{b \phi/2  } (x_1)
  \Theta (p + \beta \theta) e^{b \phi/2  } (x_2)  \\
 \times
   | \gamma - t_1 | e^{ - \frac12 b \phi } (1 )
   | \gamma - t_2 |^{- 2 l} e^{ b l \phi } (0 ) ~.
   \nonumber
\end{align}
Now that we inserted boundary operators
at $u_1 = 1$ and $u_2 = 0$, the boundary conditions for the regions
${\cal C}_1 = [ - \infty , 0]$,
${\cal C}_2 = [ 0 , 1]$, ${\cal C}_3 = [ 0, \infty ]$ are different.
Boundary conditions are related to the parameter $\lambda_B$ in
the boundary interaction term of \eqref{freeint}, and we use
$\lambda_B^i$ with $i=1,2,3$ for the parameter in the each region.
As in the same way as $c_-^{(v)}$ we  obtain
\begin{align}
 c_-^{(b)} = 2l  \sum_{i,j = 1}^3
  \frac{\lambda_B^i \lambda_B^j }{2}
 \int_{ {\cal C}_i} d x_1 \int_{ {\cal C}_j} dx_2
  \frac{| ( 1 - x_1 )(1 - x_2) |^{b^2}}
 {| x_1 x_2 |^{2 b^2 l} | x_1 - x_2 |^{b^2 + 1}}
  \left( \frac{1}{x_1} + \frac{1}{ x_2} \right) ~.
  \label{boundaryc}
\end{align}
As seen above, the coefficient $c_- (l)$ can be written in terms
of integral, and fortunately the integrals can be performed explicitly
as in \cite{FZZb,FH}.

Now the problem is to compute
the following form of the integral as
\begin{align}
  J = \int d x_1 \int dx_2
  \frac{| ( 1 - x_1 )(1 - x_2) |^{b^2}}
 {|  x_1  x_2   |^{2 b^2 l}
 | x_1 - x_2 |^{b^2 + 1}}
 \left( \frac{1}{x_1} + \frac{1}{ x_2} \right) ~.
 \label{intJ}
\end{align}
For the purpose it is  convenient to rewrite
\begin{align}
 J = \frac{1}{2 b^2 l} \partial_u I(u) |_{ u = 0 } ~, \qquad
 I (u) =   \int d x_1 \int dx_2
  \frac{| ( 1 - x_1 )(1 - x_2) |^{b^2}}
 {| ( x_1 - u ) ( x_2 - u ) |^{2 b^2 l}
 | x_1 - x_2 |^{b^2 + 1}} ~.
\end{align}
Since the integral $I(u)$ can be reduced to
\begin{align}
 I (u) = (1-u)^{b^2 - 4 b^2 l + 1} I (0) ~,
\end{align}
we obtain a simple formula as
\begin{align}
 J = \frac{ 4 b^2 l  - b^2 - 1 }{2 b^2 l} I (0 ) ~.
\end{align}
In this way we have observed that the integral $J$ is related to
another integral $I(0)$ in a simple way.
This fact is quite nice since $I(0)$ have been computed in
the last part of section 3 in \cite{FH}, therefore we can
borrow their results. It is also possible to
reproduce their results by explicit calculations.

With the above integral formula, we can obtain the explicit
expression of $c_-(l)$. The contribution with bulk interaction term
is
\begin{align}
 c_-^{(v)} =
 - \lambda \zeta I_0 \sin (\pi b^2) \sin ^2 (2 \pi b^2 l)
\end{align}
with
\begin{align}
 I_0 = 2 l  \frac{\gamma(\frac{1}{2}(1 + b^2))}
{ \pi  \sin \pi b^2} \Gamma (  - 2 b^2 l) \Gamma (2 b^2 l)
 \Gamma( \tfrac{1}{2} - \tfrac{b^2}{2} + 2 b^2 l)
  \Gamma ( \tfrac{1}{2} - \tfrac{b^2}{2} - 2 b^2 l) ~.
\end{align}
The contribution with boundary interaction terms is
\begin{align}
 c_-^{(b)}
  &= I_0 \Bigl( - ( \lambda^1_B )^2 \sin \pi b^2  \cos \tfrac{\pi b^2}{2}
   - (\lambda_B^2)^2 \sin 2 \pi b^2 l
 \cos \pi (2 b^2 l - \tfrac{b^2}{2})  \\
   &+ (\lambda_B^3)^2 \sin 2 \pi b^2 l
 \cos \pi (2 b^2 l + \tfrac{b^2}{2})
   + \lambda_B^1 \lambda_B^2 \sin \pi b^2
 \cos \pi (2 b^2 l - \tfrac{b^2}{2}) \nonumber \\
   &- \lambda_B^1 \lambda_B^3 \sin \pi b^2
 \cos \pi (2 b^2 l + \tfrac{b^2}{2})
 + \lambda^2_B \lambda_B^3 2 \sin \tfrac{ \pi b^2 }{2}
   \cos \pi (2 b^2 l - \tfrac{b^2}{2})\cos \pi (2 b^2 l + \tfrac{b^2}{2}) \Bigr) ~.\nonumber
\end{align}
We relate the parameters of boundary condition $\lambda_B$ and $\rho$
by the following formula as
\begin{align}
 ( \lambda_B^i )^2 \cos \tfrac{\pi b^2}{2} =
 \lambda  ( \sinh 2 \pi b^2 \rho_i )^2 ~, \qquad
  ( \lambda_B^i )^2 \cos \tfrac{\pi b^2}{2} =
 \lambda  ( \cosh 2 \pi b^2 \rho_i )^2 
\end{align}
for $\zeta = + 1$ and $\zeta = - 1$, respectively.
Here we set $\rho_1 = \rho , \rho_2 = \rho ' ,
\rho _3 = \rho '' = \rho ' - i/2$.
For $\zeta = + 1$, we then find
\begin{align}
 c_- (l) =& - 4 \lambda I_0 \sin \pi b^2
 \cos b^2 \pi (i \rho ' + i \rho + l)
 \sin b^2 \pi (i \rho ' - i \rho + l)  \\ & \times
 \sin b^2 \pi (i \rho ' + i \rho + l + \tfrac12)
 \cos b^2 \pi (i \rho ' - i \rho + l + \tfrac12)  ~, \nonumber
\end{align}
and for $\zeta = - 1$
\begin{align}
 c_- (l) =&  - 4 \lambda I_0 \sin \pi b^2
 \sin b^2 \pi (i \rho ' + i \rho + l)
 \sin b^2 \pi (i \rho ' - i \rho + l)  \\ & \times
 \cos b^2 \pi (i \rho ' + i \rho + l + \tfrac12)
 \cos b^2 \pi (i \rho ' - i \rho + l + \tfrac12)  ~. \nonumber
\end{align}
When we set $\rho '' = \rho ' + i/2$, we again obtain
$c_- (l)$ as above but with $\rho \to - \rho$,
$\rho ' \to - \rho '$.

Now that we know $c_+(l)$ and $c_-(l)$,
we have prepared for solving the constraint equation \eqref{b2con}.
A solution for the coefficient $d(l;\rho , \rho ')$ to the
constraint
can be written in terms of special functions
${\bf G}(x)$ and ${ \bf S }(x)$. With $Q = b + 1/b$,
they are defined by \cite{FZZb}
\begin{align}
 &\ln {\bf G} (x) = \int_0^\infty \frac{dt}{t}
 \left[ \frac{e^{- Qt/2} - e^{-xt}}{(1 - e^{-bt}) (1 - e^{ - t/b})}
  + \frac{(Q/2 -x)^2}{2} e^{- t}
 + \frac{(Q/2 -x)}{t} \right] ~, \\
 &\ln {\bf S} (x) = \int_0^\infty \frac{dt}{t}
 \left[ \frac{\sinh (Q /2 - x) t}{2 \sinh bt/2 \sinh t/2b}
 - \frac{(Q -2 x)}{ t} \right] ~,
\end{align}
and behaves under shifts as
\begin{align}
 &{\bf G} ( x + b ) = \frac{ b^{1/2 - b x} }{ \sqrt{2 \pi}}
  \Gamma (b x) {\bf G} (x) ~, \qquad
 {\bf G} ( x + 1/b ) = \frac{ b^{x/b - 1/2} }{ \sqrt{2 \pi}}
  \Gamma (x/b) {\bf G} (x) ~, \\
  & {\bf S} ( x + b ) = 2 \sin ( \pi b x ) {\bf S} (x) ~, \qquad
 {\bf S} ( x + 1/b ) = 2 \sin ( \pi x / b )  {\bf S} (x) ~.
\end{align}
It is also useful to define as in \cite{FH}
\begin{align}
 &{\bf G}_\text{NS} (x ) =  {\bf G} (\tfrac{x}{2} )
 {\bf G} (\tfrac{x+Q}{2} ) ~, \qquad
   {\bf G} _\text{R}(x) =  {\bf G} (\tfrac{x + b}{2} )
 {\bf G} (\tfrac{x+ b^{-1}}{2} ) ~, \\
 &{\bf S}_\text{NS} (x ) =  {\bf S} (\tfrac{x}{2} )
 {\bf S} (\tfrac{x+Q}{2} ) ~, \qquad
   {\bf S}_\text{R} (x) =  {\bf S} (\tfrac{x + b}{2} )
 {\bf S} (\tfrac{x+ b^{-1}}{2} ) ~.
\end{align}
For $\zeta = + 1$ a solution for the coefficient $d(l;\rho ' , \rho)$
is
\begin{align}
\label{d-}
&d(l;\rho ' , \rho)
=  \Gamma ( 1 - 2 l )
 (\pi \lambda \gamma ( \tfrac12 (1 + b^2)) b^{- 1 -b^2})^{ - 2 (l - \frac14)} {\bf G}_\text{R} (b - 4 b l)
 {\bf G}_\text{R} (4 b l - b) ^{-1} \\ &  \times
[ {\bf S}_\text{R} (2 b (l + i \rho + i \rho '))
 {\bf S}_\text{NS} (2 b (l - i \rho + i \rho '))
 {\bf S}_\text{R} (2 b (l - i \rho - i \rho '))
 {\bf S}_\text{NS} (2 b (l + i \rho - i \rho ')) ]^{-1} ~.
 \nonumber
\end{align}
For  $\zeta = - 1$ it is given by
\begin{align} \label{d+}
&d(l;\rho ' , \rho)
= \Gamma ( 1 - 2 l)
 (\pi \lambda \gamma ( \tfrac12 (1 + b^2)) b^{- 1 -b^2})
^{ - 2 (l - \frac14)} {\bf G}_\text{R} (b - 4 b l)
 {\bf G}_\text{R} (4 b l - b) ^{-1} \\ &  \times
  [ {\bf S}_\text{NS} (2 b (l + i \rho + i \rho '))
 {\bf S}_\text{NS} (2 b (l - i \rho + i \rho '))
 {\bf S}_\text{NS} (2 b (l - i \rho - i \rho '))
 {\bf S}_\text{NS} (2 b (l + i \rho - i \rho ')) ] ^{-1} .
\nonumber
\end{align}
In section \ref{sec:aaopen} we compute the spectral density of open string
by using the two point functions. In particular, the dependence
of boundary condition would be important.

\subsection{Second type of two point function}

In order to compute the spectral density of open string, we also
need the coefficient $d ' (l ; \rho , \rho ')$ in \eqref{second2}.
The function can be computed in a similar way as before.
In this case the relevant OPE is
\begin{align}
 &{\cal V}^{\rho '' , \rho ' }_{- \frac12} (t_1 | u_1)
p{\cal V}^{\rho ' , \rho }_{l} (t_2 | u_2)
 = | u_{12} |^{2 b^2 l } \tilde  c_+ (l)
 [ p{\cal V}_{l - \frac12} ^{\rho '' , \rho }
 (t_2  | u_2)]_\text{o} \\ & \qquad
 + |u_{12}|^{  b^2 - 1 } c_0 (l) [ {\cal V}_{l }^{\rho '' , \rho }
 (t_2 | u_2)]_\text{e}
  + |u_{12}|^{  b^2 - 2 b^2 l } \tilde c_- (l)
  t_{12}  [ p {\cal V}_{l + \frac12}^{\rho '' , \rho }
(t_2  | u_2) ]_\text{o} 
 \nonumber
\end{align}
with $\rho '' = \rho ' - i/2$.
Taking different limits of three point function
$ \langle {\cal V}_{ - \frac12}^{\rho '' , \rho '} p{\cal V}_{l}
 ^{\rho ' , \rho }  p {\cal V}_{l + \frac12}
 ^{\rho  , \rho  '' } \rangle$, we have
\begin{align}
  \tilde c_- (l) d '(l + \tfrac12 ; \rho '' , \rho)
 = \tilde c_+ (l + \tfrac12)
   d ' (l ; \rho ' , \rho) ~.
   \label{b2con2}
\end{align}
First problem is therefore to find out
the explicit forms of $\tilde c_+(l)$ and $\tilde c_-(l)$.
After that we solve the constraint equation
in order to obtain $d ' (l ; \rho ' , \rho)$.

As before we express the vertex operator in terms of free fields,
and now we adopt
\begin{align}
 p {\cal V}_l (t| u) = | \gamma - t |^{-2 l } p e^{bl\phi} ~.
\end{align}
This overall normalization is the same as the requirement
$\tilde  c_+ (l) = 1$ since
\begin{align}
 {\cal V}_{- \frac12} (t_1  | u_1)
 p {\cal V}_{ l} (t_2  | u_2)
  \sim | u_{12} |^{2 b^2 l } | \gamma - t_2  |^{- 2 l + 1}
  p e^{ b (l - 1/2 ) \phi} (u_2 ) + \cdots
\end{align}
for $t_1 \to t_2 , u_1 \to u_2$.
A difficult part is the computation of $\tilde c_- (l)$,
which we divide as $\tilde c_- (l) = \tilde c^{(v)}_- (l) +
\tilde c^{(b)}_- (l)$ as before.
The contribution with the bulk interaction is given by
\begin{align}
 - i \lambda \int d^2 z (p + \beta \theta)
  (\bar p - \bar \beta \bar \theta ) e^{b \phi } (z)
   | \gamma - t_1 | e^{ - \frac12 b \phi } (1 ) p
   | \gamma - t_2 |^{- 2 l} e^{ b l \phi } (0 ) ~,
\end{align}
thus we have
\begin{align}
 \tilde c_-^{(v)} =  2 l \lambda \zeta
 \int d^2 z \frac{| 1 - z |^{2 b^2}}
 {|z |^{4 b^2 l} | z - \bar z |^{b^2 + 1 }}\left(
 1 + \frac{( z - \bar z )^2}{z \bar z }
   \right) \left( \frac{1}{z} + \frac{1}{\bar z}
  \right) ~.
\end{align}
For the boundary contribution we have to calculate
\begin{align}
 \sum_{i,j = 1}^3
  \frac{\lambda_B^i \lambda_B^j }{2}
 \int_{ {\cal C}_i} d x_1 \int_{ {\cal C}_j} dx_2
 \Theta (p + \beta \theta) e^{b /2 \phi } (x_1)
  \Theta (p + \beta \theta) e^{b /2 \phi } (x_2)  \\
 \times
   | \gamma - t_1 | e^{ - \frac12 b \phi } (1 )
  p | \gamma - t_2 |^{- 2 l} e^{ b l \phi } (0 ) ~,
   \nonumber
\end{align}
which leads to
\begin{align}
\tilde  c_-^{(b)} = 2l  \sum_{i,j = 1}^3
  \frac{\lambda_B^i \lambda_B^j }{2}
 \int_{ {\cal C}_i} d x_1 \int_{ {\cal C}_j} dx_2
  \frac{| ( 1 - x_1 )(1 - x_2) |^{b^2}}
 {| x_1 x_2 |^{2 b^2 l} | x_1 - x_2 |^{b^2 + 1}}
   \left( 1 +  \frac{( x_1 - x_2 )^2 }{ x_1 x_2} \right)
  \left( \frac{1}{x_1} + \frac{1}{ x_2} \right) ~.
\end{align}
The integrals can be divided into two parts, 
and the first terms are the same as before. 
Therefore we just need to examine the second terms.

Notice that the second terms are of the form
\begin{align}
  \tilde J  = \int d x_1 \int dx_2
  \frac{| ( 1 - x_1 )(1 - x_2) |^{b^2}}
 {|  x_1  x_2   |^{2 b^2 l + 1}
 | x_1 - x_2 |^{b^2 - 1}}
 \left( \frac{1}{x_1} + \frac{1}{ x_2} \right) ~.
\end{align}
Therefore we can reduce them into more simpler integrals as
\begin{align}
\tilde J  = \frac{ 4 b^2 l - b^2 - 1 }{2 b^2 l + 1} \tilde I  (0 ) ~, \qquad
 \tilde I  (0) =   \int d x_1 \int dx_2
  \frac{| ( 1 - x_1 )(1 - x_2) |^{b^2}}
 {| x_1  x_2 |^{2 b^2 l + 1}
 | x_1 - x_2 |^{b^2 - 1}} ~.
\end{align}
The integrals of $\tilde I(0)$ have not been calculated  in \cite{FH}, 
so we have to do it by ourselves.
By combining the first term and the second term in the integral,
we obtain the explicit form of the bulk contribution as
\begin{align}
 \tilde c_-^{(v)} =
 - \lambda \zeta \tilde I_0  \sin (\pi b^2) \sin ^2 (2 \pi b^2 l)
\end{align}
with
\begin{align}
 \tilde I_0  &= 2 l  \left( 1 - \frac{b^2 ( 1 + b^2 )}{(2 b^2 l)^2 - 1 }
\right)\frac{\gamma(\frac{1}{2}(1 + b^2))}
{ \pi  \sin \pi b^2}
 \Gamma (  - 2 b^2 l) \Gamma (2 b^2 l)
 \Gamma( \tfrac{1}{2} - \tfrac{b^2}{2} + 2 b^2 l)
  \Gamma ( \tfrac{1}{2} - \tfrac{b^2}{2} - 2 b^2 l) ~.
\end{align}
The contribution with the boundary interaction terms is
\begin{align}
 \tilde c_-^{(b)}
  &= \tilde I_0  \Bigl( - ( \lambda^1_B )^2 \sin \pi b^2  \cos \tfrac{\pi b^2}{2}
   - (\lambda_B^2)^2 \sin 2 \pi b^2 l
 \cos \pi (2 b^2 l - \tfrac{b^2}{2})  \\
   &+ (\lambda_B^3)^2 \sin 2 \pi b^2 l
 \cos \pi (2 b^2 l + \tfrac{b^2}{2})
   - \lambda_B^1 \lambda_B^2 \sin \pi b^2
 \cos \pi (2 b^2 l - \tfrac{b^2}{2}) \nonumber \\
   &+ \lambda_B^1 \lambda_B^3 \sin \pi b^2
 \cos \pi (2 b^2 l + \tfrac{b^2}{2})
 + \lambda^2_B \lambda_B^3 2 \sin \tfrac{ \pi b^2 }{2}
   \cos \pi (2 b^2 l - \tfrac{b^2}{2})\cos \pi (2 b^2 l + \tfrac{b^2}{2}) \Bigr) ~.\nonumber
\end{align}
Compared to the previous result, we have opposite signs as
$\lambda_B^1 \lambda_B^2 \to - \lambda_B^1 \lambda_B^2 $ and
$\lambda_B^1 \lambda_B^3 \to - \lambda_B^1 \lambda_B^3 $,%
\footnote{If we compute the boundary contributions naively, then
we have the same signs as before. We change the signs according
to the prescription for ${\cal N}=1$ super-Liouville field theory in \cite{FH,Fukuda}.} along with the different factor in $\tilde I_0$.
For $\zeta = + 1$, we then find
\begin{align}
 \tilde c_- (l) =& - 4 \lambda \tilde I_0 \sin \pi b^2
 \sin b^2 \pi (i \rho ' + i \rho + l)
 \cos b^2 \pi (i \rho ' - i \rho + l)  \\ & \times
 \cos b^2 \pi (i \rho ' + i \rho + l + \tfrac12)
 \sin b^2 \pi (i \rho ' - i \rho + l + \tfrac12)  ~, \nonumber
\end{align}
and for $\zeta = - 1$
\begin{align}
 \tilde c_- (l) =&  - 4 \lambda \tilde I_0  \sin \pi b^2
 \cos b^2 \pi (i \rho ' + i \rho + l)
 \cos b^2 \pi (i \rho ' - i \rho + l)  \\ & \times
 \sin b^2 \pi (i \rho ' + i \rho + l + \tfrac12)
 \sin b^2 \pi (i \rho ' - i \rho + l + \tfrac12)  ~. \nonumber
\end{align}
When we set $\rho '' = \rho ' + i/2$, we again obtain
$\tilde c_- (l)$ as above but with $\rho \to - \rho$,
$\rho ' \to - \rho '$.

Using the explicit form of $\tilde c_+(l)$ and $\tilde c_-(l)$
we solve the constraint equation \eqref{b2con2} for
$d ' (l;\rho ' , \rho )$.
A solution may be given by
\begin{align}
\nonumber
&d ' (l;\rho ' , \rho)
=  g(l) \Gamma ( 1 - 2 l )
 (\pi \lambda \gamma ( \tfrac12 (1 + b^2)) b^{- 1 -b^2})^{ - 2 (l - \frac14)} {\bf G}_\text{R} (b - 4 b l)
 {\bf G}_\text{R} (4 b l - b) ^{-1} \\ & \quad \times
 [ {\bf S}_\text{NS} (2 b (l + i \rho + i \rho '))
 {\bf S}_\text{R} (2 b (l - i \rho + i \rho '))
 {\bf S}_\text{NS} (2 b (l - i \rho - i \rho '))
 {\bf S}_\text{R} (2 b (l + i \rho - i \rho ')) ]^{-1}
\label{dprime-}
\end{align}
for $\zeta = + 1$ and
\begin{align}
 \nonumber
&d ' (l;\rho ' , \rho)
=  g(l) \Gamma ( 1 - 2 l)
 (\pi \lambda \gamma ( \tfrac12 (1 + b^2)) b^{- 1 -b^2})
^{ - 2 (l - \frac14)} {\bf G}_\text{R} (b - 4 b l)
 {\bf G}_\text{R} (4 b l - b) ^{-1}  \\ &  \quad  \times
 [ {\bf S}_\text{R} (2 b (l + i \rho + i \rho '))
 {\bf S}_\text{R} (2 b (l - i \rho + i \rho '))
 {\bf S}_\text{R} (2 b (l - i \rho - i \rho '))
 {\bf S}_\text{R} (2 b (l + i \rho - i \rho ')) ]^{-1}
 \label{dprime+}
\end{align}
for $\zeta = - 1$.
The factor $g(l)$ depends on the notation, and in the
present case it is given by a solution to
\begin{align}
 \frac{g(l ) }{ g(l + 1/2) }
 = 1 - \frac{ b^2 ( 1 + b^2) }{ (2 b^2 l )^2 - 1 } ~.
\end{align}
Thus we may use
\begin{align}
 g(l) = \frac{\Gamma (2 l - b^{-2}) \Gamma (2 l + b^{-2}) }{\Gamma (2 l + \sqrt{1 + b^{-2} + b^{-4}})
\Gamma (2 l - \sqrt{1 + b^{-2} + b^{-4}})} ~.
\end{align}

\end{document}